\documentclass[aps,notitlepage,nofootinbib,superscriptaddress,twocolumn,longbibliography,10pt]{revtex4-2}

\usepackage[normalem]{ulem}

\newcommand{\CE}{\mathcal{E}}

\newcommand{\beq}{\begin{eqnarray}}
\newcommand{\eeq}{\end{eqnarray}}
\newcommand{\beqq}{\begin{eqnarray*}}
\newcommand{\eeqq}{\end{eqnarray*}}

\newcommand{\be}{\begin{equation}}
\newcommand{\ee}{\end{equation}}
\newcommand{\barr}{\begin{array}}
\newcommand{\earr}{\end{array}}
\newcommand{\Tr}{\text{Tr}}
\newcommand{\tr}{\text{tr}}

\newcommand{\bra}[1]{{\langle #1|}}
\newcommand{\ket}[1]{{|#1 \rangle}}
\def\braket#1#2{\left\langle #1|#2\right\rangle}

\def\({\left(}
\def\){\right)}
\newcommand{\figref}[1]{Fig.\,\ref{#1}}
\newcommand{\eqnref}[1]{Eq.\,\eqref{#1}}
\newcommand{\secref}[1]{Sec.\,\ref{#1}}
\newcommand{\supp}{\mathop{\operatorname{supp}}}
\newcommand{\Bra}[1]{{\langle\!\langle #1|}}
\newcommand{\Ket}[1]{{|#1 \rangle\!\rangle}}
\newcommand{\eq}[1]{\begin{equation}#1\end{equation}}
\newcommand{\eqs}[1]{\begin{equation}\begin{split}#1\end{split}\end{equation}}
\newcommand{\arcosh}{\mathrm{arcosh}\,}

\newcommand{\bulkC}{\mathsf{bulkC}}
\newcommand{\bdryC}{\mathsf{bdryC}}
\newcommand{\minC}{\mathsf{minC}}
\def\parfig#1#2{
\parbox{#1\textwidth}
{\includegraphics[width=#1\textwidth]{#2}}
}

\usepackage{booktabs}

\usepackage{amsmath}
\usepackage{amsthm, amssymb}

\usepackage[colorlinks=true,citecolor=blue,linkcolor=blue]{hyperref}
\usepackage{graphicx}
\usepackage{dcolumn}
\usepackage{bm}
\usepackage{color}
\usepackage{tabularx}
\usepackage{comment}
\usepackage{float}
\usepackage{amsmath}
\usepackage{gensymb}
\usepackage{mathtools}
\usepackage{tikz-cd}
\usepackage{adjustbox}
\usepackage{dsfont}
\usepackage[small]{caption}
\usepackage{subcaption}
\usepackage{multirow}
\usepackage{xcolor}

\begin{document}

\begin{titlepage}

\widetext

\title{Holographic Classical Shadow Tomography}

\author{Shuhan Zhang}
\affiliation{Department of Physics, University of California, San Diego, California 92093, USA}

\author{Xiaozhou Feng}
\affiliation{Department of Physics, The University of Texas at Austin, Austin, TX 78712, USA}
\author{Matteo Ippoliti}
\email[]{ippoliti@utexas.edu}
\affiliation{Department of Physics, The University of Texas at Austin, Austin, TX 78712, USA}
\author{Yi-Zhuang You}
\email[]{yzyou@physics.ucsd.edu}
\affiliation{Department of Physics, University of California, San Diego, California 92093, USA}

\setcounter{equation}{0}
\setcounter{figure}{0}
\setcounter{table}{0}

\makeatletter
\renewcommand{\theequation}{S\arabic{equation}}
\renewcommand{\thefigure}{S\arabic{figure}}
\renewcommand{\thetable}{S\Roman{table}}
\renewcommand{\bibnumfmt}[1]{[S#1]}
\renewcommand{\citenumfont}[1]{S#1}

\date{\today}

\begin{abstract}
We introduce ``holographic shadows,'' a new class of randomized measurement schemes for classical shadow tomography that achieves the optimal scaling of sample complexity for learning geometrically local Pauli operators at any length scale, without the need for fine-tuning protocol parameters such as circuit depth or measurement rate. Our approach utilizes hierarchical quantum circuits, such as tree quantum circuits or holographic random tensor networks. 
Measurements within the holographic bulk correspond to measurements at different scales on the boundary (i.e. the physical system of interests), facilitating efficient quantum state estimation across observable at all scales. 
Considering the task of estimating string-like Pauli observables supported on contiguous intervals of $k$ sites in a 1D system, our method achieves an optimal sample complexity scaling of $\sim d^k\mathrm{poly}(k)$, with $d$ the local Hilbert space dimension. We present a holographic minimal cut framework to demonstrate the universality of this sample complexity scaling and validate it with numerical simulations, illustrating the efficacy of holographic shadows in enhancing quantum state learning capabilities.
\end{abstract}

\pacs{}

\maketitle

\draft

\vspace{2mm}

\end{titlepage} 

\section{Introduction}

The efficient extraction of classical information from quantum systems is critical for many tasks in quantum technology. Classical shadow tomography, introduced by Huang {\it et al}\,\cite{Huang_2020} (building on prior work by Aaronson~\cite{Aaronson_shadow_2017}), serves as an essential technique in this context \cite{Chen2021R2011.09636, Enshan-Koh2020C2011.11580, Huang2021E2103.07510, Zhao2021F2010.16094, Hu2022H2102.10132, Hu2023C2107.04817, Bu2024C2202.03272, Akhtar2023S2209.02093, Bertoni2022S2209.12924, Nguyen2022O, Zhou2023P2212.11068, Zhou2023E2309.01258, Wu2023E2310.12726,2022arXiv221109835A,Ippoliti_2023,Ippoliti2024C2305.10723,PhysRevLett.120.050406,doi:10.1126/science.aau4963,2021arXiv210505992A,2022arXiv220808416Z,PhysRevResearch.3.033155,PhysRevLett.131.240602,2023CMaPh.404..629W,PhysRevX.13.011049,PhysRevLett.131.160601,2023arXiv230912933D,2023arXiv230910745I,Van-Kirk2022H2212.06084,2023arXiv231100695L,PhysRevLett.125.200501,2024arXiv240116922F,2024arXiv240118071F,Ippoliti2024L2307.15011,Akhtar2024M2308.01653,Akhtar2024D2404.01068}. This method facilitates a practical bridge between quantum and classical realms by enabling sample-efficient estimation of various quantum state properties simutaneously. It significantly reduces the complexity and resource demands compared to the traditional full quantum state reconstruction \cite{Haah2015S1508.01797, ODonnell2015E1508.01907, Flammia2012Q1205.2300}, offering experimentally feasible quantum state tomography approaches on quantum devices \cite{doi:10.1126/science.abn7293,CrossPlatform,Struchalin2020E2008.05234, Zhang2021E2106.10190,2023arXiv230716882V,Hu2024D2402.17911}. This efficiency makes classical shadow tomography particularly valuable in various quantum computing applications where rapid and accurate state estimation is paramount, including state verification \cite{Lukens2021A2012.08997, Morris2021Q2109.03860}, device benchmarking \cite{Levy2024C2110.02965, Kunjummen2023S2110.03629, Helsen2023S2110.13178,PRXQuantum.2.010102,PhysRevLett.124.010504}, Hamiltonian learning \cite{Hadfield2020M2006.15788, McNulty2023E2206.08912, Dutt2023P2312.07497,EntanglementHamiltonian,PhysRevLett.127.170501}, error mitigation \cite{Hu2022L2203.07263,Seif2023S2203.07309,2023arXiv230504956J,2023arXiv231003071Z}, and quantum machine learning \cite{2023arXiv230600061J,2023arXiv230614838Z,doi:10.1126/science.abk3333,2023MLS&T...4a5005H}.

Central to classical shadow tomography is the selection of randomized measurement schemes \cite{Notarnicola2023A2112.11046, Elben2023T2203.11374}. Possible variations of measurement schemes encompass the choice of the structure of the unitary circuit in which the state evolves, the random unitary ensemble from which the unitary gates are sampled, and the locations where measurements are performed. Over recent years, researchers have explored various randomized measurement schemes for classical shadow tomography. These include Hamiltonian-driven dynamics \cite{Hu2022H2102.10132,PhysRevX.13.011049,2023arXiv231100695L}, locally-scrambled quantum dynamics \cite{Hu2023C2107.04817,Bu2024C2202.03272,Zhou2023E2309.01258}, shallow quantum circuits \cite{Akhtar2023S2209.02093,Bertoni2022S2209.12924,2022arXiv221109835A,Ippoliti_2023}, hybrid quantum circuits \cite{Ippoliti2024L2307.15011,Akhtar2024M2308.01653}, and dual unitary circuits \cite{Akhtar2024D2404.01068}. The objective is to investigate these measurement schemes across various estimation tasks to optimize sample complexity scaling.

In these studies, a central question to address is the number of measurement samples (i.e., sample complexity) required to accurately estimate the expectation value of a linear observable. As introduced in the foundational work Ref.\,\cite{Huang_2020}, the sample complexity is bounded above by 
a quantity called the (squared) {\it shadow norm} $\Vert O\Vert^2_{\CE_\sigma}$ of the observable $O$, which depends on the given randomized measurement scheme $\CE_\sigma$. 
For geometrically-local measurement schemes, 
the shadow norm typically scales exponentially with the operator weight $k$ (the number of qudits on which $O$ acts nontrivially), expressed as $\Vert O\Vert^2_{\CE_\sigma}\sim \beta^k$. 
The optimal performance across all operators supported inside a contiguous system\footnote{Improvements over this scaling are possible by restricting the set of operators under consideration~\cite{Ippoliti2024C2305.10723}.} 
approaches $\beta\to d$ in systems composed of $d$-dimensional \emph{qudit} Hilbert spaces, with $d=2$ representing \emph{qubit} systems (see Appendix~\ref{app:optimality}).

Measurement schemes that approach the $d^k$-scaling of the shadow norm include shallow shadows \cite{Akhtar2023S2209.02093,Bertoni2022S2209.12924,2022arXiv221109835A,Ippoliti_2023}, dual unitary shadows \cite{Akhtar2024D2404.01068}, and hybrid circuit shadows \cite{Ippoliti2024L2307.15011,Akhtar2024M2308.01653}. In shallow shadows, optimal scaling is achieved by tailoring the circuit depth to the size of observable, with the optimal depth increasing logarithmically with operator size \cite{Ippoliti_2023}. Dual unitary shadows also require similar adaptive circuit depth adjustments but achieve optimal performance with a parametrically shallower circuit. Hybrid circuit shadows eliminate the need for circuit depth by tuning the rate $p$ of mid-circuit measurements, achieving optimal scaling across all observable sizes simultaneously at the measurement-induced phase transition. However, they still require fine-tuning of the measurement rate $p$ to the critical point $p_c$. 

\begin{figure}[htbp]
    \begin{subfigure}[t]{0.18\textwidth}
        \caption{\hspace{1.3cm} (a)}
        \includegraphics[width=0.8\textwidth]{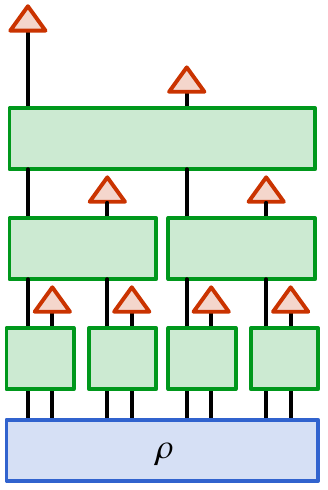}
    \end{subfigure}
    \hfill
    \begin{subfigure}[t]{0.27\textwidth}
        \caption{\hspace{2.12cm} (b)}
        \includegraphics[width=\textwidth]{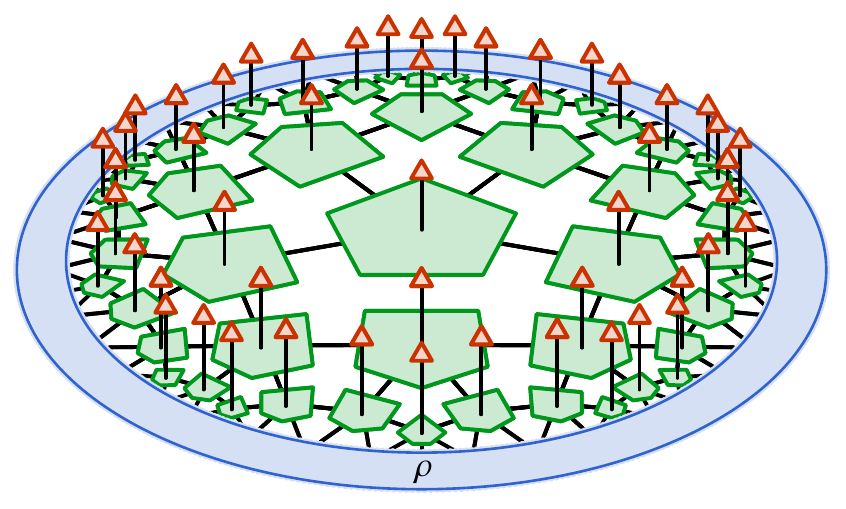}
    \end{subfigure}
    \caption{Schematic of (a) tree circuit and (b) holographic circuit.}\label{fig:schematic}
\end{figure}

In this work, we introduce a new class of randomized measurement schemes, termed \emph{holographic shadows}, which achieves the desired $d^k$-scaling of sample complexity for contiguous operators without the need for adaptive changes in circuit depth or fine-tuning of the measurement rate. The core concept involves employing a hierarchical random unitary circuit, similar to multi-scale entanglement renormalization ansatz (MERA) tensor networks \cite{Vidal2007E,Vidal2008Cquant-ph/0610099,Evenbly2011T1106.1082} or holographic quantum error-correcting codes \cite{Pastawski2015H1503.06237,Hayden2016H1601.01694}, to implement a ``holographic mapping'' of the quantum state into the holographic bulk, where each qubit is measured independently. Local measurements at varying depths within the holographic bulk correspond to the measurement of non-local observables on the holographic boundary in a scale-free manner. This approach enables optimal sample complexity scaling without the necessity of fine-tuning the measurement scheme to criticality.

In particular, we will study two randomized measurement schemes, sketched in Fig.~\ref{fig:schematic}: the tree circuit and the holographic circuit, both achieving the optimal sample complexity scaling without adapting circuit depths or tuning measurement rates. Both circuits display a hierarchical structure. 
The tree circuit, sketched in Fig.~\ref{fig:schematic}(a), comprises unitary gates and projective measurements arranged in the configuration of a binary tree, such that at every layer, half of the qubits are measured out (except the very last layer, where both qubits are measured). The layers of the tree can be labeled by a depth $t$, $1\leq t \leq \log_2(N)$, so that measurements taking place at depth $t$ capture information about operators at length scale $\sim 2^t$. 
The holographic circuit, sketched in Fig.~\ref{fig:schematic}(b), utilizes the structure of a holographic random tensor network (RTN), where the unknown state $\rho$ of the system of interest occupies the 1D ``boundary'' of a hyperbolic space while the randomized measurements take place in the 2D bulk, and the radial distance from the boundary plays a role analogous to the depth $t$ in the tree. Notably, the circular layers of the holographic RTN are generically not unitary operators, but rather matrix product operators (MPO). Thus the experimental implementation of the holographic circuit is less straightforward than the tree, but as we will show, the greater connectivity of the holographic network can yield superior performance.

In both schemes, Pauli estimation proceeds according to the standard classical shadow protocol: snapshots are collected experimentally, then ``back-evolved'' in classical simulation with the appropriate (tree or holographic) tensor network, and used to compute the desired Pauli expectation value. By exploiting properties of these networks, we are able to analytically and numerically compute the shadow norms of Pauli operators and thus characterize the scaling of sample complexity for these practically relevant learning tasks.

The rest of the paper is structured as follows. In \secref{sec:review}, we review the concept of Pauli learning rate, a key tool we use for shadow norm calculation. In \secref{TC}, we introduce a recursion method to calculate the shadow norm of tree circuits, bound the sample complexity close to its optimal $\sim d^k$ scaling at general $d$, and solve it exactly in the $d\to\infty$ limit.
In section \secref{HC}, we discuss holographic circuits based on holographic random tensor networks (RTN) with \{3,7\} and \{5,4\} tilings. We introduce a method where the Pauli learning rate is mapped onto a statistical mechanics model to facilitate the calculation of the sample complexity scaling. In \secref{summary}, we summarize the main results in this paper, compare tree circuits to holographic circuits, and suggest some directions for future work.

\section{Review of Pauli estimation with classical shadows \label{sec:review}}

One of the most practically important applications of classical shadows is learning the expectation values of Pauli operators, $\langle P \rangle = {\rm Tr}(\rho P)$. To do so, one performs a randomized measurement on the system and classically constructs a `snapshot' state $\sigma$ based on the outcome; the average of such snapshot states over experimental repetitions is related to the true state $\rho$ of the system by a quantum channel~\cite{Huang_2020}
\begin{equation}
    \mathbb{E}_{\sigma\sim p(\sigma|\rho)}[\sigma] = \hat{\mathcal{M}}(\rho),
\end{equation}
dubbed the measurement channel or shadow channel. With knowledge of $\hat{\mathcal M}$ one can compute the `inverted snapshots' $\hat{\rho} = \hat{\mathcal M}^{-1}(\sigma)$, and thus the unbiased estimators ${\rm Tr}(P\hat{\rho})$ of the desired expectation value ${\rm Tr}(\rho P)$. 

A key advantage of {\it locally scrambled} measurement ensembles~\cite{Bu2024C2202.03272,Hu2023C2107.04817} is that Pauli operators are eigenmodes of the channel $\mathcal{M}$. 
Using the approach developed in Refs.~\cite{akhtar2023measurementinduced,Ippoliti2024L2307.15011}, 
this fact can be expressed as:
\eq{
\hat{\mathcal M}=\sum_P \Ket{P} w_{\mathcal{E_\sigma}}(P) \Bra{P}
}
where $\Ket{P}$ is the Choi representation for Pauli operator $P$ (normalized as $\langle\!\langle P\Ket{P} = 1$) and the eigenvalue $w_{\mathcal{E}_\sigma}(P)$ is given by
\eq{
w_{\mathcal{E}_\sigma}(P)=\mathbb{E}_{\sigma \sim p(\sigma)}{(\text{Tr}~P\sigma)^2 \over (\text{Tr}~\sigma)^2},
\label{eq:plr_def}
} 
where $\CE_\sigma=\{\sigma|\sigma \sim p(\sigma)\}$ is the prior classical snapshot ensemble.
Given its importance in the present work, we name the eigenvalue $w_{\mathcal{E}_\sigma}(P)$ the {\it Pauli learning rate} (PLR) of operator $P$. 
The PLR represents the averge amount of information learned about $\langle P\rangle$ in each run of the experiment. Specifically, if the randomized measurement is Clifford, then the PLR is the probability to measure $P$ in each shot of the randomized measurement. In this sense it represents the rate of learning as the measurement protocol is iterated. 

The expectation value of $P$ can be expressed as:
\begin{align}
\langle P\rangle 
& = \text{Tr}(P\rho) = \mathbb{E}_{\sigma \sim p(\sigma|\rho)} \text{Tr}(\hat{\mathcal M}^{-1}(P)\sigma) \nonumber \\
& 
= \mathbb{E}_{\sigma \sim p(\sigma|\rho)} \frac{\text{Tr}(P\sigma) }{w_{\mathcal{E_\sigma}}(P)}
\end{align}
where we used Hermiticity of $\hat{\mathcal{M}}$ in the first line.
The sample complexity for estimating $\langle P\rangle$ is proportional to the variance of single-shot estimation for $P$ over $\CE_\sigma$, characterized by the squared shadow norm \cite{Hu2023C2107.04817}:
\eq{
\Vert P\Vert^2_{\mathcal{E}_\sigma}={1 \over w_{\mathcal{E}_\sigma}(P)}.
\label{eq:shadownorm}}

In summary, we have that---provided our tree and holographic circuits are locally scrambled---the sample complexity for learning the expectation of an arbitrary Pauli operator $P$ is inversely proportional to its learning rate $w_{\mathcal{E}_\sigma}(P)$, given in Eq.~\eqref{eq:plr_def}. 

\section{Tree Circuit \label{TC}}
In this section, we examine the classical shadow tomography protocol based on tree quantum circuits, which employ two-qudit random unitary gates and projective measurements arranged in a binary tree configuration, as illustrated in \figref{fig:schematic}(a). Starting with a one-dimensional chain of qudits, each having a local Hilbert space dimension $d$, this setup applies two-qudit unitary operators across successive layers. At each layer, projective measurements are performed to project half of the qudits to product states in the computational basis. In the final layer, measurements are performed on both qudits. Using the measurement outcomes and classical simulation of the tree tensor network, we aim to learn the expectation of Pauli operators on the unknown initial state $\rho$. Our goal in this section will be to quantify the sample complexity (via the PLR, as reviewed in Sec.~\ref{sec:review}) of this task for Pauli operators of variable weight $k$. 

\subsection{Recursive Construction}
\label{sec:recursion_tree}
Here we discuss how to compute the squared shadow norm in the tree quantum circuit by taking advantage of its recursive structure. 
We assume the distribution of each random unitary gate in the tree quantum circuit is \emph{locally scrambled}. This means that for each random gate $U\in \mathcal{U}(d^2)$ in the circuit, the probability distribution $p(U)$ is invariant under local basis transformations, i.e., $p(U)=p(VU)=p(UV)$ for any $V=V_1\otimes V_2\in \mathcal{U}(d)^{\otimes 2}$. This in particular ensures that the measurement ensemble as a whole is locally-scrambled, allowing the application of established methods for this case \cite{Hu2023C2107.04817,Bu2024C2202.03272,Zhou2023E2309.01258}.

As mentioned in Sec.~\ref{sec:review}, Pauli operators form an eigenbasis of the channel $\hat{\mathcal M}$, and their shadow norm is fully determined by the associated eigenvalue, the Pauli learning rate $w_{\mathcal{E}_\sigma}(P)$. This depends upon the chosen ensemble of snapshots $\mathcal{E}_\sigma$---in this case, instances of the tree circuit. 
The PLR can be computed recursively for a tree circuit. Here we outline the main logic, leaving the details of the derivation in Appendix~\ref{sec:recursion_derivation}. 
The idea is that the tree structure allows us to view a tree of depth $T \equiv \log_2(N)$ (we assume $N$ is a power of 2) as a combination of two subtrees of depth $T-1$ joined by a unitary gate as shown in Fig.~\ref{fig:schematic}(a). This hierarchical structure allows us to first compute the PLRs for the two subtrees of depth $T-1$, and then use them as inputs to calculate the PLR of the whole depth-$T$ tree. By iterating this process, we can obtain the PLRs $w_{\mathcal{E}_{\sigma}}(P)$ recursively. 
Under the local scrambling condition, the PLR will only depend on the support $A$ of the Pauli operator $P$. Therefore, we will focus on tracking the PLR associated to each region $A$, rather than considering specific Pauli operators.

From the definition of the Pauli learning rate Eq.~\eqref{eq:plr_def}, we see that $w_{\mathcal{E}_\sigma}(P)$ is a two-replica (i.e. quadratic) function of the ensemble of snapshots $\sigma$, which in this case are instances of the tree tensor network, with Haar-random unitary gates that must be averaged over. This average gives a projection onto permutations of the two replicas, $\mathds{1}$ (identity permutation) and $c$ (swap). It is convenient to use these replica permutation variables to express a recursive relation for the PLR. 
In particular, we have
\begin{equation}
w_{\mathcal{E}_\sigma}(A) = 
\bar{w}_{\mathds{1}}(A) + \bar{w}_{c}(A),
\end{equation}
with each term representing a possible replica permutation ($\mathds{1}$ or $c$) assigned to the top of the tree.
We can now define analogous coefficients $\bar{w}_{\mathds{1}}^s(A)$ and $\bar{w}_{c}^s(A)$ for each subtree $s$; these depend on the intersection of the Pauli support $A$ with the leaves of subtree $s$. 
Each time two neighboring subtrees $s_1$ and $s_2$ are fused into a tree $s$, the two-component $\bar{w}^s$ vector can be expressed recursively as 
\be
\begin{pmatrix}
    \bar{w}^s_{\mathds{1}}(A)\\\\
    \bar{w}^s_{c}(A)
\end{pmatrix}
=\begin{pmatrix}
1&a&a&0\\
0&a&a&1
\end{pmatrix}
\begin{pmatrix}
    \bar{w}^{s_1}_{\mathds{1}}(A)\\\\
    \bar{w}^{s_1}_{c}(A)
\end{pmatrix}\otimes 
\begin{pmatrix}
    \bar{w}^{s_2}_{\mathds{1}}(A)\\\\
    \bar{w}^{s_2}_{c}(A)
\end{pmatrix}.
\label{eq:PLR_recursion}
\ee
with $a=d/(d^2+1)$. Using \eqnref{eq:PLR_recursion}, we can calculate the corresponding $\bar{w}$ vector of the whole system recursively and sum the two components to get the overall PLR $w_{\mathcal{E}_{\sigma}}(P)=\bar{w}_{\mathds{1}}(A)+\bar{w}_{c}(A)$. 

The remaining question is the initial condition of this recursive construction, which is fully determined by the support $A$. We observe that all single-layer (i.e. two-qudit) subtrees that intersect the support --- which will be referred to as {\it particle-like} and denoted by $\bullet$ --- give the same $\bar{w}^s$ vector
\be
\frac{1}{(d^4-1)}
\begin{pmatrix}
    -1\\\\
    d^2
\end{pmatrix}.
\label{eq:particle_initial}
\ee
This result arises from the fact that all non-identity Pauli strings defined on the two qudits are equivalent to each other under Haar random gates. 
On the other hand, single-layer subtrees that do not intersect the support --- which will be referred to as {\it hole-like} and denoted by $\circ$ --- have $\bar{w}^s$ vector
\be
\begin{pmatrix}
    1\\\\0
\end{pmatrix}.
\label{eq:hole_initial}
\ee
This follows from unitary invariance of the identity operator. 

With these ingredients, we have a practical approach for calculating the Pauli learning rate within the whole tree circuit. We can first determine the $\bar{w}^s$ vector for all depth-1 subtrees, categorizing them as ``particle-like'' or ``hole-like'' depending on whether they intersect the Pauli support $A$. These vectors are then treated as leaves of a new effective tree, which is half the size of the original, and can be given as input to the recursion Eq.~\eqref{eq:PLR_recursion}. 
More details on the derivation of these rules are provided in Appendix~\ref{sec:recursion_derivation}.

\subsection{Contiguous Case \label{sec:treecont}}
A particularly simple case arises when the Pauli support $A$ is a contiguous segment\footnote{It is worth noting that certain non-contiguous choices of $A$ could also be mapped to contiguous segments by exchanging two subtrees that connect at a node, which is a symmetry of our measurement ensemble.} whose length $k = |A|$ is a power of 2, so that it exactly fills the leaves of a subtree. 
Under these conditions, we demonstrate below that the sample complexity is parametrically reduced 
compared to the local measurement scenario as the support size $k$ increases.

Suppose the size of the support is $k = 2^m$, so we have a a depth-$m$ subtree $s_m$ on top of $A$. The initial condition on all leaves in $A$ is particle-like. Consequently, for every step $t\le m$, the two inputs are identical. This leads to the recursive relation
\be
\begin{pmatrix}
    \bar{w}_{\mathds{1}}^{s_t} \\\\
    \bar{w}_{c}^{s_t}
\end{pmatrix}
=
\begin{pmatrix}
    (\bar{w}_{\mathds{1}}^{s_{t-1}})^2+2a\bar{w}_{\mathds{1}}^{s_{t-1}} \bar{w}_{c}^{s_{t-1}} \\\\
    (\bar{w}_{c}^{s_{t-1}} )^2+2a\bar{w}_{\mathds{1}}^{s_{t-1}} \bar{w}_{c}^{s_{t-1}}
\end{pmatrix}.\label{eq:contiguous_recursion}
\ee
Here $\bar{w}^{s_t}$ represents the vector of a depth-$t$ subtree with $1\le t\le m$ ($t=0$ represents the initial condition). 
Although this recursive relation is quadratic and generally lacks an analytical solution, insights can still be gleaned. 

Consider the ratio $g_t = \bar{w}_{\mathds{1}}^{s_t} /\bar{w}_c^{s_t}$; this obeys the recursive relation
\be
g_t = g_{t-1} \frac{g_{t-1}+2a }{1+2ag_{t-1}}. 
\ee
It has fixed points $g^{\ast} = 0,1$ with the stable one being $g^{\ast}=0$. Starting from the initial condition $g_0 = -1/d$, $g_t$ tends to zero as $t$ increases. When $t$ is sufficiently large, $\bar{w}^{s_t}_{\mathds{1}}$ becomes negligible relative to $\bar{w}^{s_t}_c$ and the main problem is to calculate $\bar{w}^{s_t}_c$. Using \eqnref{eq:contiguous_recursion}, we obtain
\begin{align}
\bar{w}_c^{s_m} & =\frac{d^{k}}{(d^2-1)^{k}}\exp\left[ k\sum_{i=0}^{m-1} \frac{\ln(1+2ag_i) }{2^{i+1}} \right]\label{eq:sum}
\end{align}
where again $m = \log_2(k)$.
The summation above involves only $g_i$, which we know decays exponentially towards zero. Then, the summation of the series in the exponent of \eqnref{eq:sum} converges to a constant as $k\to\infty$.  
Therefore, when the support size $k$ is sufficiently large, the upper limit of the summation can effectively be extended to infinity with a small error. By defining 
\begin{equation}
    Q(d) = \sum_{i=0}^\infty \frac{\ln(1+2ag_i)}{2^{i+1}},
    \label{eq:Qd_def}
\end{equation}
we have, in the large-$k$ limit,
\begin{equation}
\bar{w}^{s_m}_{c}(P) \simeq 
\left( \frac{d}{d^2-1} e^{Q(d)} \right)^k 
\end{equation}
So far, we have focused only on the subtree that sits on top of $A$; when $k<N$, there will be other subtrees with a hole-like initial condition, which evolves trivially. Taking the rest of the system into account only changes the result above by an $O(1)$ factor, therefore we can conclude that the Pauli leraning rate obeys
\begin{equation}
w_{\mathcal{E}_\sigma}(P) \propto \beta^k,\qquad
\beta = \frac{d}{d^2-1} e^{Q(d)}
\label{eq:Pauli_largek}.
\end{equation}

\begin{table}
    \centering
    \begin{tabular}{cccc} \toprule
         $d$ & $Q(d)$ & $\beta(d)$  \\ \midrule 
         2 & $-0.3402$ & $2.1079$ \\
         3 & $-0.1350$ & $3.0521$ \\
         4 & $-0.0720$ & $4.0301$ \\
         5 & $-0.0447$ & $5.0196$ \\
         10 & $-0.0105$ & $10.0050$ \\
         20 & $-0.0026$ & $20.0012$ \\
         \bottomrule 
    \end{tabular}
    \caption{Numerically computed values of the series $Q(d)$, Eq.~\eqref{eq:Qd_def}, and of the parameter $\beta$ that determines the sample complexity scaling $\| P\|_{\mathcal{E}_\sigma}^2 \sim \beta^k$ in tree circuits, Eq.~\eqref{eq:Pauli_largek}. We find that $\beta$ is always close to $d$ and scales as $d+\frac{1}{2}d^{-2} + O(d^{-3})$ as $d$ becomes large.}
    \label{tab:tree_betas}
\end{table}

The unknown factor $Q$ can be calculated numerically for finite $d$; results reported in Table~\ref{tab:tree_betas} show that it is always close to $d$, though slightly larger. In addition, we can derive analytical upper and lower bounds for the shadow norm. 
Using the monotonicity of $g$ --- which is guaranteed when $-1/d\le g<0$ --- we can establish the following bound on $Q(d)$:
\begin{equation}
    Q(d) \geq \sum_{i=0}^\infty \frac{\ln(1+2ag_0)}{2^{i+1}} = \ln \left(\frac{d^2-1}{d^2+1} \right),
\end{equation}
It follows that
\begin{equation}
w_{\mathcal{E}_{\sigma}}(P) \gtrsim  
\left( \frac{d}{d^2+1} \right)^k,
\end{equation}
and thus the squared shadow norm $\Vert P\Vert_{\CE_\sigma}^2=1/w_{\mathcal{E}_{\sigma}}(P)$ is bounded above by
\be
\Vert P\Vert_{\CE_\sigma}^2\lesssim\left(d+\frac{1}{d}\right)^{k}.\label{eq:contiguous_tree_bound}
\ee
In conclusion, when $k$ is sufficiently large, \eqnref{eq:contiguous_tree_bound} ensures that the shadow norm in the tree is parametrically smaller than the local measurement case, where it scales as $\Vert P\Vert_{\CE_\sigma}^2\sim (d+1)^{k}$. We also remark that this is a rather loose upper bound on $\beta$; at large $d$, a series expansion in $1/d$ yields $\beta = d + \frac{1}{2} d^{-2} + O(d^{-3})$, which is in excellent agreement with the numerical values in Table~\ref{tab:tree_betas} already for $d \gtrsim 5$.

\begin{figure}
    \centering
    \includegraphics[width=1.0\columnwidth]{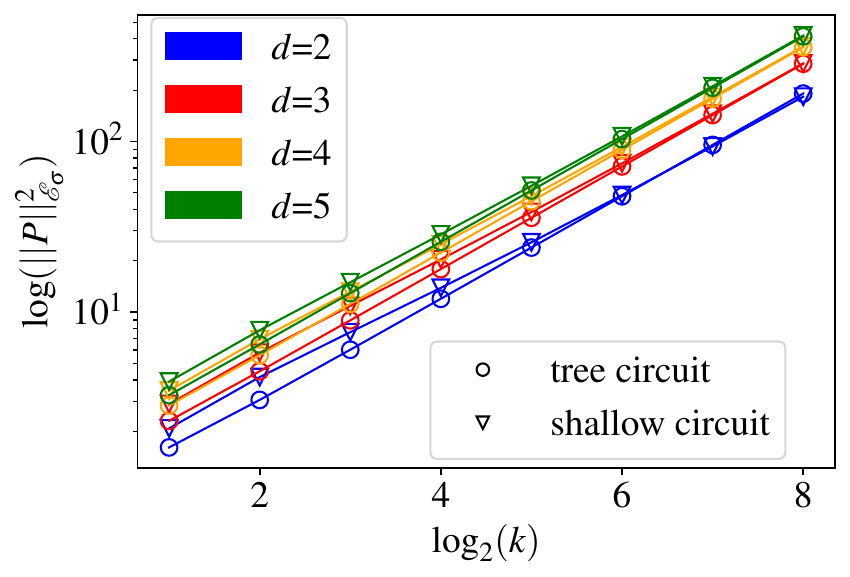}
    \caption{Comparison of Pauli learning rates in tree circuits and shallow circuits with optimal depth. The support is chosen to be contiguous, with $k = 2^m$. It can be seen that the tree circuit gives a smaller shadow norm when the support size is small, but the growth of the shadow norm with support size is slower in the shallow circuit.}
    \label{fig:tree_shallow_comparison}
\end{figure}

It is instructive to compare the shadow norm in tree circuits with that in shallow circuits \cite{Hu2023C2107.04817,Ippoliti_2023}, which have been previously demonstrated to offer parametric improvements over local measurements. When the circuit is tuned to a $k$-dependent optimal depth $\sim \log(k)$, the Pauli learning rate for operators with contiguous support scales as $\sim k d^k$. 
Our calculation for tree circuits (see Appendix~\ref{sec:recursion_derivation} for details) yields a slightly larger base of the exponent $\beta\gtrsim d$, but lacks the factor of $k$, implying that tree circuits will perform better (for operators that fit a subtree exactly) at sufficiently small $k$, while shallow quantum circuits will perform better asymptotically in large $k$. 
Numerical results shown in Fig.~\ref{fig:tree_shallow_comparison} show that there is a critical support size $k^{\ast}$ such that the tree circuit exhibits a smaller sample complexity when $k\le k^{\ast}$, while if the support size $k$ is larger than $k^{\ast}$, the shallow circuit begins to outperform the tree circuit. 

To understand this difference, we use the operator spreading and relaxation picture introduced in Ref.~\cite{Ippoliti_2023}. In shallow circuits, the relaxation of operators is dominant at the early time, which leads to the reduction of squared shadow norm. At the same time, the spreading of the operator near the boundary of the support increases the shadow norm and eventually becomes dominant when the circuit depth is large enough. 
The balance of these two effects gives the optimal depth for shallow circuits, $\sim\log(k)$. 
In the tree circuit, operator relaxation is weaker due to the diluted gates and the fact that many measurements happen at low depth; however operator spreading is also suppressed (it is entirely absent until depth $\log(k)$ when the operator's support matches a subtree perfectly). 
Consequently, when the support size $k$ is small, the latter (boundary) effect is more important, making tree circuits more efficient; while when $k$ is very large, the bulk effect dominates, and shallow circuits have lower sample complexity due to their stronger scrambling dynamics.

\subsection{Large $d$ Limit}
We conclude this section by discussing the shadow norm of general Pauli operators (i.e., not necessarily contiguous) in the limit $d\to\infty$. 
Given an arbitrary Pauli string $P$ with support $A$, we first apply the coarse-graining discussed in Sec.~\ref{sec:recursion_tree} to map the original tree $s$ to a new tree $s'$ with half of the leaves. In the new tree, each hole has the same $\bar{w}^{\circ} = (1,0)^T$ vector as before, while each particle assumes a vector 
\be
\bar{w}^{\bullet} = 
\begin{pmatrix}
    -1/(d^4-1)\\\\ d^2/(d^4-1)
\end{pmatrix},
\ee

In the large-$d$ limit, we will call a subtree $s$ `particle-like' if the $\bar{w}^s_c$ component is much larger (i.e. of a higher order in $d$) than the $\bar{w}^s_{\mathds{1}}$ component. Similarly, if the $\bar{w}_{\mathds{1}}^s$ component is the dominant one, we call the subtree `hole-like'. Additionally, we introduce a third kind, labeled $\oplus$, for subtrees where the two components are of the same order in $d$. Then, Eq.~\eqref{eq:PLR_recursion} gives the following fusion rules:
\begin{align}
    \bullet + \bullet\to\bullet,\quad \circ+\circ\to\circ,\quad\oplus+\oplus\to\oplus,\nonumber\\
    \bullet+\circ\to\oplus,\quad \bullet+\oplus\to\bullet,\quad \circ+\oplus\to \circ.
\end{align}
Each time the rule is applied, the dominant term after the fusion equals the product of the dominant terms of the two input subtrees, except in the case $\bullet+\circ\to\oplus$, where an extra term $1/d$ arises from the coefficient $a=d/(d^2+1)$. 

Based on these rules, it is easy to keep track of the powers of $d$ that will control the Pauli learning rate. Namely, every particle-like leaf in $s'$ contributes a factor of $1/d^2$, and each occurrence of the $\bullet+\circ\to \oplus$ fusion contributes an extra factor of $1/d$. 
In all, the leading term in the PLR is given by
\begin{equation}
    w_{\mathcal{E}_\sigma}(P) \sim d^{-\bdryC(A)-\bulkC(A)},
\end{equation}
where we introduced the notation $\bdryC(A)$ to denote twice the number of particle-like leaves in $s'$, and $\bulkC(A)$ to denote the number of $\bullet + \circ \to \oplus$ fusions. This notation, whose motivation will be cleared in Sec.~\ref{HC}, refers to the fact that each $\bullet + \circ \to \oplus$ fusion connects a hole-like and particle-like subtree, so $\bulkC(A)$ is the minimal number of bonds to cut in the bulk of the tree in order to separate particle-like and hole-like regions. Similarly $\bdryC(A)$ measures the length of a ``boundary cut'' that separates the boundary subsystem $A$ of interest from the rest of the network. The sum $\bdryC(A) + \bulkC(A)$ thus measures the total minimal cut length
\footnote{The reason we cannot directly apply the minimal cut picture in the original tree is that $a$ is of the order of $1/d$.
Then, the scenario where a particle-like tree and a hole-like tree combine to yield another particle-like tree can happen. This probability is effectively nullified in the new tree since each particle's dominant component is at least $d^2$ times larger than the smaller component. }. Thus in the large-$d$ limit the scaling of shadow norms for arbitrary Paulis is given by the minimal cut result:
\be
\log_d(\Vert P\Vert_{\mathcal{E}_{\sigma}}^2) = \bulkC(A) + \bdryC(A)
\ee
with $A$ the support of $P$.
This result is analogous to that in holographic circuits, which is discussed in the next section. It also recovers the result of the contiguous case, Sec.~\ref{sec:treecont}: there $\bdryC(A) = k$ (the support occupies a subtree perfectly so there are $k/2$ particle-like leaves in $s'$) and $\bulkC(A) = 1$ (the unique particle-like subtree can be excised from the rest of the network by a single cut at depth $\log_2(k)$), giving $\Vert P\Vert_{\mathcal{E}_{\sigma}}^2 \sim d^{k+1}$.

\section{Holographic Circuit \label{HC}}
As discussed in \secref{sec:treecont}, the tree circuit is tomographically inefficient when the support of the operator is not commensurate with a subtree. 
This is due to the branching structure of the tree, which causes operators with support covering different branches to be mapped to measurement rather deep in the circuit. Thus, in order to overcome this problem and achieve similar tomographic efficiency on a wider set of operators, we generalize the tree circuit to MERA-like holographic circuits. Such circuits preserve the hierarchical structure of tree circuits while introducing ``disentanglers" inducing connections between different branches within each layer, allowing local operators to be resolved at lower depths. 

\begin{figure}[htbp]
    \includegraphics[width=0.35\textwidth]{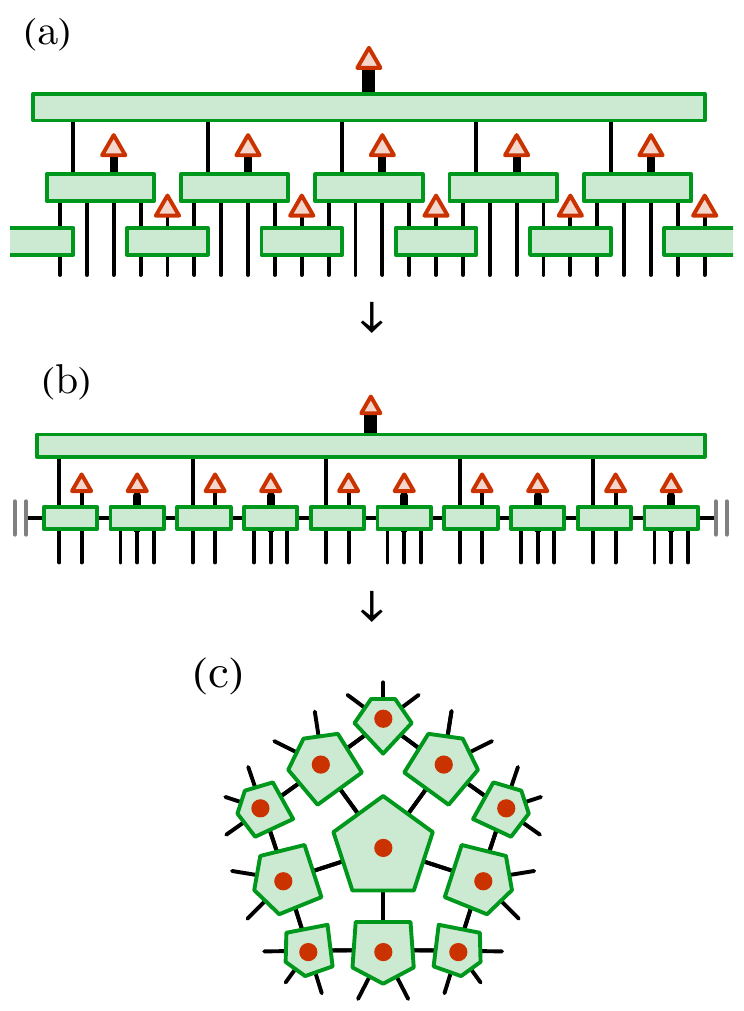}
    \caption{Schematic of (a) unitary picture, where each gate is unitary; (b) MPO picture, where each MPO layer can be viewed as a layer of unitary gates. Note that this requires measurements represented by red triangles/dots to have different bond dimensions, as indicated by different line thickness; (c) RTN picture, where each MPO tensor is a random tensor. }
    \label{fig:pics}
\end{figure}

However, the absence of a tree structure means that the recursion method cannot be efficiently applied to calculate the shadow norm of holographic circuits. Consequently, the computational complexity of determining the Pauli learning rate scales exponentially with system size, hindering our ability to carry out the post-processing step of the protocol (involving the inverse map $\hat{\mathcal{M}}^{-1}$) and to understand the universal behavior of such circuits. To address this challenge, we decompose unitary layers into Matrix Product Operators (MPOs) and relax the unitarity condition, assuming each MPO tensor to be a random tensor with elements independently drawn from a Gaussian distribution. This approach naturally leads to our construction of holographic circuits based on holographic Random Tensor Networks (RTN). \figref{fig:pics} shows an example of a 2-layer 2 layers of the \{5,4\} holographic circuit from the unitary picture, MPO picture, and RTN picture. 

In this section, we start by reviewing the construction of holographic RTN in \secref{sec: HRTNconstruction}. Next, we discuss a method for mapping Pauli learning rate directly to a statistical mechanical model, facilitating the calculation of the shadow norm within the holographic circuit, as detailed in \secref{sec:HCPLR} and \secref{sec:large d}. Subsequently, we present results that demonstrate the optimal sample complexity scaling of the holographic circuit in \secref{sec:results}. We also explore how the sub-leading term in this scaling is related to the curvature of the space tiled by the holographic RTN in \secref{sec:effcc}.

\subsection{Construction of Holographic RTN \label{sec: HRTNconstruction}}

\begin{figure}[htbp]
    \begin{subfigure}[t]{0.23\textwidth}
        \caption{\hspace{1.75cm} (a)}
        \includegraphics[width=\textwidth]{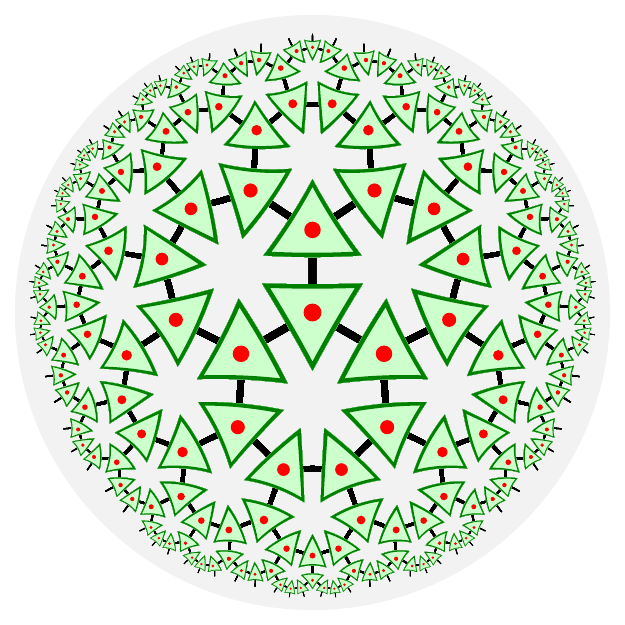}
    \end{subfigure}
    \hfill
    \begin{subfigure}[t]{0.23\textwidth}
        \caption{\hspace{1.75cm} (b)}
        \includegraphics[width=\textwidth]{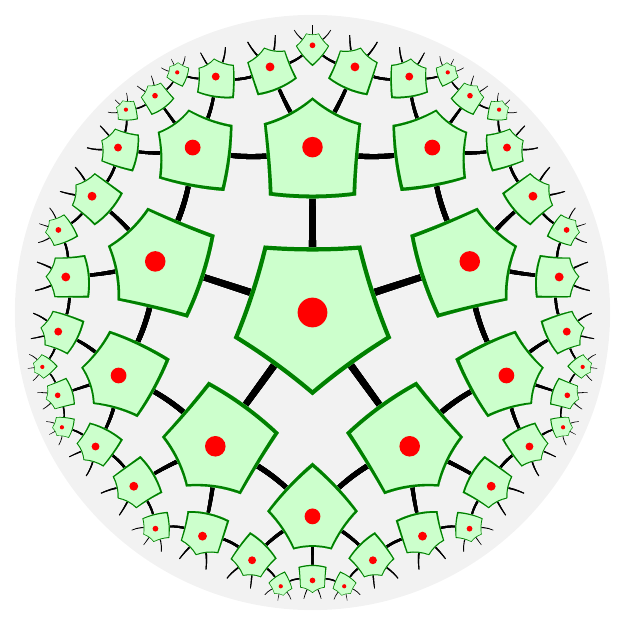}
    \end{subfigure}
    \caption{Holographic RTN with (a)\{3,7\} tiling and (b) \{5,4\} tiling. The green triangles and pentagons represent random tensors and the red dots represent measurement on bulk legs.}\label{fig:tiling}
\end{figure}

To construct the holographic RTN, we implement hyperbolic geometry based on regular hyperbolic tilings. As discussed in \cite{Basteiro_2022}, two-dimensional hyperbolic space can be discretized using $\{p,q\}$ tilings, where $q$ regular $p$-gons meet at each vertex, provided that $(p-2)(q-2)>4$. We develop holographic RTN models based on \{3,7\} and \{5,4\} tilings by positioning a random tensor on each face of the tiling and connecting tensors on adjacent faces with an edge, as illustrated in \figref{fig:tiling}. 

Each tensor network structure can be described by a graph $\mathcal{G}=(\mathcal{V},\mathcal{E})$, comprises the vertex set $\mathcal{V}$ (the set of tensors) and the edge set $\mathcal{E}$ (the set of tensor legs). We will distinguish the edges (tensor legs) by the bulk legs contracted between random tensors in the bulk, and the boundary legs dangling from the out-most layer of tensors on the boundary. We assume all tensor legs are of the bond dimension $d$ to simplify the problem. 
The red dot at the center of each tensor represents the final step measurement on the bulk degrees of freedoms. 

Networks with these geometries can be created by unitary circuits in the MERA architecture (Fig.~\ref{fig:pics}), however this generally places some constraints on the bulk and boundary bond dimensions. Arbitrary bond dimensions can be achieved by a ``fusion'' approach using at most $O(N)$ auxiliary qudits, see Appendix \ref{app:fusion}.

\subsection{Pauli Learning Rate of Holographic Circuit \label{sec:HCPLR}}





Following the Pauli learning rate approach reviewed in \secref{sec:review}, our primary task is to calculate $w_{\mathcal{E}_\sigma}(P)$ for an ensemble $\mathcal{E}_\sigma$ of holographic RTN states. Inspired by the mapping of the entanglement feature (EF) for RTN state ensembles to the Ising model in the large bond dimension limit as presented in \cite{You_2018}, we construct a statistical mechanics model to compute the Pauli learning rate. We first assume large bond dimension limit condition, i.e.  $D_e \rightarrow \infty$ so that 

\begin{align}
\label{eq:PLR}
w_{\mathcal{E}_\sigma}(P)=\mathbb{E}_{\sigma \sim p(\sigma)}{(\text{Tr}~P\sigma)^2 \over (\text{Tr}~\sigma)^2} \simeq {\mathbb{E}_{\sigma \sim p(\sigma)}(\text{Tr}~P\sigma)^2 \over \mathbb{E}_{\sigma \sim p(\sigma)}(\text{Tr}~\sigma)^2}
\end{align}

Subsequently, we place an Ising spin $\sigma_v=\pm1$ on each vertex $v \in \mathcal{V}$ of the holographic RTN and define the energy of a specific Ising configuration $\sigma=\{\sigma_v\}$ to be:

\begin{align}
    E(\sigma)=-\sum_{e \in \mathcal{E}} J \prod_{v \in \partial e} \sigma_v - h\sum_{v \in \mathcal{V}_{\partial}}\sigma_v
    \label{eq:energy}
\end{align}
where $\partial e$ denotes the set of vertices that are endpoints of edge $e$, $\mathcal{V}_\partial$ denotes the set of vertices that connect to boundary legs, and $J=h={1 \over 2}\ln d$. We then map $w_{\mathcal{E}_\sigma}(P)$ to a statistical mechanics model:
\eqs{w_{\mathcal{E}_\sigma}(P)
    \simeq {{\sum\limits_{[\sigma_v]} \prod\limits_{dv \in \supp P}}\delta_{\sigma_v, -1} e^{-E(\sigma_v)} \over  \sum\limits_{[\sigma_v]} e^{-E(\sigma_v)}}\label{eq:PLR stat model}}
where $dv$ denotes boundary regions that are connected to vertex $v$ through boundary legs.

Unlike the Ising model for EF which includes a boundary pinning field, the statistical mechanics model for PLR imposes a stricter boundary condition by pinning the boundary spin. The derivation of \eqnref{eq:PLR stat model} is detailed in Appendix~\ref{sec:PLRderivation}. It is important to note that \eqnref{eq:PLR stat model} requires summation over all vertex configurations, which poses a significant computational challenge for large systems. In the subsequent section, we will introduce two simplifications to facilitate the calculation of PLR for large systems.

\subsection{Large $d$ limit of Pauli learning rate \label{sec:large d}}
We assume the bond dimension of RTN to be $d$, and consider the asymptotic limit where $d \rightarrow \infty$. The equality of bulk and boundary dimensions implies $J=h$ in \eqnref{eq:PLR stat model}, and the asymptotic limit $d \rightarrow \infty$ allows us to focus only on the leading-order term of d in both the denominator and numerator of \eqnref{eq:PLR stat model}. In the statistical mechanics picture, this is equivalent to considering only the lowest energy vertex configuration in both the denominator and numerator.

The leading order term in the denominator comes from the configuration with all vertices having spin 1, which gives a contribution of $d^{2n_b}$, where $n_b$ represents the total number of bonds. In the numerator, the deviation from ground state energy comes from (i) interaction between vertices with different spins across bulk domain wall and (ii) boundary vertices that are fixed to $-1$. These deviations correspond to the two sums in \eqnref{eq:energy}, respectively. 
Specifically, the type (i) deviation contributes a factor of $d^{-\bulkC(A)}$, where $\bulkC(A)$ is the ``bulk cut" representing the length of the domain wall which encloses the boundary subsystem $A$. The type (ii) deviation yields a factor of $d^{-\bdryC(A)}$, where $\bdryC(A)$ is the ``boundary cut" representing the number of boundary vertices with spin $-1$. Thus, the leading order term in the numerator comes from the configuration where the sum of $\bulkC(A)$ and $\bdryC(A)$ is minimized. We define the minimum cut as \eqs{
\minC(A)=\text{min}\(\bulkC(A)+\bdryC(A)\)
}
Then, to leading order in $d$, the PLR can be expressed as
\eqs{
w_{\mathcal{E}_\sigma}(P)& \simeq {d^{2n_b-\minC(\supp P)}\over d^{2n_b}}=d^{-\minC(\supp P)}\label{eq:PLR simp}
}

In the subsequent analysis, we will focus on Pauli operators with continuous support. Since all boundary vertices with connection to $\supp P$ are fixed to $-1$, $\bdryC(\supp P) \geq |\supp P|=k$. There, $\minC(\supp P)$ will be achieved with $\bdryC(\supp P) \simeq k$\footnote{this is not a strict equality when $\supp P$ doesn't contain all boundary legs connecting to a boundary vertex}, with $\bulkC(\supp P)$ being bulk geodesic starting from one end of $\supp P$ and ending on the other end. 

To determine the minimum bulk cut, we leverage the fact that the holographic RTN's are planar graphs to construct their dual graphs. The minimal $\bulkC(\supp P)$ in a holographic RTN corresponds to the shortest path connecting the endpoints of $\supp P$ on its dual graph. This shortest path can be efficiently identified through a breadth-first search. To simplify the notation and reduce redundancy, in the following sections, in the following sections, the terms $\bdryC(\supp P)$ and $\bulkC(\supp P)$ will specifically refer to the minimal values of $\bdryC(\supp P)$ and $\bulkC(\supp P)$, respectively. 

\begin{figure}[htbp]
    \begin{subfigure}[t]{0.4\textwidth}
        \caption{\hspace{0.5\textwidth} (a)}
        \hspace*{-0.6cm}
        \includegraphics[width=\textwidth]{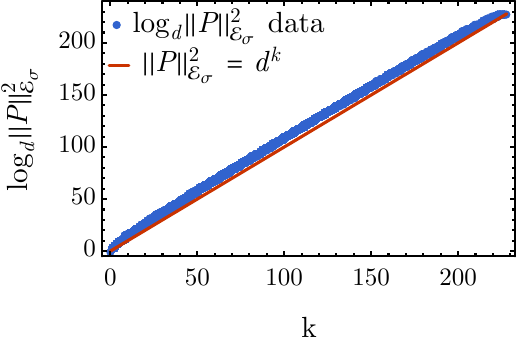}
    \end{subfigure}
    \hfill
    \begin{subfigure}[t]{0.4\textwidth}
        \caption{\hspace{0.5\textwidth} (b)}
        \hspace*{-0.6cm}
        \includegraphics[width=\textwidth]{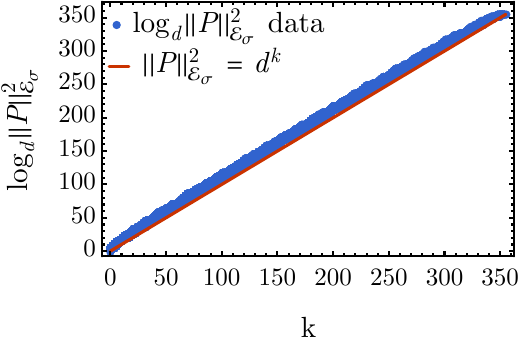} 
    \end{subfigure}
    \caption{Scaling of $\Vert P\Vert_{\CE_\sigma}^2$ in holographic circuit with (a) \{3,7\} tiling and (b) \{5,4\} tiling.}\label{fig:HRTNSN}
\end{figure}

\subsection{Results and Interpretation \label{sec:results}}
We use the inflation rule described in \cite{Basteiro_2022} to generate graphs of holographic RTN of various layers with \{3,7\} and \{5,4\} tilings. Following the methodology outlined in \secref{sec:large d}, we determine the minimum cuts for all consecutive boundary regions. Then, we use \eqnref{eq:PLR simp} to calculate PLR for Pauli operators supported on these boundary regions and calculate their shadow norm according to \eqnref{eq:shadownorm}. The results are illustrated in \figref{fig:HRTNSN}. 

Note that the y-axis is $\log_d \Vert P\Vert^2_{\CE_\sigma}=-\log_d w_{\CE_\sigma} \simeq \minC(\supp P)$. So the slope of this graph illustrates how $\minC(\supp P)$ scales with $k=|\supp P|$. For both \{3,7\} and \{5,4\} tilings, the shadow norm is lower-bounded by the gray line
$\Vert P\Vert_{\CE_\sigma}^2 = d^{k}$. 
This observation is consistent with the intuition that the $\minC(\supp P)$ is lower-bounded by $\bdryC(\supp P)\simeq k$. Furthermore, the shadow norm for any Pauli operators supported on consecutive boundary regions does not significantly deviate from this lower bound. This suggests that the contribution from $\bdryC(\supp P)$ predominates over $\bulkC(\supp P)$ in $\minC(\supp P)$. 

Now we aim to better understand the contribution from $\bulkC(\supp P)$. As discussed in \secref{sec:large d}, in the minimum energy configuration, $\bulkC(\supp P)$ is interpreted as the length of the geodesic that connects the two ends of $\supp P$. This suggests a connection to entanglement entropy $S(\supp P)$, which is related to the area of the minimal surface bonding the region $\supp P$ in the holographic bulk \cite{Ryu_2006}. In fact, in the $d \rightarrow \infty$ limit, 
\eqs{e^{-S(\supp P)}\simeq d^{-\bulkC(\supp P)}\label{eq:EEtoBC}}
where $S(\supp P)$ is the second Renyi entropy over the region $\supp P$. This relationship is derived in Appendix~\ref{sec:EEtoBCderivation} by taking the large $d$ limit in the Ising model for the EF.

In discretized holographic bulk constructed from \{p,q\} tiling, the entanglement entropy of a continuous boundary subregion A takes the form \cite{Basteiro_2022}:
\eq{S(A)\simeq{c_{\text{eff}}(p,q)}\ln(\text{min}(|A|,|\overline{A}|))\label{eq:EEtoCC}}
where $c_{\text{eff}}$ is the tiling dependent effective central charge, and we absorb the factor 1/3 originally present in \cite{Basteiro_2022} in our definition of $c_{\text{eff}}$. Physically, $c_{\text{eff}}$ characterizes the curvature of the holographic bulk and will be discussed in more detail in the next section. 

Combining \eqnref{eq:EEtoBC} and \eqnref{eq:EEtoCC}, we obtain \eqs{\bulkC(\supp P) \simeq {c_{\text{eff}}(p,q)}\ln(\text{min}(k,N-k))/\ln d} 
We further absorb the factor of 1/$\ln d$ into $c_{\text{eff}}(p,q)$, yielding 
\eq{\bulkC(\supp P) \simeq {c_{\text{eff}}(p,q)}\ln(\text{min}(k,N-k))\label{eq:BCtoCC}}
where N represents the total system size. To verify \eqnref{eq:BCtoCC} and to gain a deeper understanding of how the tiling $\{p,q\}$ influences $c_{\text{eff}}(p,q)$, we treat ${c_{\text{eff}}(p,q)}$ as an unknown fitting parameter and fit the shadow norm data to the function
\eqs{\Vert P\Vert^2_{\CE_\sigma}\simeq d^{\minC(\supp P)}\simeq d^{ k+{c_{\text{eff}}(p,q)}\ln(\text{min}(k,N-k))}\label{eq:fit}}

\begin{figure}[htbp]
    \centering
    \begin{subfigure}[t]{0.4\textwidth}
        \caption{\hspace{0.5\textwidth} (a)}
        \hspace*{-0.5cm}
        \includegraphics[width=\textwidth]{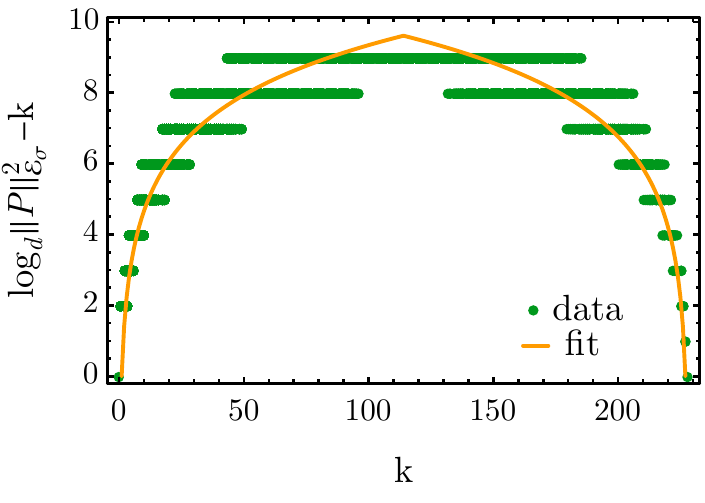}
    \end{subfigure}
    \hfill
    \begin{subfigure}[t]{0.4\textwidth}
        \caption{\hspace{0.5\textwidth} (b)}
        \hspace*{-0.5cm}
        \includegraphics[width=\textwidth]{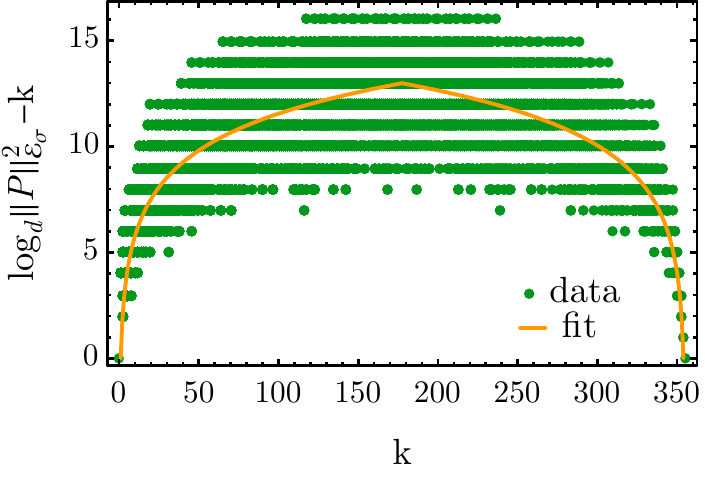}
    \end{subfigure}
\caption{Fitting for $c_{\text{eff}}(p,q)$ in holographic RTN using the function ${c_{\text{eff}}(p,q)}\ln(\text{min}(k,N-k))$. (a) 4-layer \{3,7\} tiling gives $c_{\text{eff}}(3,7)=2.029 \pm 0.003$ and (b) 3-layer \{5,4\} tiling gives $c_{\text{eff}}(5,4)=2.154\pm0.008$. Note that for the \{5,4\} tiling we omit boundary regions that don’t contain all boundary legs connecting to a boundary vertex to ensure $\bdryC(\supp P)=|\supp P|=k$, which allows us to obtain a more accurate fitting for $c_{\text{eff}}(5,4)$.}\label{fig:HRTNfit}
\end{figure}
As illustrated in \figref{fig:HRTNfit}, the function described by \eqnref{eq:fit} provides a good fit to the data, resulting in distinct values of $c_{\text{eff}}$ for the \{3,7\} and \{5,4\} tilings. This confirms that in the $d \rightarrow \infty$ limit, \eqnref{eq:fit} is valid. 

In the following section, we further explore $c_{\text{eff}}$ and investigate its relationship with bulk geometry.  

\subsection{Effective central charge \label{sec:effcc}}
According to \eqnref{eq:BCtoCC}, the bulk geodesic separating a boundary region $A$ from $\overline{A}$ is proportional to $\ln(\text{min}(|A|,|\overline{A}|))$, with $c_{\text{eff}}$ serving as the constant of proportionality. Intuitively,  $c_{\text{eff}}$ should reflect intrinsic properties of the tensor network that depend on the specific tiling. Thus, we expect the values of $c_{\text{eff}}(p,q)$ to converge to a continuous limit as the size of the tensor network increases. 

However, the method of determining $c_{\text{eff}}(p,q)$ by fitting data from the minimum cut across every continuous boundary region, as described in \secref{sec:results}, becomes impractical in large tensor networks. To simplify the fitting process for $c_{\text{eff}}(p,q)$, we choose A to be half of the total boundary region, so $\bulkC(A)$ would be $2l+1$ 
for the \{3,7\} tiling, with $l$ the layer number, and approximately\footnote{For \{5,4\} tiling, $\bulkC(A)$ ranges from 3l+1 and 4l for different regions covering half of the boundary, so we average these values to ${{7 \over 2}l+{1 \over 2} }$ here.} 
$(7l+1)/2$ 
for the \{5,4\} tiling. Given that regions covering half of the boundary predominate in the fitting, this simplification is justified. Implementing this strategy, we use \eqnref{eq:BCtoCC} to obtain 
\eq{
\begin{split}
    c_{\text{eff}}(3,7) &\simeq ({2l+1})/\ln\({N(l,3,7) \over 2}\)\\
    c_{\text{eff}}(5,4) &\simeq \({{7 \over 2}l+{1 \over 2} }\) / \ln\({N(l,5,4) \over 2}\)
\end{split}
\label{eq:rCdiscrete}}
where $N(l,p,q)$ denotes the total boundary region of an l-layer hologrphic RTN with $\{p,q\}$ tiling. Since $N(l,p,q)$ can be easily obtained through the inflation rule, \eqnref{eq:rCdiscrete} facilitates an efficient computation of $c_{\text{eff}}(p,q)$ in large tensor networks.

\begin{figure}[htbp]
    \centering
    \begin{subfigure}[t]{0.4\textwidth}
        \caption{\hspace{0.5\textwidth} (a)}
        \hspace*{-0.6cm}
        \includegraphics[width=\textwidth]{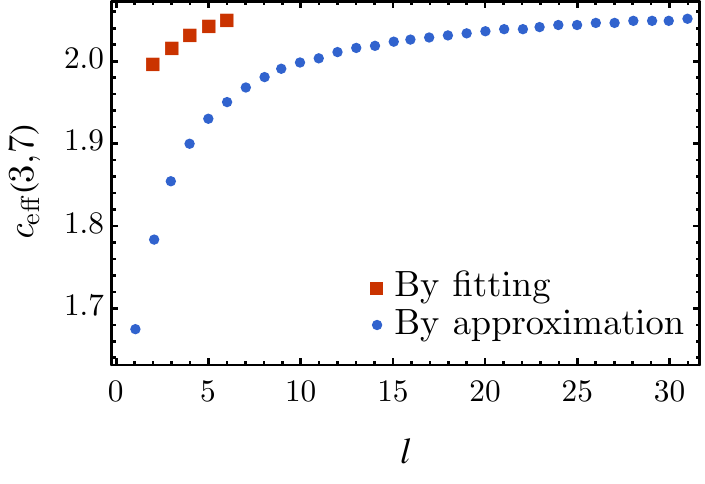}
    \end{subfigure}
    \hfill
    \begin{subfigure}[t]{0.4\textwidth}
        \caption{\hspace{0.5\textwidth} (b)}
        \hspace*{-0.6cm}
        \includegraphics[width=\textwidth]{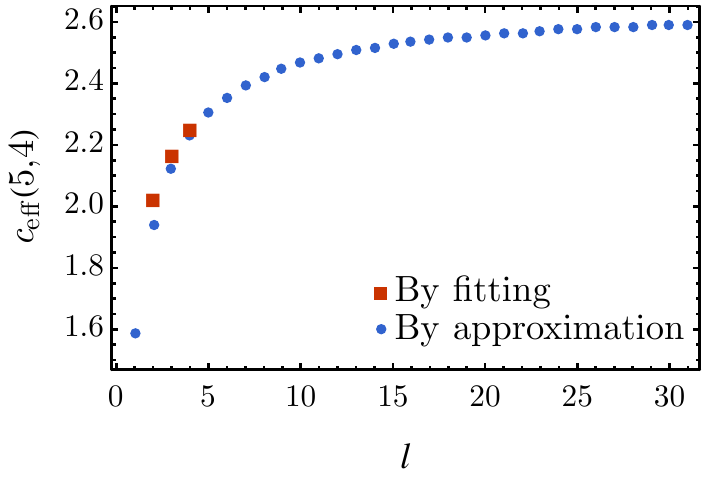}
    \end{subfigure}
    \caption{(a) $c_{\text{eff}}(3,7)$ and (b) $c_{\text{eff}}(5,4)$ calculated by fitting using method in \secref{sec:results} and by approximation in \eqnref{eq:rCdiscrete}. Note that the gap between $c_{\text{eff}}(3,7)$ obtained by the two methods is due to the small boundary size in the first few layers of RTN with \{3,7\} tiling.}\label{fig:HRTNcc}
\end{figure}

In \figref{fig:HRTNcc}, we display plots of $c_{\text{eff}}(3,7)$ and $c_{\text{eff}} (5,4)$ for layer 2 to 31 using \eqnref{eq:rCdiscrete} and compare them with $c_{\text{eff}}(3,7)$ for layer 2 to 6 and $c_{\text{eff}} (5,4)$ for layer 2 to 4 obtained from fitting using the method in \secref{sec:results}. $c_{\text{eff}}$ values derived from both methods exhibit convergence towards certain limits as $l$ increases, and their large-$l$ limits are approximately aligned. This alignment illustrates a finite size effect stemming from the discretization of the bulk by the tensor networks, suggesting that $c_{\text{eff}}$ is likely related to the curvature of the bulk as the space transitions towards a continuous limit. To explore this relationship more rigorously, we will transition from discrete to continuous space, where an analytical expression for $c_{\text{eff}}$ can be derived.

In the continuous space, the metric of the holographic bulk is described by the Poincarcé model:
\eqs{ds^2=(2R)^2{d\rho^2+\rho^2 d\theta^2 \over (1-\rho^2)^2}\label{eq:PDmetric}}
where R is the AdS radius. R characterizes the curvature of the bulk space and is related to the Gaussian curvature $K$ by $K=-{1 \over R^2}$ and to the Ricci scalar $Sc$ by $Sc=-{2 \over R^2}$. 

Consider an arc of length $L$ that spans an angle $\phi$ on a circle centered at the origin of the Poincaré disk. Let $d_L$ denote the geodesic that connects the two endpoints of this arc. In this setting, $c_{\text{eff}}$ is defined as the proportionality constant relating $\ln{L}$ to $d_L$, analogous to \eqref{eq:rCdiscrete} in continuous space.
\eq{c_{\text{eff}} \simeq d_L/ \ln L \label{eq:rC}}
In the context of discrete space, approaching the large-$l$ limit corresponds to the $\rho \rightarrow 1$ limit within continuous space. Similarly, selecting regions that encompass half of the boundary in discrete space corresponds to approaching the limit where $\phi\simeq \pi$ in continuous space. Considering these limits and following the derivation in Appendix~\ref{sec:eff}, we find that the effective central charge is proportional to the AdS radius
\eq{c_{\text{eff}} \simeq 2R\label{eq:cR}}
This demonstrates that the effective central charge is related to the curvature of space by $c_{\text{eff}} \simeq {2 \over \sqrt{-K}} \simeq {2 \sqrt{2} \over \sqrt{-Sc}}$.

Thus, employing a tiling that corresponds to a holographic RTN with more negative curvature results in a smaller $c_{\text{eff}}$, which leads to a smaller deviation from the optimal scaling of the shadow norm. It is important to recognize that the curvature is not solely an intrinsic property of the tiling $\{p,q\}$, but it also depends on the size of each tile. However, in the context of holographic RTN, we assume that each bulk edge being cut corresponds to a unit length, effectively fixing the side length of the tiles. With this fixed side length, the curvature of the holographic RTN is then completely determined by the tiling $\{p,q\}$.

\section{Conclusion \label{summary}}
In this work, we introduced a novel approach to classical shadow tomography that leverages hierarchical quantum circuits to achieve optimal scaling of sample complexity for the estimation of geometrically local Pauli operators. We first explore the tree quantum circuit consisting of two-qudit random unitary gates arranged in a binary tree structure, where half of the qudits are measured out at each layer. We use a recursion method to calculate the shadow norms of Pauli operators under the tree circuit protocol and find that for Pauli operators with a contiguous support of length $k = 2^m$ that exactly fits a subtree, the sample complexity has a favorable scaling, bounded above by
\be
\Vert P\Vert_{\CE_\sigma}^2 < \left(d+\frac{1}{d}\right)^{|A|},
\ee
where $d$ is the local Hilbert space dimension of the qudits. This proves a parametric improvement over random Pauli measurements, where the analogous scaling is $\Vert P\Vert_{\CE_\sigma}^2 \sim \beta^k$ with $\beta = d+1$. It is however a loose bound; values of $\beta$ for finite $d$ are reported in Table~\ref{tab:tree_betas}, and the large-$d$ asymptotic is $\beta = d - \frac{1}{2} d^{-2} + O(d^{-3})$. 

Then, to extend these results to operators with more general support, we explore classical shadow protocols based on holographic RTNs, with measurement performed on the bulk legs of random tensors arranged on a hyperbolic tiling. We use a statistical mechanics model to obtain an approximation of the Pauli learning rate in the large-$d$ limit. We find that at the leading order in $d$, the squared shadow norm for Pauli operators with a contiguous support of length $k$ scales as
\eq{
\Vert P\Vert^2_{\CE_\sigma} \simeq d^{k} f(k) ,
}
where $f(k)$ is a polynomial in $k$ given by
\eq{
f(k) = \text{min}(k,N-k)^{ {c_{\text{eff}}(p,q)}\ln d},
}
$N$ is the system size, and $c_{\text{eff}}(p,q)$ is the effective central charge related to the curvature determined by the tiling of holographic RTN. Therefore, we demonstrate that a randomized measurement scheme for classical shadow tomography based on holographic RTN allows for sample complexity with optimal scaling. 

The optimal scaling of sample complexity is achieved because hierarchical quantum circuits have an inherent self-similar structure characteristic of a critical system --- different layers of the circuit describe correlations at different length scales in a recursive manner, so that measurements performed on each layer extract information with different degrees of locality. Since these circuits are scale-invariant in nature, they treat observables with different sizes equally, so there is no need to adjust circuit depth for different observables or to fine-tune measurement rate, unlike other schemes based on shallow circuits or mid-circuit measurements. 

Holographic circuits are more advantaguous than the tree circuits due to their enhanced translational symmetry at the boundary. However, random tensors present challenges in implementation compared to unitary gates---e.g., one general-purpose approach reviewed in App.~\ref{app:fusion} requires $O(N)$ auxiliary qudits. One future research direction points to exploring more physically realizable tensor network models of hyperbolic space, such as generalized MERA \cite{Anand2023genMERA}, to leverage their theoretical benefits in practical applications. Another possible approach to achieve a holographic entanglement structure with local unitary circuits and measurements is by space-time modulation of the measurement rate to induce desired holographic bulk geometries~\cite{cowsik_entanglementmetric_2023}. We leave these directions to future work. 

{\it Note added.}
While writing this manuscript we became aware of related upcoming work~\cite{harvard_umd_multiscale} on classical shadow tomography via hierarchical circuits. Our works are complementary and our results agree where they overlap. 

\acknowledgments

We acknowledge the helpful discussions with Ahmed A Akhtar, Hong-Ye Hu, Xiao-Liang Qi, and John Preskill. SZ and YZY are supported by the NSF Grant No. DMR-2238360. Numerical simulations were performed in part on HPC resources provided by the Texas Advanced Computing Center (TACC) at the University of Texas at Austin. 

\bibliographystyle{apsrev4-1} 
\bibliography{ref-1.bib}

\begin{thebibliography}{87}%
\makeatletter
\providecommand \@ifxundefined [1]{%
 \@ifx{#1\undefined}
}%
\providecommand \@ifnum [1]{%
 \ifnum #1\expandafter \@firstoftwo
 \else \expandafter \@secondoftwo
 \fi
}%
\providecommand \@ifx [1]{%
 \ifx #1\expandafter \@firstoftwo
 \else \expandafter \@secondoftwo
 \fi
}%
\providecommand \natexlab [1]{#1}%
\providecommand \enquote  [1]{``#1''}%
\providecommand \bibnamefont  [1]{#1}%
\providecommand \bibfnamefont [1]{#1}%
\providecommand \citenamefont [1]{#1}%
\providecommand \href@noop [0]{\@secondoftwo}%
\providecommand \href [0]{\begingroup \@sanitize@url \@href}%
\providecommand \@href[1]{\@@startlink{#1}\@@href}%
\providecommand \@@href[1]{\endgroup#1\@@endlink}%
\providecommand \@sanitize@url [0]{\catcode `\\12\catcode `\$12\catcode
  `\&12\catcode `\#12\catcode `\^12\catcode `\_12\catcode `\%12\relax}%
\providecommand \@@startlink[1]{}%
\providecommand \@@endlink[0]{}%
\providecommand \url  [0]{\begingroup\@sanitize@url \@url }%
\providecommand \@url [1]{\endgroup\@href {#1}{\urlprefix }}%
\providecommand \urlprefix  [0]{URL }%
\providecommand \Eprint [0]{\href }%
\providecommand \doibase [0]{http://dx.doi.org/}%
\providecommand \selectlanguage [0]{\@gobble}%
\providecommand \bibinfo  [0]{\@secondoftwo}%
\providecommand \bibfield  [0]{\@secondoftwo}%
\providecommand \translation [1]{[#1]}%
\providecommand \BibitemOpen [0]{}%
\providecommand \bibitemStop [0]{}%
\providecommand \bibitemNoStop [0]{.\EOS\space}%
\providecommand \EOS [0]{\spacefactor3000\relax}%
\providecommand \BibitemShut  [1]{\csname bibitem#1\endcsname}%
\let\auto@bib@innerbib\@empty
\bibitem [{\citenamefont {Huang}\ \emph {et~al.}(2020)\citenamefont {Huang},
  \citenamefont {Kueng},\ and\ \citenamefont {Preskill}}]{Huang_2020}%
  \BibitemOpen
  \bibfield  {author} {\bibinfo {author} {\bibfnamefont {H.-Y.}\ \bibnamefont
  {Huang}}, \bibinfo {author} {\bibfnamefont {R.}~\bibnamefont {Kueng}}, \ and\
  \bibinfo {author} {\bibfnamefont {J.}~\bibnamefont {Preskill}},\ }\href
  {\doibase 10.1038/s41567-020-0932-7} {\bibfield  {journal} {\bibinfo
  {journal} {Nature Physics}\ }\textbf {\bibinfo {volume} {16}},\ \bibinfo
  {pages} {1050} (\bibinfo {year} {2020})}\BibitemShut {NoStop}%
\bibitem [{\citenamefont {Aaronson}(2020)}]{Aaronson_shadow_2017}%
  \BibitemOpen
  \bibfield  {author} {\bibinfo {author} {\bibfnamefont {S.}~\bibnamefont
  {Aaronson}},\ }\href {\doibase 10.1137/18M120275X} {\bibfield  {journal}
  {\bibinfo  {journal} {SIAM Journal on Computing}\ }\textbf {\bibinfo {volume}
  {49}},\ \bibinfo {pages} {STOC18} (\bibinfo {year} {2020})},\ \Eprint
  {http://arxiv.org/abs/https://doi.org/10.1137/18M120275X}
  {https://doi.org/10.1137/18M120275X} \BibitemShut {NoStop}%
\bibitem [{\citenamefont {{Chen}}\ \emph {et~al.}(2021)\citenamefont {{Chen}},
  \citenamefont {{Yu}}, \citenamefont {{Zeng}},\ and\ \citenamefont
  {{Flammia}}}]{Chen2021R2011.09636}%
  \BibitemOpen
  \bibfield  {author} {\bibinfo {author} {\bibfnamefont {S.}~\bibnamefont
  {{Chen}}}, \bibinfo {author} {\bibfnamefont {W.}~\bibnamefont {{Yu}}},
  \bibinfo {author} {\bibfnamefont {P.}~\bibnamefont {{Zeng}}}, \ and\ \bibinfo
  {author} {\bibfnamefont {S.~T.}\ \bibnamefont {{Flammia}}},\ }\href {\doibase
  10.1103/PRXQuantum.2.030348} {\bibfield  {journal} {\bibinfo  {journal} {PRX
  Quantum}\ }\textbf {\bibinfo {volume} {2}},\ \bibinfo {pages} {030348}
  (\bibinfo {year} {2021})}\BibitemShut {NoStop}%
\bibitem [{\citenamefont {{Enshan Koh}}\ and\ \citenamefont
  {{Grewal}}(2022)}]{Enshan-Koh2020C2011.11580}%
  \BibitemOpen
  \bibfield  {author} {\bibinfo {author} {\bibfnamefont {D.}~\bibnamefont
  {{Enshan Koh}}}\ and\ \bibinfo {author} {\bibfnamefont {S.}~\bibnamefont
  {{Grewal}}},\ }\href {\doibase 10.22331/q-2022-08-16-776} {\bibfield
  {journal} {\bibinfo  {journal} {{Quantum}}\ }\textbf {\bibinfo {volume}
  {6}},\ \bibinfo {pages} {776} (\bibinfo {year} {2022})}\BibitemShut {NoStop}%
\bibitem [{\citenamefont {{Huang}}\ \emph {et~al.}(2021)\citenamefont
  {{Huang}}, \citenamefont {{Kueng}},\ and\ \citenamefont
  {{Preskill}}}]{Huang2021E2103.07510}%
  \BibitemOpen
  \bibfield  {author} {\bibinfo {author} {\bibfnamefont {H.-Y.}\ \bibnamefont
  {{Huang}}}, \bibinfo {author} {\bibfnamefont {R.}~\bibnamefont {{Kueng}}}, \
  and\ \bibinfo {author} {\bibfnamefont {J.}~\bibnamefont {{Preskill}}},\
  }\href {\doibase 10.1103/PhysRevLett.127.030503} {\bibfield  {journal}
  {\bibinfo  {journal} {\prl}\ }\textbf {\bibinfo {volume} {127}},\ \bibinfo
  {pages} {030503} (\bibinfo {year} {2021})}\BibitemShut {NoStop}%
\bibitem [{\citenamefont {{Zhao}}\ \emph {et~al.}(2021)\citenamefont {{Zhao}},
  \citenamefont {{Rubin}},\ and\ \citenamefont
  {{Miyake}}}]{Zhao2021F2010.16094}%
  \BibitemOpen
  \bibfield  {author} {\bibinfo {author} {\bibfnamefont {A.}~\bibnamefont
  {{Zhao}}}, \bibinfo {author} {\bibfnamefont {N.~C.}\ \bibnamefont {{Rubin}}},
  \ and\ \bibinfo {author} {\bibfnamefont {A.}~\bibnamefont {{Miyake}}},\
  }\href {\doibase 10.1103/PhysRevLett.127.110504} {\bibfield  {journal}
  {\bibinfo  {journal} {\prl}\ }\textbf {\bibinfo {volume} {127}},\ \bibinfo
  {pages} {110504} (\bibinfo {year} {2021})}\BibitemShut {NoStop}%
\bibitem [{\citenamefont {{Hu}}\ and\ \citenamefont
  {{You}}(2022)}]{Hu2022H2102.10132}%
  \BibitemOpen
  \bibfield  {author} {\bibinfo {author} {\bibfnamefont {H.-Y.}\ \bibnamefont
  {{Hu}}}\ and\ \bibinfo {author} {\bibfnamefont {Y.-Z.}\ \bibnamefont
  {{You}}},\ }\href {\doibase 10.1103/PhysRevResearch.4.013054} {\bibfield
  {journal} {\bibinfo  {journal} {Physical Review Research}\ }\textbf {\bibinfo
  {volume} {4}},\ \bibinfo {pages} {013054} (\bibinfo {year}
  {2022})}\BibitemShut {NoStop}%
\bibitem [{\citenamefont {{Hu}}\ \emph {et~al.}(2023)\citenamefont {{Hu}},
  \citenamefont {{Choi}},\ and\ \citenamefont {{You}}}]{Hu2023C2107.04817}%
  \BibitemOpen
  \bibfield  {author} {\bibinfo {author} {\bibfnamefont {H.-Y.}\ \bibnamefont
  {{Hu}}}, \bibinfo {author} {\bibfnamefont {S.}~\bibnamefont {{Choi}}}, \ and\
  \bibinfo {author} {\bibfnamefont {Y.-Z.}\ \bibnamefont {{You}}},\ }\href
  {\doibase 10.1103/PhysRevResearch.5.023027} {\bibfield  {journal} {\bibinfo
  {journal} {Physical Review Research}\ }\textbf {\bibinfo {volume} {5}},\
  \bibinfo {pages} {023027} (\bibinfo {year} {2023})}\BibitemShut {NoStop}%
\bibitem [{\citenamefont {{Bu}}\ \emph {et~al.}(2024)\citenamefont {{Bu}},
  \citenamefont {{Koh}}, \citenamefont {{Garcia}},\ and\ \citenamefont
  {{Jaffe}}}]{Bu2024C2202.03272}%
  \BibitemOpen
  \bibfield  {author} {\bibinfo {author} {\bibfnamefont {K.}~\bibnamefont
  {{Bu}}}, \bibinfo {author} {\bibfnamefont {D.~E.}\ \bibnamefont {{Koh}}},
  \bibinfo {author} {\bibfnamefont {R.~J.}\ \bibnamefont {{Garcia}}}, \ and\
  \bibinfo {author} {\bibfnamefont {A.}~\bibnamefont {{Jaffe}}},\ }\href
  {\doibase 10.1038/s41534-023-00801-w} {\bibfield  {journal} {\bibinfo
  {journal} {npj Quantum Information}\ }\textbf {\bibinfo {volume} {10}},\
  \bibinfo {pages} {6} (\bibinfo {year} {2024})}\BibitemShut {NoStop}%
\bibitem [{\citenamefont {{Akhtar}}\ \emph {et~al.}(2023)\citenamefont
  {{Akhtar}}, \citenamefont {{Hu}},\ and\ \citenamefont
  {{You}}}]{Akhtar2023S2209.02093}%
  \BibitemOpen
  \bibfield  {author} {\bibinfo {author} {\bibfnamefont {A.~A.}\ \bibnamefont
  {{Akhtar}}}, \bibinfo {author} {\bibfnamefont {H.-Y.}\ \bibnamefont {{Hu}}},
  \ and\ \bibinfo {author} {\bibfnamefont {Y.-Z.}\ \bibnamefont {{You}}},\
  }\href {\doibase 10.22331/q-2023-06-01-1026} {\bibfield  {journal} {\bibinfo
  {journal} {Quantum}\ }\textbf {\bibinfo {volume} {7}},\ \bibinfo {pages}
  {1026} (\bibinfo {year} {2023})}\BibitemShut {NoStop}%
\bibitem [{\citenamefont {Bertoni}\ \emph {et~al.}(2023)\citenamefont
  {Bertoni}, \citenamefont {Haferkamp}, \citenamefont {Hinsche}, \citenamefont
  {Ioannou}, \citenamefont {Eisert},\ and\ \citenamefont
  {Pashayan}}]{Bertoni2022S2209.12924}%
  \BibitemOpen
  \bibfield  {author} {\bibinfo {author} {\bibfnamefont {C.}~\bibnamefont
  {Bertoni}}, \bibinfo {author} {\bibfnamefont {J.}~\bibnamefont {Haferkamp}},
  \bibinfo {author} {\bibfnamefont {M.}~\bibnamefont {Hinsche}}, \bibinfo
  {author} {\bibfnamefont {M.}~\bibnamefont {Ioannou}}, \bibinfo {author}
  {\bibfnamefont {J.}~\bibnamefont {Eisert}}, \ and\ \bibinfo {author}
  {\bibfnamefont {H.}~\bibnamefont {Pashayan}},\ }\href@noop {} {\enquote
  {\bibinfo {title} {Shallow shadows: Expectation estimation using low-depth
  random clifford circuits},}\ } (\bibinfo {year} {2023}),\ \Eprint
  {http://arxiv.org/abs/2209.12924} {arXiv:2209.12924 [quant-ph]} \BibitemShut
  {NoStop}%
\bibitem [{\citenamefont {{Nguyen}}\ \emph {et~al.}(2022)\citenamefont
  {{Nguyen}}, \citenamefont {{B{\"o}nsel}}, \citenamefont {{Steinberg}},\ and\
  \citenamefont {{G{\"u}hne}}}]{Nguyen2022O}%
  \BibitemOpen
  \bibfield  {author} {\bibinfo {author} {\bibfnamefont {H.~C.}\ \bibnamefont
  {{Nguyen}}}, \bibinfo {author} {\bibfnamefont {J.~L.}\ \bibnamefont
  {{B{\"o}nsel}}}, \bibinfo {author} {\bibfnamefont {J.}~\bibnamefont
  {{Steinberg}}}, \ and\ \bibinfo {author} {\bibfnamefont {O.}~\bibnamefont
  {{G{\"u}hne}}},\ }\href {\doibase 10.1103/PhysRevLett.129.220502} {\bibfield
  {journal} {\bibinfo  {journal} {\prl}\ }\textbf {\bibinfo {volume} {129}},\
  \bibinfo {eid} {220502} (\bibinfo {year} {2022})}\BibitemShut {NoStop}%
\bibitem [{\citenamefont {{Zhou}}\ and\ \citenamefont
  {{Liu}}(2023)}]{Zhou2023P2212.11068}%
  \BibitemOpen
  \bibfield  {author} {\bibinfo {author} {\bibfnamefont {Y.}~\bibnamefont
  {{Zhou}}}\ and\ \bibinfo {author} {\bibfnamefont {Q.}~\bibnamefont {{Liu}}},\
  }\href {\doibase 10.22331/q-2023-06-29-1044} {\bibfield  {journal} {\bibinfo
  {journal} {Quantum}\ }\textbf {\bibinfo {volume} {7}},\ \bibinfo {pages}
  {1044} (\bibinfo {year} {2023})}\BibitemShut {NoStop}%
\bibitem [{\citenamefont {Zhou}\ and\ \citenamefont
  {Zhang}(2023)}]{Zhou2023E2309.01258}%
  \BibitemOpen
  \bibfield  {author} {\bibinfo {author} {\bibfnamefont {T.-G.}\ \bibnamefont
  {Zhou}}\ and\ \bibinfo {author} {\bibfnamefont {P.}~\bibnamefont {Zhang}},\
  }\href@noop {} {\enquote {\bibinfo {title} {Efficient classical shadow
  tomography through many-body localization dynamics},}\ } (\bibinfo {year}
  {2023}),\ \Eprint {http://arxiv.org/abs/2309.01258} {arXiv:2309.01258
  [quant-ph]} \BibitemShut {NoStop}%
\bibitem [{\citenamefont {Wu}\ and\ \citenamefont
  {Koh}(2023)}]{Wu2023E2310.12726}%
  \BibitemOpen
  \bibfield  {author} {\bibinfo {author} {\bibfnamefont {B.}~\bibnamefont
  {Wu}}\ and\ \bibinfo {author} {\bibfnamefont {D.~E.}\ \bibnamefont {Koh}},\
  }\href@noop {} {\enquote {\bibinfo {title} {Error-mitigated fermionic
  classical shadows on noisy quantum devices},}\ } (\bibinfo {year} {2023}),\
  \Eprint {http://arxiv.org/abs/2310.12726} {arXiv:2310.12726 [quant-ph]}
  \BibitemShut {NoStop}%
\bibitem [{\citenamefont {{Arienzo}}\ \emph {et~al.}(2022)\citenamefont
  {{Arienzo}}, \citenamefont {{Heinrich}}, \citenamefont {{Roth}},\ and\
  \citenamefont {{Kliesch}}}]{2022arXiv221109835A}%
  \BibitemOpen
  \bibfield  {author} {\bibinfo {author} {\bibfnamefont {M.}~\bibnamefont
  {{Arienzo}}}, \bibinfo {author} {\bibfnamefont {M.}~\bibnamefont
  {{Heinrich}}}, \bibinfo {author} {\bibfnamefont {I.}~\bibnamefont {{Roth}}},
  \ and\ \bibinfo {author} {\bibfnamefont {M.}~\bibnamefont {{Kliesch}}},\
  }\href {\doibase 10.48550/arXiv.2211.09835} {\bibfield  {journal} {\bibinfo
  {journal} {arXiv e-prints}\ ,\ \bibinfo {eid} {arXiv:2211.09835}} (\bibinfo
  {year} {2022})},\ \Eprint {http://arxiv.org/abs/2211.09835} {arXiv:2211.09835
  [quant-ph]} \BibitemShut {NoStop}%
\bibitem [{\citenamefont {Ippoliti}\ \emph {et~al.}(2023)\citenamefont
  {Ippoliti}, \citenamefont {Li}, \citenamefont {Rakovszky},\ and\
  \citenamefont {Khemani}}]{Ippoliti_2023}%
  \BibitemOpen
  \bibfield  {author} {\bibinfo {author} {\bibfnamefont {M.}~\bibnamefont
  {Ippoliti}}, \bibinfo {author} {\bibfnamefont {Y.}~\bibnamefont {Li}},
  \bibinfo {author} {\bibfnamefont {T.}~\bibnamefont {Rakovszky}}, \ and\
  \bibinfo {author} {\bibfnamefont {V.}~\bibnamefont {Khemani}},\ }\href
  {\doibase 10.1103/physrevlett.130.230403} {\bibfield  {journal} {\bibinfo
  {journal} {Physical Review Letters}\ }\textbf {\bibinfo {volume} {130}}
  (\bibinfo {year} {2023}),\ 10.1103/physrevlett.130.230403}\BibitemShut
  {NoStop}%
\bibitem [{\citenamefont {{Ippoliti}}(2024)}]{Ippoliti2024C2305.10723}%
  \BibitemOpen
  \bibfield  {author} {\bibinfo {author} {\bibfnamefont {M.}~\bibnamefont
  {{Ippoliti}}},\ }\href {\doibase 10.22331/q-2024-03-21-1293} {\bibfield
  {journal} {\bibinfo  {journal} {Quantum}\ }\textbf {\bibinfo {volume} {8}},\
  \bibinfo {pages} {1293} (\bibinfo {year} {2024})},\ \Eprint
  {http://arxiv.org/abs/2305.10723} {arXiv:2305.10723 [quant-ph]} \BibitemShut
  {NoStop}%
\bibitem [{\citenamefont {Elben}\ \emph {et~al.}(2018)\citenamefont {Elben},
  \citenamefont {Vermersch}, \citenamefont {Dalmonte}, \citenamefont {Cirac},\
  and\ \citenamefont {Zoller}}]{PhysRevLett.120.050406}%
  \BibitemOpen
  \bibfield  {author} {\bibinfo {author} {\bibfnamefont {A.}~\bibnamefont
  {Elben}}, \bibinfo {author} {\bibfnamefont {B.}~\bibnamefont {Vermersch}},
  \bibinfo {author} {\bibfnamefont {M.}~\bibnamefont {Dalmonte}}, \bibinfo
  {author} {\bibfnamefont {J.~I.}\ \bibnamefont {Cirac}}, \ and\ \bibinfo
  {author} {\bibfnamefont {P.}~\bibnamefont {Zoller}},\ }\href {\doibase
  10.1103/PhysRevLett.120.050406} {\bibfield  {journal} {\bibinfo  {journal}
  {Phys. Rev. Lett.}\ }\textbf {\bibinfo {volume} {120}},\ \bibinfo {pages}
  {050406} (\bibinfo {year} {2018})}\BibitemShut {NoStop}%
\bibitem [{\citenamefont {Brydges}\ \emph {et~al.}(2019)\citenamefont
  {Brydges}, \citenamefont {Elben}, \citenamefont {Jurcevic}, \citenamefont
  {Vermersch}, \citenamefont {Maier}, \citenamefont {Lanyon}, \citenamefont
  {Zoller}, \citenamefont {Blatt},\ and\ \citenamefont
  {Roos}}]{doi:10.1126/science.aau4963}%
  \BibitemOpen
  \bibfield  {author} {\bibinfo {author} {\bibfnamefont {T.}~\bibnamefont
  {Brydges}}, \bibinfo {author} {\bibfnamefont {A.}~\bibnamefont {Elben}},
  \bibinfo {author} {\bibfnamefont {P.}~\bibnamefont {Jurcevic}}, \bibinfo
  {author} {\bibfnamefont {B.}~\bibnamefont {Vermersch}}, \bibinfo {author}
  {\bibfnamefont {C.}~\bibnamefont {Maier}}, \bibinfo {author} {\bibfnamefont
  {B.~P.}\ \bibnamefont {Lanyon}}, \bibinfo {author} {\bibfnamefont
  {P.}~\bibnamefont {Zoller}}, \bibinfo {author} {\bibfnamefont
  {R.}~\bibnamefont {Blatt}}, \ and\ \bibinfo {author} {\bibfnamefont {C.~F.}\
  \bibnamefont {Roos}},\ }\href {\doibase 10.1126/science.aau4963} {\bibfield
  {journal} {\bibinfo  {journal} {Science}\ }\textbf {\bibinfo {volume}
  {364}},\ \bibinfo {pages} {260} (\bibinfo {year} {2019})}\BibitemShut
  {NoStop}%
\bibitem [{\citenamefont {Acharya}\ \emph {et~al.}(2021)\citenamefont
  {Acharya}, \citenamefont {Saha},\ and\ \citenamefont
  {Sengupta}}]{2021arXiv210505992A}%
  \BibitemOpen
  \bibfield  {author} {\bibinfo {author} {\bibfnamefont {A.}~\bibnamefont
  {Acharya}}, \bibinfo {author} {\bibfnamefont {S.}~\bibnamefont {Saha}}, \
  and\ \bibinfo {author} {\bibfnamefont {A.~M.}\ \bibnamefont {Sengupta}},\
  }\href@noop {} {\enquote {\bibinfo {title} {Informationally complete
  povm-based shadow tomography},}\ } (\bibinfo {year} {2021}),\ \Eprint
  {http://arxiv.org/abs/2105.05992} {arXiv:2105.05992 [quant-ph]} \BibitemShut
  {NoStop}%
\bibitem [{\citenamefont {Zhou}\ and\ \citenamefont
  {Liu}(2023)}]{2022arXiv220808416Z}%
  \BibitemOpen
  \bibfield  {author} {\bibinfo {author} {\bibfnamefont {Y.}~\bibnamefont
  {Zhou}}\ and\ \bibinfo {author} {\bibfnamefont {Z.}~\bibnamefont {Liu}},\
  }\href@noop {} {\enquote {\bibinfo {title} {A hybrid framework for estimating
  nonlinear functions of quantum states},}\ } (\bibinfo {year} {2023}),\
  \Eprint {http://arxiv.org/abs/2208.08416} {arXiv:2208.08416 [quant-ph]}
  \BibitemShut {NoStop}%
\bibitem [{\citenamefont {Garcia}\ \emph {et~al.}(2021)\citenamefont {Garcia},
  \citenamefont {Zhou},\ and\ \citenamefont
  {Jaffe}}]{PhysRevResearch.3.033155}%
  \BibitemOpen
  \bibfield  {author} {\bibinfo {author} {\bibfnamefont {R.~J.}\ \bibnamefont
  {Garcia}}, \bibinfo {author} {\bibfnamefont {Y.}~\bibnamefont {Zhou}}, \ and\
  \bibinfo {author} {\bibfnamefont {A.}~\bibnamefont {Jaffe}},\ }\href
  {\doibase 10.1103/PhysRevResearch.3.033155} {\bibfield  {journal} {\bibinfo
  {journal} {Phys. Rev. Res.}\ }\textbf {\bibinfo {volume} {3}},\ \bibinfo
  {pages} {033155} (\bibinfo {year} {2021})}\BibitemShut {NoStop}%
\bibitem [{\citenamefont {Helsen}\ and\ \citenamefont
  {Walter}(2023)}]{PhysRevLett.131.240602}%
  \BibitemOpen
  \bibfield  {author} {\bibinfo {author} {\bibfnamefont {J.}~\bibnamefont
  {Helsen}}\ and\ \bibinfo {author} {\bibfnamefont {M.}~\bibnamefont
  {Walter}},\ }\href {\doibase 10.1103/PhysRevLett.131.240602} {\bibfield
  {journal} {\bibinfo  {journal} {Phys. Rev. Lett.}\ }\textbf {\bibinfo
  {volume} {131}},\ \bibinfo {pages} {240602} (\bibinfo {year}
  {2023})}\BibitemShut {NoStop}%
\bibitem [{\citenamefont {{Wan}}\ \emph {et~al.}(2023)\citenamefont {{Wan}},
  \citenamefont {{Huggins}}, \citenamefont {{Lee}},\ and\ \citenamefont
  {{Babbush}}}]{2023CMaPh.404..629W}%
  \BibitemOpen
  \bibfield  {author} {\bibinfo {author} {\bibfnamefont {K.}~\bibnamefont
  {{Wan}}}, \bibinfo {author} {\bibfnamefont {W.~J.}\ \bibnamefont
  {{Huggins}}}, \bibinfo {author} {\bibfnamefont {J.}~\bibnamefont {{Lee}}}, \
  and\ \bibinfo {author} {\bibfnamefont {R.}~\bibnamefont {{Babbush}}},\ }\href
  {\doibase 10.1007/s00220-023-04844-0} {\bibfield  {journal} {\bibinfo
  {journal} {Communications in Mathematical Physics}\ }\textbf {\bibinfo
  {volume} {404}},\ \bibinfo {pages} {629} (\bibinfo {year} {2023})},\ \Eprint
  {http://arxiv.org/abs/2207.13723} {arXiv:2207.13723 [quant-ph]} \BibitemShut
  {NoStop}%
\bibitem [{\citenamefont {Tran}\ \emph {et~al.}(2023)\citenamefont {Tran},
  \citenamefont {Mark}, \citenamefont {Ho},\ and\ \citenamefont
  {Choi}}]{PhysRevX.13.011049}%
  \BibitemOpen
  \bibfield  {author} {\bibinfo {author} {\bibfnamefont {M.~C.}\ \bibnamefont
  {Tran}}, \bibinfo {author} {\bibfnamefont {D.~K.}\ \bibnamefont {Mark}},
  \bibinfo {author} {\bibfnamefont {W.~W.}\ \bibnamefont {Ho}}, \ and\ \bibinfo
  {author} {\bibfnamefont {S.}~\bibnamefont {Choi}},\ }\href {\doibase
  10.1103/PhysRevX.13.011049} {\bibfield  {journal} {\bibinfo  {journal} {Phys.
  Rev. X}\ }\textbf {\bibinfo {volume} {13}},\ \bibinfo {pages} {011049}
  (\bibinfo {year} {2023})}\BibitemShut {NoStop}%
\bibitem [{\citenamefont {McGinley}\ and\ \citenamefont
  {Fava}(2023)}]{PhysRevLett.131.160601}%
  \BibitemOpen
  \bibfield  {author} {\bibinfo {author} {\bibfnamefont {M.}~\bibnamefont
  {McGinley}}\ and\ \bibinfo {author} {\bibfnamefont {M.}~\bibnamefont
  {Fava}},\ }\href {\doibase 10.1103/PhysRevLett.131.160601} {\bibfield
  {journal} {\bibinfo  {journal} {Phys. Rev. Lett.}\ }\textbf {\bibinfo
  {volume} {131}},\ \bibinfo {pages} {160601} (\bibinfo {year}
  {2023})}\BibitemShut {NoStop}%
\bibitem [{\citenamefont {Denzler}\ \emph {et~al.}(2023)\citenamefont
  {Denzler}, \citenamefont {Mele}, \citenamefont {Derbyshire}, \citenamefont
  {Guaita},\ and\ \citenamefont {Eisert}}]{2023arXiv230912933D}%
  \BibitemOpen
  \bibfield  {author} {\bibinfo {author} {\bibfnamefont {J.}~\bibnamefont
  {Denzler}}, \bibinfo {author} {\bibfnamefont {A.~A.}\ \bibnamefont {Mele}},
  \bibinfo {author} {\bibfnamefont {E.}~\bibnamefont {Derbyshire}}, \bibinfo
  {author} {\bibfnamefont {T.}~\bibnamefont {Guaita}}, \ and\ \bibinfo {author}
  {\bibfnamefont {J.}~\bibnamefont {Eisert}},\ }\href@noop {} {\enquote
  {\bibinfo {title} {Learning fermionic correlations by evolving with random
  translationally invariant hamiltonians},}\ } (\bibinfo {year} {2023}),\
  \Eprint {http://arxiv.org/abs/2309.12933} {arXiv:2309.12933 [quant-ph]}
  \BibitemShut {NoStop}%
\bibitem [{\citenamefont {Imai}\ \emph {et~al.}(2023)\citenamefont {Imai},
  \citenamefont {T{\'o}th},\ and\ \citenamefont
  {G{\"u}hne}}]{2023arXiv230910745I}%
  \BibitemOpen
  \bibfield  {author} {\bibinfo {author} {\bibfnamefont {S.}~\bibnamefont
  {Imai}}, \bibinfo {author} {\bibfnamefont {G.}~\bibnamefont {T{\'o}th}}, \
  and\ \bibinfo {author} {\bibfnamefont {O.}~\bibnamefont {G{\"u}hne}},\
  }\href@noop {} {\enquote {\bibinfo {title} {Collective randomized
  measurements in quantum information processing},}\ } (\bibinfo {year}
  {2023}),\ \Eprint {http://arxiv.org/abs/2309.10745} {arXiv:2309.10745
  [quant-ph]} \BibitemShut {NoStop}%
\bibitem [{\citenamefont {Kirk}\ \emph {et~al.}(2022)\citenamefont {Kirk},
  \citenamefont {Cotler}, \citenamefont {Huang},\ and\ \citenamefont
  {Lukin}}]{Van-Kirk2022H2212.06084}%
  \BibitemOpen
  \bibfield  {author} {\bibinfo {author} {\bibfnamefont {K.~V.}\ \bibnamefont
  {Kirk}}, \bibinfo {author} {\bibfnamefont {J.}~\bibnamefont {Cotler}},
  \bibinfo {author} {\bibfnamefont {H.-Y.}\ \bibnamefont {Huang}}, \ and\
  \bibinfo {author} {\bibfnamefont {M.~D.}\ \bibnamefont {Lukin}},\ }\href@noop
  {} {\enquote {\bibinfo {title} {Hardware-efficient learning of quantum
  many-body states},}\ } (\bibinfo {year} {2022}),\ \Eprint
  {http://arxiv.org/abs/2212.06084} {arXiv:2212.06084 [quant-ph]} \BibitemShut
  {NoStop}%
\bibitem [{\citenamefont {Liu}\ \emph {et~al.}(2023)\citenamefont {Liu},
  \citenamefont {Hao},\ and\ \citenamefont {Hu}}]{2023arXiv231100695L}%
  \BibitemOpen
  \bibfield  {author} {\bibinfo {author} {\bibfnamefont {Z.}~\bibnamefont
  {Liu}}, \bibinfo {author} {\bibfnamefont {Z.}~\bibnamefont {Hao}}, \ and\
  \bibinfo {author} {\bibfnamefont {H.-Y.}\ \bibnamefont {Hu}},\ }\href@noop {}
  {\enquote {\bibinfo {title} {Predicting arbitrary state properties from
  single hamiltonian quench dynamics},}\ } (\bibinfo {year} {2023}),\ \Eprint
  {http://arxiv.org/abs/2311.00695} {arXiv:2311.00695 [quant-ph]} \BibitemShut
  {NoStop}%
\bibitem [{\citenamefont {Elben}\ \emph
  {et~al.}(2020{\natexlab{a}})\citenamefont {Elben}, \citenamefont {Kueng},
  \citenamefont {Huang}, \citenamefont {van Bijnen}, \citenamefont {Kokail},
  \citenamefont {Dalmonte}, \citenamefont {Calabrese}, \citenamefont {Kraus},
  \citenamefont {Preskill}, \citenamefont {Zoller},\ and\ \citenamefont
  {Vermersch}}]{PhysRevLett.125.200501}%
  \BibitemOpen
  \bibfield  {author} {\bibinfo {author} {\bibfnamefont {A.}~\bibnamefont
  {Elben}}, \bibinfo {author} {\bibfnamefont {R.}~\bibnamefont {Kueng}},
  \bibinfo {author} {\bibfnamefont {H.-Y.~R.}\ \bibnamefont {Huang}}, \bibinfo
  {author} {\bibfnamefont {R.}~\bibnamefont {van Bijnen}}, \bibinfo {author}
  {\bibfnamefont {C.}~\bibnamefont {Kokail}}, \bibinfo {author} {\bibfnamefont
  {M.}~\bibnamefont {Dalmonte}}, \bibinfo {author} {\bibfnamefont
  {P.}~\bibnamefont {Calabrese}}, \bibinfo {author} {\bibfnamefont
  {B.}~\bibnamefont {Kraus}}, \bibinfo {author} {\bibfnamefont
  {J.}~\bibnamefont {Preskill}}, \bibinfo {author} {\bibfnamefont
  {P.}~\bibnamefont {Zoller}}, \ and\ \bibinfo {author} {\bibfnamefont
  {B.}~\bibnamefont {Vermersch}},\ }\href {\doibase
  10.1103/PhysRevLett.125.200501} {\bibfield  {journal} {\bibinfo  {journal}
  {Phys. Rev. Lett.}\ }\textbf {\bibinfo {volume} {125}},\ \bibinfo {pages}
  {200501} (\bibinfo {year} {2020}{\natexlab{a}})}\BibitemShut {NoStop}%
\bibitem [{\citenamefont {Fawzi}\ \emph {et~al.}(2024)\citenamefont {Fawzi},
  \citenamefont {Kueng}, \citenamefont {Markham},\ and\ \citenamefont
  {Oufkir}}]{2024arXiv240116922F}%
  \BibitemOpen
  \bibfield  {author} {\bibinfo {author} {\bibfnamefont {O.}~\bibnamefont
  {Fawzi}}, \bibinfo {author} {\bibfnamefont {R.}~\bibnamefont {Kueng}},
  \bibinfo {author} {\bibfnamefont {D.}~\bibnamefont {Markham}}, \ and\
  \bibinfo {author} {\bibfnamefont {A.}~\bibnamefont {Oufkir}},\ }\href@noop {}
  {\enquote {\bibinfo {title} {Learning properties of quantum states without
  the i.i.d. assumption},}\ } (\bibinfo {year} {2024}),\ \Eprint
  {http://arxiv.org/abs/2401.16922} {arXiv:2401.16922 [quant-ph]} \BibitemShut
  {NoStop}%
\bibitem [{\citenamefont {Fischer}\ \emph {et~al.}(2024)\citenamefont
  {Fischer}, \citenamefont {Dao}, \citenamefont {Tavernelli},\ and\
  \citenamefont {Tacchino}}]{2024arXiv240118071F}%
  \BibitemOpen
  \bibfield  {author} {\bibinfo {author} {\bibfnamefont {L.~E.}\ \bibnamefont
  {Fischer}}, \bibinfo {author} {\bibfnamefont {T.}~\bibnamefont {Dao}},
  \bibinfo {author} {\bibfnamefont {I.}~\bibnamefont {Tavernelli}}, \ and\
  \bibinfo {author} {\bibfnamefont {F.}~\bibnamefont {Tacchino}},\ }\href@noop
  {} {\enquote {\bibinfo {title} {Dual frame optimization for informationally
  complete quantum measurements},}\ } (\bibinfo {year} {2024}),\ \Eprint
  {http://arxiv.org/abs/2401.18071} {arXiv:2401.18071 [quant-ph]} \BibitemShut
  {NoStop}%
\bibitem [{\citenamefont {{Ippoliti}}\ and\ \citenamefont
  {{Khemani}}(2024)}]{Ippoliti2024L2307.15011}%
  \BibitemOpen
  \bibfield  {author} {\bibinfo {author} {\bibfnamefont {M.}~\bibnamefont
  {{Ippoliti}}}\ and\ \bibinfo {author} {\bibfnamefont {V.}~\bibnamefont
  {{Khemani}}},\ }\href {\doibase 10.1103/PRXQuantum.5.020304} {\bibfield
  {journal} {\bibinfo  {journal} {PRX Quantum}\ }\textbf {\bibinfo {volume}
  {5}},\ \bibinfo {eid} {020304} (\bibinfo {year} {2024})},\ \Eprint
  {http://arxiv.org/abs/2307.15011} {arXiv:2307.15011 [quant-ph]} \BibitemShut
  {NoStop}%
\bibitem [{\citenamefont {{Akhtar}}\ \emph
  {et~al.}(2024{\natexlab{a}})\citenamefont {{Akhtar}}, \citenamefont {{Hu}},\
  and\ \citenamefont {{You}}}]{Akhtar2024M2308.01653}%
  \BibitemOpen
  \bibfield  {author} {\bibinfo {author} {\bibfnamefont {A.~A.}\ \bibnamefont
  {{Akhtar}}}, \bibinfo {author} {\bibfnamefont {H.-Y.}\ \bibnamefont {{Hu}}},
  \ and\ \bibinfo {author} {\bibfnamefont {Y.-Z.}\ \bibnamefont {{You}}},\
  }\href {\doibase 10.1103/PhysRevB.109.094209} {\bibfield  {journal} {\bibinfo
   {journal} {\prb}\ }\textbf {\bibinfo {volume} {109}},\ \bibinfo {eid}
  {094209} (\bibinfo {year} {2024}{\natexlab{a}})},\ \Eprint
  {http://arxiv.org/abs/2308.01653} {arXiv:2308.01653 [quant-ph]} \BibitemShut
  {NoStop}%
\bibitem [{\citenamefont {{Akhtar}}\ \emph
  {et~al.}(2024{\natexlab{b}})\citenamefont {{Akhtar}}, \citenamefont
  {{Anand}}, \citenamefont {{Marshall}},\ and\ \citenamefont
  {{You}}}]{Akhtar2024D2404.01068}%
  \BibitemOpen
  \bibfield  {author} {\bibinfo {author} {\bibfnamefont {A.~A.}\ \bibnamefont
  {{Akhtar}}}, \bibinfo {author} {\bibfnamefont {N.}~\bibnamefont {{Anand}}},
  \bibinfo {author} {\bibfnamefont {J.}~\bibnamefont {{Marshall}}}, \ and\
  \bibinfo {author} {\bibfnamefont {Y.-Z.}\ \bibnamefont {{You}}},\ }\href
  {\doibase 10.48550/arXiv.2404.01068} {\bibfield  {journal} {\bibinfo
  {journal} {arXiv e-prints}\ ,\ \bibinfo {eid} {arXiv:2404.01068}} (\bibinfo
  {year} {2024}{\natexlab{b}})},\ \Eprint {http://arxiv.org/abs/2404.01068}
  {arXiv:2404.01068 [quant-ph]} \BibitemShut {NoStop}%
\bibitem [{\citenamefont {Haah}\ \emph {et~al.}(2017)\citenamefont {Haah},
  \citenamefont {Harrow}, \citenamefont {Ji}, \citenamefont {Wu},\ and\
  \citenamefont {Yu}}]{Haah2015S1508.01797}%
  \BibitemOpen
  \bibfield  {author} {\bibinfo {author} {\bibfnamefont {J.}~\bibnamefont
  {Haah}}, \bibinfo {author} {\bibfnamefont {A.~W.}\ \bibnamefont {Harrow}},
  \bibinfo {author} {\bibfnamefont {Z.}~\bibnamefont {Ji}}, \bibinfo {author}
  {\bibfnamefont {X.}~\bibnamefont {Wu}}, \ and\ \bibinfo {author}
  {\bibfnamefont {N.}~\bibnamefont {Yu}},\ }\href {\doibase
  10.1109/tit.2017.2719044} {\bibfield  {journal} {\bibinfo  {journal} {IEEE
  Transactions on Information Theory}\ } (\bibinfo {year} {2017}),\
  10.1109/tit.2017.2719044}\BibitemShut {NoStop}%
\bibitem [{\citenamefont {O'Donnell}\ and\ \citenamefont
  {Wright}(2015)}]{ODonnell2015E1508.01907}%
  \BibitemOpen
  \bibfield  {author} {\bibinfo {author} {\bibfnamefont {R.}~\bibnamefont
  {O'Donnell}}\ and\ \bibinfo {author} {\bibfnamefont {J.}~\bibnamefont
  {Wright}},\ }\href@noop {} {\enquote {\bibinfo {title} {Efficient quantum
  tomography},}\ } (\bibinfo {year} {2015}),\ \Eprint
  {http://arxiv.org/abs/1508.01907} {arXiv:1508.01907 [quant-ph]} \BibitemShut
  {NoStop}%
\bibitem [{\citenamefont {{Flammia}}\ \emph {et~al.}(2012)\citenamefont
  {{Flammia}}, \citenamefont {{Gross}}, \citenamefont {{Liu}},\ and\
  \citenamefont {{Eisert}}}]{Flammia2012Q1205.2300}%
  \BibitemOpen
  \bibfield  {author} {\bibinfo {author} {\bibfnamefont {S.~T.}\ \bibnamefont
  {{Flammia}}}, \bibinfo {author} {\bibfnamefont {D.}~\bibnamefont {{Gross}}},
  \bibinfo {author} {\bibfnamefont {Y.-K.}\ \bibnamefont {{Liu}}}, \ and\
  \bibinfo {author} {\bibfnamefont {J.}~\bibnamefont {{Eisert}}},\ }\href
  {\doibase 10.1088/1367-2630/14/9/095022} {\bibfield  {journal} {\bibinfo
  {journal} {New Journal of Physics}\ }\textbf {\bibinfo {volume} {14}},\
  \bibinfo {pages} {095022} (\bibinfo {year} {2012})}\BibitemShut {NoStop}%
\bibitem [{\citenamefont {Huang}\ \emph
  {et~al.}(2022{\natexlab{a}})\citenamefont {Huang}, \citenamefont {Broughton},
  \citenamefont {Cotler}, \citenamefont {Chen}, \citenamefont {Li},
  \citenamefont {Mohseni}, \citenamefont {Neven}, \citenamefont {Babbush},
  \citenamefont {Kueng}, \citenamefont {Preskill},\ and\ \citenamefont
  {McClean}}]{doi:10.1126/science.abn7293}%
  \BibitemOpen
  \bibfield  {author} {\bibinfo {author} {\bibfnamefont {H.-Y.}\ \bibnamefont
  {Huang}}, \bibinfo {author} {\bibfnamefont {M.}~\bibnamefont {Broughton}},
  \bibinfo {author} {\bibfnamefont {J.}~\bibnamefont {Cotler}}, \bibinfo
  {author} {\bibfnamefont {S.}~\bibnamefont {Chen}}, \bibinfo {author}
  {\bibfnamefont {J.}~\bibnamefont {Li}}, \bibinfo {author} {\bibfnamefont
  {M.}~\bibnamefont {Mohseni}}, \bibinfo {author} {\bibfnamefont
  {H.}~\bibnamefont {Neven}}, \bibinfo {author} {\bibfnamefont
  {R.}~\bibnamefont {Babbush}}, \bibinfo {author} {\bibfnamefont
  {R.}~\bibnamefont {Kueng}}, \bibinfo {author} {\bibfnamefont
  {J.}~\bibnamefont {Preskill}}, \ and\ \bibinfo {author} {\bibfnamefont
  {J.~R.}\ \bibnamefont {McClean}},\ }\href {\doibase 10.1126/science.abn7293}
  {\bibfield  {journal} {\bibinfo  {journal} {Science}\ }\textbf {\bibinfo
  {volume} {376}},\ \bibinfo {pages} {1182} (\bibinfo {year}
  {2022}{\natexlab{a}})}\BibitemShut {NoStop}%
\bibitem [{\citenamefont {Zhu}\ \emph {et~al.}(2022)\citenamefont {Zhu},
  \citenamefont {Cian}, \citenamefont {Noel}, \citenamefont {Risinger},
  \citenamefont {Biswas}, \citenamefont {Egan}, \citenamefont {Zhu},
  \citenamefont {Green}, \citenamefont {Alderete}, \citenamefont {Nguyen},
  \citenamefont {Wang}, \citenamefont {Maksymov}, \citenamefont {Nam},
  \citenamefont {Cetina}, \citenamefont {Linke}, \citenamefont {Hafezi},\ and\
  \citenamefont {Monroe}}]{CrossPlatform}%
  \BibitemOpen
  \bibfield  {author} {\bibinfo {author} {\bibfnamefont {D.}~\bibnamefont
  {Zhu}}, \bibinfo {author} {\bibfnamefont {Z.~P.}\ \bibnamefont {Cian}},
  \bibinfo {author} {\bibfnamefont {C.}~\bibnamefont {Noel}}, \bibinfo {author}
  {\bibfnamefont {A.}~\bibnamefont {Risinger}}, \bibinfo {author}
  {\bibfnamefont {D.}~\bibnamefont {Biswas}}, \bibinfo {author} {\bibfnamefont
  {L.}~\bibnamefont {Egan}}, \bibinfo {author} {\bibfnamefont {Y.}~\bibnamefont
  {Zhu}}, \bibinfo {author} {\bibfnamefont {A.~M.}\ \bibnamefont {Green}},
  \bibinfo {author} {\bibfnamefont {C.~H.}\ \bibnamefont {Alderete}}, \bibinfo
  {author} {\bibfnamefont {N.~H.}\ \bibnamefont {Nguyen}}, \bibinfo {author}
  {\bibfnamefont {Q.}~\bibnamefont {Wang}}, \bibinfo {author} {\bibfnamefont
  {A.}~\bibnamefont {Maksymov}}, \bibinfo {author} {\bibfnamefont
  {Y.}~\bibnamefont {Nam}}, \bibinfo {author} {\bibfnamefont {M.}~\bibnamefont
  {Cetina}}, \bibinfo {author} {\bibfnamefont {N.~M.}\ \bibnamefont {Linke}},
  \bibinfo {author} {\bibfnamefont {M.}~\bibnamefont {Hafezi}}, \ and\ \bibinfo
  {author} {\bibfnamefont {C.}~\bibnamefont {Monroe}},\ }\href {\doibase
  10.1038/s41467-022-34279-5} {\bibfield  {journal} {\bibinfo  {journal}
  {Nature Communications}\ }\textbf {\bibinfo {volume} {13}},\ \bibinfo {pages}
  {6620} (\bibinfo {year} {2022})}\BibitemShut {NoStop}%
\bibitem [{\citenamefont {Struchalin}\ \emph {et~al.}(2021)\citenamefont
  {Struchalin}, \citenamefont {Zagorovskii}, \citenamefont {Kovlakov},
  \citenamefont {Straupe},\ and\ \citenamefont
  {Kulik}}]{Struchalin2020E2008.05234}%
  \BibitemOpen
  \bibfield  {author} {\bibinfo {author} {\bibfnamefont {G.}~\bibnamefont
  {Struchalin}}, \bibinfo {author} {\bibfnamefont {Y.~A.}\ \bibnamefont
  {Zagorovskii}}, \bibinfo {author} {\bibfnamefont {E.}~\bibnamefont
  {Kovlakov}}, \bibinfo {author} {\bibfnamefont {S.}~\bibnamefont {Straupe}}, \
  and\ \bibinfo {author} {\bibfnamefont {S.}~\bibnamefont {Kulik}},\ }\href
  {\doibase 10.1103/prxquantum.2.010307} {\bibfield  {journal} {\bibinfo
  {journal} {PRX Quantum}\ }\textbf {\bibinfo {volume} {2}} (\bibinfo {year}
  {2021}),\ 10.1103/prxquantum.2.010307}\BibitemShut {NoStop}%
\bibitem [{\citenamefont {{Zhang}}\ \emph {et~al.}(2021)\citenamefont
  {{Zhang}}, \citenamefont {{Sun}}, \citenamefont {{Fang}}, \citenamefont
  {{Zhang}}, \citenamefont {{Yuan}},\ and\ \citenamefont
  {{Lu}}}]{Zhang2021E2106.10190}%
  \BibitemOpen
  \bibfield  {author} {\bibinfo {author} {\bibfnamefont {T.}~\bibnamefont
  {{Zhang}}}, \bibinfo {author} {\bibfnamefont {J.}~\bibnamefont {{Sun}}},
  \bibinfo {author} {\bibfnamefont {X.-X.}\ \bibnamefont {{Fang}}}, \bibinfo
  {author} {\bibfnamefont {X.-M.}\ \bibnamefont {{Zhang}}}, \bibinfo {author}
  {\bibfnamefont {X.}~\bibnamefont {{Yuan}}}, \ and\ \bibinfo {author}
  {\bibfnamefont {H.}~\bibnamefont {{Lu}}},\ }\href {\doibase
  10.1103/PhysRevLett.127.200501} {\bibfield  {journal} {\bibinfo  {journal}
  {\prl}\ }\textbf {\bibinfo {volume} {127}},\ \bibinfo {pages} {200501}
  (\bibinfo {year} {2021})}\BibitemShut {NoStop}%
\bibitem [{\citenamefont {{Vitale}}\ \emph {et~al.}(2023)\citenamefont
  {{Vitale}}, \citenamefont {{Rath}}, \citenamefont {{Jurcevic}}, \citenamefont
  {{Elben}}, \citenamefont {{Branciard}},\ and\ \citenamefont
  {{Vermersch}}}]{2023arXiv230716882V}%
  \BibitemOpen
  \bibfield  {author} {\bibinfo {author} {\bibfnamefont {V.}~\bibnamefont
  {{Vitale}}}, \bibinfo {author} {\bibfnamefont {A.}~\bibnamefont {{Rath}}},
  \bibinfo {author} {\bibfnamefont {P.}~\bibnamefont {{Jurcevic}}}, \bibinfo
  {author} {\bibfnamefont {A.}~\bibnamefont {{Elben}}}, \bibinfo {author}
  {\bibfnamefont {C.}~\bibnamefont {{Branciard}}}, \ and\ \bibinfo {author}
  {\bibfnamefont {B.}~\bibnamefont {{Vermersch}}},\ }\href {\doibase
  10.48550/arXiv.2307.16882} {\bibfield  {journal} {\bibinfo  {journal} {arXiv
  e-prints}\ ,\ \bibinfo {eid} {arXiv:2307.16882}} (\bibinfo {year} {2023})},\
  \Eprint {http://arxiv.org/abs/2307.16882} {arXiv:2307.16882 [quant-ph]}
  \BibitemShut {NoStop}%
\bibitem [{\citenamefont {{Hu}}\ \emph {et~al.}(2024)\citenamefont {{Hu}},
  \citenamefont {{Gu}}, \citenamefont {{Majumder}}, \citenamefont {{Ren}},
  \citenamefont {{Zhang}}, \citenamefont {{Wang}}, \citenamefont {{You}},
  \citenamefont {{Minev}}, \citenamefont {{Yelin}},\ and\ \citenamefont
  {{Seif}}}]{Hu2024D2402.17911}%
  \BibitemOpen
  \bibfield  {author} {\bibinfo {author} {\bibfnamefont {H.-Y.}\ \bibnamefont
  {{Hu}}}, \bibinfo {author} {\bibfnamefont {A.}~\bibnamefont {{Gu}}}, \bibinfo
  {author} {\bibfnamefont {S.}~\bibnamefont {{Majumder}}}, \bibinfo {author}
  {\bibfnamefont {H.}~\bibnamefont {{Ren}}}, \bibinfo {author} {\bibfnamefont
  {Y.}~\bibnamefont {{Zhang}}}, \bibinfo {author} {\bibfnamefont {D.~S.}\
  \bibnamefont {{Wang}}}, \bibinfo {author} {\bibfnamefont {Y.-Z.}\
  \bibnamefont {{You}}}, \bibinfo {author} {\bibfnamefont {Z.}~\bibnamefont
  {{Minev}}}, \bibinfo {author} {\bibfnamefont {S.~F.}\ \bibnamefont
  {{Yelin}}}, \ and\ \bibinfo {author} {\bibfnamefont {A.}~\bibnamefont
  {{Seif}}},\ }\href {\doibase 10.48550/arXiv.2402.17911} {\bibfield  {journal}
  {\bibinfo  {journal} {arXiv e-prints}\ ,\ \bibinfo {eid} {arXiv:2402.17911}}
  (\bibinfo {year} {2024})},\ \Eprint {http://arxiv.org/abs/2402.17911}
  {arXiv:2402.17911 [quant-ph]} \BibitemShut {NoStop}%
\bibitem [{\citenamefont {{Lukens}}\ \emph {et~al.}(2021)\citenamefont
  {{Lukens}}, \citenamefont {{Law}},\ and\ \citenamefont
  {{Bennink}}}]{Lukens2021A2012.08997}%
  \BibitemOpen
  \bibfield  {author} {\bibinfo {author} {\bibfnamefont {J.~M.}\ \bibnamefont
  {{Lukens}}}, \bibinfo {author} {\bibfnamefont {K.~J.~H.}\ \bibnamefont
  {{Law}}}, \ and\ \bibinfo {author} {\bibfnamefont {R.~S.}\ \bibnamefont
  {{Bennink}}},\ }\href {\doibase 10.1038/s41534-021-00447-6} {\bibfield
  {journal} {\bibinfo  {journal} {npj Quantum Information}\ }\textbf {\bibinfo
  {volume} {7}},\ \bibinfo {pages} {113} (\bibinfo {year} {2021})}\BibitemShut
  {NoStop}%
\bibitem [{\citenamefont {Morris}\ \emph {et~al.}(2022)\citenamefont {Morris},
  \citenamefont {Saggio}, \citenamefont {Go{\v c}anin},\ and\ \citenamefont
  {Daki{\'c}}}]{Morris2021Q2109.03860}%
  \BibitemOpen
  \bibfield  {author} {\bibinfo {author} {\bibfnamefont {J.}~\bibnamefont
  {Morris}}, \bibinfo {author} {\bibfnamefont {V.}~\bibnamefont {Saggio}},
  \bibinfo {author} {\bibfnamefont {A.}~\bibnamefont {Go{\v c}anin}}, \ and\
  \bibinfo {author} {\bibfnamefont {B.}~\bibnamefont {Daki{\'c}}},\ }\href
  {\doibase 10.1002/qute.202100118} {\bibfield  {journal} {\bibinfo  {journal}
  {Advanced Quantum Technologies}\ }\textbf {\bibinfo {volume} {5}} (\bibinfo
  {year} {2022}),\ 10.1002/qute.202100118}\BibitemShut {NoStop}%
\bibitem [{\citenamefont {{Levy}}\ \emph {et~al.}(2024)\citenamefont {{Levy}},
  \citenamefont {{Luo}},\ and\ \citenamefont {{Clark}}}]{Levy2024C2110.02965}%
  \BibitemOpen
  \bibfield  {author} {\bibinfo {author} {\bibfnamefont {R.}~\bibnamefont
  {{Levy}}}, \bibinfo {author} {\bibfnamefont {D.}~\bibnamefont {{Luo}}}, \
  and\ \bibinfo {author} {\bibfnamefont {B.~K.}\ \bibnamefont {{Clark}}},\
  }\href {\doibase 10.1103/PhysRevResearch.6.013029} {\bibfield  {journal}
  {\bibinfo  {journal} {Physical Review Research}\ }\textbf {\bibinfo {volume}
  {6}},\ \bibinfo {pages} {013029} (\bibinfo {year} {2024})}\BibitemShut
  {NoStop}%
\bibitem [{\citenamefont {{Kunjummen}}\ \emph {et~al.}(2023)\citenamefont
  {{Kunjummen}}, \citenamefont {{Tran}}, \citenamefont {{Carney}},\ and\
  \citenamefont {{Taylor}}}]{Kunjummen2023S2110.03629}%
  \BibitemOpen
  \bibfield  {author} {\bibinfo {author} {\bibfnamefont {J.}~\bibnamefont
  {{Kunjummen}}}, \bibinfo {author} {\bibfnamefont {M.~C.}\ \bibnamefont
  {{Tran}}}, \bibinfo {author} {\bibfnamefont {D.}~\bibnamefont {{Carney}}}, \
  and\ \bibinfo {author} {\bibfnamefont {J.~M.}\ \bibnamefont {{Taylor}}},\
  }\href {\doibase 10.1103/PhysRevA.107.042403} {\bibfield  {journal} {\bibinfo
   {journal} {\pra}\ }\textbf {\bibinfo {volume} {107}},\ \bibinfo {pages}
  {042403} (\bibinfo {year} {2023})}\BibitemShut {NoStop}%
\bibitem [{\citenamefont {{Helsen}}\ \emph {et~al.}(2023)\citenamefont
  {{Helsen}}, \citenamefont {{Ioannou}}, \citenamefont {{Kitzinger}},
  \citenamefont {{Onorati}}, \citenamefont {{Werner}}, \citenamefont
  {{Eisert}},\ and\ \citenamefont {{Roth}}}]{Helsen2023S2110.13178}%
  \BibitemOpen
  \bibfield  {author} {\bibinfo {author} {\bibfnamefont {J.}~\bibnamefont
  {{Helsen}}}, \bibinfo {author} {\bibfnamefont {M.}~\bibnamefont {{Ioannou}}},
  \bibinfo {author} {\bibfnamefont {J.}~\bibnamefont {{Kitzinger}}}, \bibinfo
  {author} {\bibfnamefont {E.}~\bibnamefont {{Onorati}}}, \bibinfo {author}
  {\bibfnamefont {A.~H.}\ \bibnamefont {{Werner}}}, \bibinfo {author}
  {\bibfnamefont {J.}~\bibnamefont {{Eisert}}}, \ and\ \bibinfo {author}
  {\bibfnamefont {I.}~\bibnamefont {{Roth}}},\ }\href {\doibase
  10.1038/s41467-023-39382-9} {\bibfield  {journal} {\bibinfo  {journal}
  {Nature Communications}\ }\textbf {\bibinfo {volume} {14}},\ \bibinfo {pages}
  {5039} (\bibinfo {year} {2023})}\BibitemShut {NoStop}%
\bibitem [{\citenamefont {Carrasco}\ \emph {et~al.}(2021)\citenamefont
  {Carrasco}, \citenamefont {Elben}, \citenamefont {Kokail}, \citenamefont
  {Kraus},\ and\ \citenamefont {Zoller}}]{PRXQuantum.2.010102}%
  \BibitemOpen
  \bibfield  {author} {\bibinfo {author} {\bibfnamefont {J.}~\bibnamefont
  {Carrasco}}, \bibinfo {author} {\bibfnamefont {A.}~\bibnamefont {Elben}},
  \bibinfo {author} {\bibfnamefont {C.}~\bibnamefont {Kokail}}, \bibinfo
  {author} {\bibfnamefont {B.}~\bibnamefont {Kraus}}, \ and\ \bibinfo {author}
  {\bibfnamefont {P.}~\bibnamefont {Zoller}},\ }\href {\doibase
  10.1103/PRXQuantum.2.010102} {\bibfield  {journal} {\bibinfo  {journal} {PRX
  Quantum}\ }\textbf {\bibinfo {volume} {2}},\ \bibinfo {pages} {010102}
  (\bibinfo {year} {2021})}\BibitemShut {NoStop}%
\bibitem [{\citenamefont {Elben}\ \emph
  {et~al.}(2020{\natexlab{b}})\citenamefont {Elben}, \citenamefont {Vermersch},
  \citenamefont {van Bijnen}, \citenamefont {Kokail}, \citenamefont {Brydges},
  \citenamefont {Maier}, \citenamefont {Joshi}, \citenamefont {Blatt},
  \citenamefont {Roos},\ and\ \citenamefont {Zoller}}]{PhysRevLett.124.010504}%
  \BibitemOpen
  \bibfield  {author} {\bibinfo {author} {\bibfnamefont {A.}~\bibnamefont
  {Elben}}, \bibinfo {author} {\bibfnamefont {B.}~\bibnamefont {Vermersch}},
  \bibinfo {author} {\bibfnamefont {R.}~\bibnamefont {van Bijnen}}, \bibinfo
  {author} {\bibfnamefont {C.}~\bibnamefont {Kokail}}, \bibinfo {author}
  {\bibfnamefont {T.}~\bibnamefont {Brydges}}, \bibinfo {author} {\bibfnamefont
  {C.}~\bibnamefont {Maier}}, \bibinfo {author} {\bibfnamefont {M.~K.}\
  \bibnamefont {Joshi}}, \bibinfo {author} {\bibfnamefont {R.}~\bibnamefont
  {Blatt}}, \bibinfo {author} {\bibfnamefont {C.~F.}\ \bibnamefont {Roos}}, \
  and\ \bibinfo {author} {\bibfnamefont {P.}~\bibnamefont {Zoller}},\ }\href
  {\doibase 10.1103/PhysRevLett.124.010504} {\bibfield  {journal} {\bibinfo
  {journal} {Phys. Rev. Lett.}\ }\textbf {\bibinfo {volume} {124}},\ \bibinfo
  {pages} {010504} (\bibinfo {year} {2020}{\natexlab{b}})}\BibitemShut
  {NoStop}%
\bibitem [{\citenamefont {Hadfield}\ \emph {et~al.}(2022)\citenamefont
  {Hadfield}, \citenamefont {Bravyi}, \citenamefont {Raymond},\ and\
  \citenamefont {Mezzacapo}}]{Hadfield2020M2006.15788}%
  \BibitemOpen
  \bibfield  {author} {\bibinfo {author} {\bibfnamefont {C.}~\bibnamefont
  {Hadfield}}, \bibinfo {author} {\bibfnamefont {S.}~\bibnamefont {Bravyi}},
  \bibinfo {author} {\bibfnamefont {R.}~\bibnamefont {Raymond}}, \ and\
  \bibinfo {author} {\bibfnamefont {A.}~\bibnamefont {Mezzacapo}},\ }\href
  {\doibase 10.1007/s00220-022-04343-8} {\bibfield  {journal} {\bibinfo
  {journal} {Communications in Mathematical Physics}\ }\textbf {\bibinfo
  {volume} {391}},\ \bibinfo {pages} {951} (\bibinfo {year}
  {2022})}\BibitemShut {NoStop}%
\bibitem [{\citenamefont {{McNulty}}\ \emph {et~al.}(2023)\citenamefont
  {{McNulty}}, \citenamefont {{Maciejewski}},\ and\ \citenamefont
  {{Oszmaniec}}}]{McNulty2023E2206.08912}%
  \BibitemOpen
  \bibfield  {author} {\bibinfo {author} {\bibfnamefont {D.}~\bibnamefont
  {{McNulty}}}, \bibinfo {author} {\bibfnamefont {F.~B.}\ \bibnamefont
  {{Maciejewski}}}, \ and\ \bibinfo {author} {\bibfnamefont {M.}~\bibnamefont
  {{Oszmaniec}}},\ }\href {\doibase 10.1103/PhysRevLett.130.100801} {\bibfield
  {journal} {\bibinfo  {journal} {\prl}\ }\textbf {\bibinfo {volume} {130}},\
  \bibinfo {pages} {100801} (\bibinfo {year} {2023})}\BibitemShut {NoStop}%
\bibitem [{\citenamefont {Dutt}\ \emph {et~al.}(2023)\citenamefont {Dutt},
  \citenamefont {Kirby}, \citenamefont {Raymond}, \citenamefont {Hadfield},
  \citenamefont {Sheldon}, \citenamefont {Chuang},\ and\ \citenamefont
  {Mezzacapo}}]{Dutt2023P2312.07497}%
  \BibitemOpen
  \bibfield  {author} {\bibinfo {author} {\bibfnamefont {A.}~\bibnamefont
  {Dutt}}, \bibinfo {author} {\bibfnamefont {W.}~\bibnamefont {Kirby}},
  \bibinfo {author} {\bibfnamefont {R.}~\bibnamefont {Raymond}}, \bibinfo
  {author} {\bibfnamefont {C.}~\bibnamefont {Hadfield}}, \bibinfo {author}
  {\bibfnamefont {S.}~\bibnamefont {Sheldon}}, \bibinfo {author} {\bibfnamefont
  {I.~L.}\ \bibnamefont {Chuang}}, \ and\ \bibinfo {author} {\bibfnamefont
  {A.}~\bibnamefont {Mezzacapo}},\ }\href@noop {} {\enquote {\bibinfo {title}
  {Practical benchmarking of randomized measurement methods for quantum
  chemistry hamiltonians},}\ } (\bibinfo {year} {2023}),\ \Eprint
  {http://arxiv.org/abs/2312.07497} {arXiv:2312.07497 [quant-ph]} \BibitemShut
  {NoStop}%
\bibitem [{\citenamefont {Kokail}\ \emph
  {et~al.}(2021{\natexlab{a}})\citenamefont {Kokail}, \citenamefont {van
  Bijnen}, \citenamefont {Elben}, \citenamefont {Vermersch},\ and\
  \citenamefont {Zoller}}]{EntanglementHamiltonian}%
  \BibitemOpen
  \bibfield  {author} {\bibinfo {author} {\bibfnamefont {C.}~\bibnamefont
  {Kokail}}, \bibinfo {author} {\bibfnamefont {R.}~\bibnamefont {van Bijnen}},
  \bibinfo {author} {\bibfnamefont {A.}~\bibnamefont {Elben}}, \bibinfo
  {author} {\bibfnamefont {B.}~\bibnamefont {Vermersch}}, \ and\ \bibinfo
  {author} {\bibfnamefont {P.}~\bibnamefont {Zoller}},\ }\href {\doibase
  10.1038/s41567-021-01260-w} {\bibfield  {journal} {\bibinfo  {journal}
  {Nature Physics}\ }\textbf {\bibinfo {volume} {17}},\ \bibinfo {pages} {936}
  (\bibinfo {year} {2021}{\natexlab{a}})}\BibitemShut {NoStop}%
\bibitem [{\citenamefont {Kokail}\ \emph
  {et~al.}(2021{\natexlab{b}})\citenamefont {Kokail}, \citenamefont {Sundar},
  \citenamefont {Zache}, \citenamefont {Elben}, \citenamefont {Vermersch},
  \citenamefont {Dalmonte}, \citenamefont {van Bijnen},\ and\ \citenamefont
  {Zoller}}]{PhysRevLett.127.170501}%
  \BibitemOpen
  \bibfield  {author} {\bibinfo {author} {\bibfnamefont {C.}~\bibnamefont
  {Kokail}}, \bibinfo {author} {\bibfnamefont {B.}~\bibnamefont {Sundar}},
  \bibinfo {author} {\bibfnamefont {T.~V.}\ \bibnamefont {Zache}}, \bibinfo
  {author} {\bibfnamefont {A.}~\bibnamefont {Elben}}, \bibinfo {author}
  {\bibfnamefont {B.}~\bibnamefont {Vermersch}}, \bibinfo {author}
  {\bibfnamefont {M.}~\bibnamefont {Dalmonte}}, \bibinfo {author}
  {\bibfnamefont {R.}~\bibnamefont {van Bijnen}}, \ and\ \bibinfo {author}
  {\bibfnamefont {P.}~\bibnamefont {Zoller}},\ }\href {\doibase
  10.1103/PhysRevLett.127.170501} {\bibfield  {journal} {\bibinfo  {journal}
  {Phys. Rev. Lett.}\ }\textbf {\bibinfo {volume} {127}},\ \bibinfo {pages}
  {170501} (\bibinfo {year} {2021}{\natexlab{b}})}\BibitemShut {NoStop}%
\bibitem [{\citenamefont {Hu}\ \emph {et~al.}(2022)\citenamefont {Hu},
  \citenamefont {LaRose}, \citenamefont {You}, \citenamefont {Rieffel},\ and\
  \citenamefont {Wang}}]{Hu2022L2203.07263}%
  \BibitemOpen
  \bibfield  {author} {\bibinfo {author} {\bibfnamefont {H.-Y.}\ \bibnamefont
  {Hu}}, \bibinfo {author} {\bibfnamefont {R.}~\bibnamefont {LaRose}}, \bibinfo
  {author} {\bibfnamefont {Y.-Z.}\ \bibnamefont {You}}, \bibinfo {author}
  {\bibfnamefont {E.}~\bibnamefont {Rieffel}}, \ and\ \bibinfo {author}
  {\bibfnamefont {Z.}~\bibnamefont {Wang}},\ }\href@noop {} {\enquote {\bibinfo
  {title} {Logical shadow tomography: Efficient estimation of error-mitigated
  observables},}\ } (\bibinfo {year} {2022}),\ \Eprint
  {http://arxiv.org/abs/2203.07263} {arXiv:2203.07263 [quant-ph]} \BibitemShut
  {NoStop}%
\bibitem [{\citenamefont {{Seif}}\ \emph {et~al.}(2023)\citenamefont {{Seif}},
  \citenamefont {{Cian}}, \citenamefont {{Zhou}}, \citenamefont {{Chen}},\ and\
  \citenamefont {{Jiang}}}]{Seif2023S2203.07309}%
  \BibitemOpen
  \bibfield  {author} {\bibinfo {author} {\bibfnamefont {A.}~\bibnamefont
  {{Seif}}}, \bibinfo {author} {\bibfnamefont {Z.-P.}\ \bibnamefont {{Cian}}},
  \bibinfo {author} {\bibfnamefont {S.}~\bibnamefont {{Zhou}}}, \bibinfo
  {author} {\bibfnamefont {S.}~\bibnamefont {{Chen}}}, \ and\ \bibinfo {author}
  {\bibfnamefont {L.}~\bibnamefont {{Jiang}}},\ }\href {\doibase
  10.1103/PRXQuantum.4.010303} {\bibfield  {journal} {\bibinfo  {journal} {PRX
  Quantum}\ }\textbf {\bibinfo {volume} {4}},\ \bibinfo {pages} {010303}
  (\bibinfo {year} {2023})}\BibitemShut {NoStop}%
\bibitem [{\citenamefont {Jnane}\ \emph {et~al.}(2023)\citenamefont {Jnane},
  \citenamefont {Steinberg}, \citenamefont {Cai}, \citenamefont {Nguyen},\ and\
  \citenamefont {Koczor}}]{2023arXiv230504956J}%
  \BibitemOpen
  \bibfield  {author} {\bibinfo {author} {\bibfnamefont {H.}~\bibnamefont
  {Jnane}}, \bibinfo {author} {\bibfnamefont {J.}~\bibnamefont {Steinberg}},
  \bibinfo {author} {\bibfnamefont {Z.}~\bibnamefont {Cai}}, \bibinfo {author}
  {\bibfnamefont {H.~C.}\ \bibnamefont {Nguyen}}, \ and\ \bibinfo {author}
  {\bibfnamefont {B.}~\bibnamefont {Koczor}},\ }\href@noop {} {\enquote
  {\bibinfo {title} {Quantum error mitigated classical shadows},}\ } (\bibinfo
  {year} {2023}),\ \Eprint {http://arxiv.org/abs/2305.04956} {arXiv:2305.04956
  [quant-ph]} \BibitemShut {NoStop}%
\bibitem [{\citenamefont {Zhao}\ and\ \citenamefont
  {Miyake}(2023)}]{2023arXiv231003071Z}%
  \BibitemOpen
  \bibfield  {author} {\bibinfo {author} {\bibfnamefont {A.}~\bibnamefont
  {Zhao}}\ and\ \bibinfo {author} {\bibfnamefont {A.}~\bibnamefont {Miyake}},\
  }\href@noop {} {\enquote {\bibinfo {title} {Group-theoretic error mitigation
  enabled by classical shadows and symmetries},}\ } (\bibinfo {year} {2023}),\
  \Eprint {http://arxiv.org/abs/2310.03071} {arXiv:2310.03071 [quant-ph]}
  \BibitemShut {NoStop}%
\bibitem [{\citenamefont {Jerbi}\ \emph {et~al.}(2023)\citenamefont {Jerbi},
  \citenamefont {Gyurik}, \citenamefont {Marshall}, \citenamefont {Molteni},\
  and\ \citenamefont {Dunjko}}]{2023arXiv230600061J}%
  \BibitemOpen
  \bibfield  {author} {\bibinfo {author} {\bibfnamefont {S.}~\bibnamefont
  {Jerbi}}, \bibinfo {author} {\bibfnamefont {C.}~\bibnamefont {Gyurik}},
  \bibinfo {author} {\bibfnamefont {S.~C.}\ \bibnamefont {Marshall}}, \bibinfo
  {author} {\bibfnamefont {R.}~\bibnamefont {Molteni}}, \ and\ \bibinfo
  {author} {\bibfnamefont {V.}~\bibnamefont {Dunjko}},\ }\href@noop {}
  {\enquote {\bibinfo {title} {Shadows of quantum machine learning},}\ }
  (\bibinfo {year} {2023}),\ \Eprint {http://arxiv.org/abs/2306.00061}
  {arXiv:2306.00061 [quant-ph]} \BibitemShut {NoStop}%
\bibitem [{\citenamefont {{Zhang}}\ and\ \citenamefont
  {{You}}(2023)}]{2023arXiv230614838Z}%
  \BibitemOpen
  \bibfield  {author} {\bibinfo {author} {\bibfnamefont {Z.}~\bibnamefont
  {{Zhang}}}\ and\ \bibinfo {author} {\bibfnamefont {Y.-Z.}\ \bibnamefont
  {{You}}},\ }\href@noop {} {\enquote {\bibinfo {title} {Observing
  schr{\"o}dinger's cat with artificial intelligence: Emergent classicality
  from information bottleneck},}\ } (\bibinfo {year} {2023}),\ \Eprint
  {http://arxiv.org/abs/2306.14838} {arXiv:2306.14838 [quant-ph]} \BibitemShut
  {NoStop}%
\bibitem [{\citenamefont {Huang}\ \emph
  {et~al.}(2022{\natexlab{b}})\citenamefont {Huang}, \citenamefont {Kueng},
  \citenamefont {Torlai}, \citenamefont {Albert},\ and\ \citenamefont
  {Preskill}}]{doi:10.1126/science.abk3333}%
  \BibitemOpen
  \bibfield  {author} {\bibinfo {author} {\bibfnamefont {H.-Y.}\ \bibnamefont
  {Huang}}, \bibinfo {author} {\bibfnamefont {R.}~\bibnamefont {Kueng}},
  \bibinfo {author} {\bibfnamefont {G.}~\bibnamefont {Torlai}}, \bibinfo
  {author} {\bibfnamefont {V.~V.}\ \bibnamefont {Albert}}, \ and\ \bibinfo
  {author} {\bibfnamefont {J.}~\bibnamefont {Preskill}},\ }\href {\doibase
  10.1126/science.abk3333} {\bibfield  {journal} {\bibinfo  {journal}
  {Science}\ }\textbf {\bibinfo {volume} {377}} (\bibinfo {year}
  {2022}{\natexlab{b}}),\ 10.1126/science.abk3333}\BibitemShut {NoStop}%
\bibitem [{\citenamefont {Haug}\ \emph {et~al.}(2023)\citenamefont {Haug},
  \citenamefont {Self},\ and\ \citenamefont {Kim}}]{2023MLS&T...4a5005H}%
  \BibitemOpen
  \bibfield  {author} {\bibinfo {author} {\bibfnamefont {T.}~\bibnamefont
  {Haug}}, \bibinfo {author} {\bibfnamefont {C.~N.}\ \bibnamefont {Self}}, \
  and\ \bibinfo {author} {\bibfnamefont {M.~S.}\ \bibnamefont {Kim}},\ }\href
  {\doibase 10.1088/2632-2153/acb0b4} {\bibfield  {journal} {\bibinfo
  {journal} {Machine Learning: Science and Technology}\ }\textbf {\bibinfo
  {volume} {4}},\ \bibinfo {pages} {015005} (\bibinfo {year}
  {2023})}\BibitemShut {NoStop}%
\bibitem [{\citenamefont {{Notarnicola}}\ \emph {et~al.}(2023)\citenamefont
  {{Notarnicola}}, \citenamefont {{Elben}}, \citenamefont {{Lahaye}},
  \citenamefont {{Browaeys}}, \citenamefont {{Montangero}},\ and\ \citenamefont
  {{Vermersch}}}]{Notarnicola2023A2112.11046}%
  \BibitemOpen
  \bibfield  {author} {\bibinfo {author} {\bibfnamefont {S.}~\bibnamefont
  {{Notarnicola}}}, \bibinfo {author} {\bibfnamefont {A.}~\bibnamefont
  {{Elben}}}, \bibinfo {author} {\bibfnamefont {T.}~\bibnamefont {{Lahaye}}},
  \bibinfo {author} {\bibfnamefont {A.}~\bibnamefont {{Browaeys}}}, \bibinfo
  {author} {\bibfnamefont {S.}~\bibnamefont {{Montangero}}}, \ and\ \bibinfo
  {author} {\bibfnamefont {B.}~\bibnamefont {{Vermersch}}},\ }\href {\doibase
  10.1088/1367-2630/acfcd3} {\bibfield  {journal} {\bibinfo  {journal} {New
  Journal of Physics}\ }\textbf {\bibinfo {volume} {25}} (\bibinfo {year}
  {2023}),\ 10.1088/1367-2630/acfcd3}\BibitemShut {NoStop}%
\bibitem [{\citenamefont {{Elben}}\ \emph {et~al.}(2023)\citenamefont
  {{Elben}}, \citenamefont {{Flammia}}, \citenamefont {{Huang}}, \citenamefont
  {{Kueng}}, \citenamefont {{Preskill}}, \citenamefont {{Vermersch}},\ and\
  \citenamefont {{Zoller}}}]{Elben2023T2203.11374}%
  \BibitemOpen
  \bibfield  {author} {\bibinfo {author} {\bibfnamefont {A.}~\bibnamefont
  {{Elben}}}, \bibinfo {author} {\bibfnamefont {S.~T.}\ \bibnamefont
  {{Flammia}}}, \bibinfo {author} {\bibfnamefont {H.-Y.}\ \bibnamefont
  {{Huang}}}, \bibinfo {author} {\bibfnamefont {R.}~\bibnamefont {{Kueng}}},
  \bibinfo {author} {\bibfnamefont {J.}~\bibnamefont {{Preskill}}}, \bibinfo
  {author} {\bibfnamefont {B.}~\bibnamefont {{Vermersch}}}, \ and\ \bibinfo
  {author} {\bibfnamefont {P.}~\bibnamefont {{Zoller}}},\ }\href {\doibase
  10.1038/s42254-022-00535-2} {\bibfield  {journal} {\bibinfo  {journal}
  {Nature Reviews Physics}\ }\textbf {\bibinfo {volume} {5}},\ \bibinfo {pages}
  {9} (\bibinfo {year} {2023})}\BibitemShut {NoStop}%
\bibitem [{\citenamefont {Vidal}(2007)}]{Vidal2007E}%
  \BibitemOpen
  \bibfield  {author} {\bibinfo {author} {\bibfnamefont {G.}~\bibnamefont
  {Vidal}},\ }\href {\doibase 10.1103/PhysRevLett.99.220405} {\bibfield
  {journal} {\bibinfo  {journal} {Phys. Rev. Lett.}\ }\textbf {\bibinfo
  {volume} {99}},\ \bibinfo {pages} {220405} (\bibinfo {year}
  {2007})}\BibitemShut {NoStop}%
\bibitem [{\citenamefont {{Vidal}}(2008)}]{Vidal2008Cquant-ph/0610099}%
  \BibitemOpen
  \bibfield  {author} {\bibinfo {author} {\bibfnamefont {G.}~\bibnamefont
  {{Vidal}}},\ }\href {\doibase 10.1103/PhysRevLett.101.110501} {\bibfield
  {journal} {\bibinfo  {journal} {\prl}\ }\textbf {\bibinfo {volume} {101}},\
  \bibinfo {eid} {110501} (\bibinfo {year} {2008})},\ \Eprint
  {http://arxiv.org/abs/quant-ph/0610099} {arXiv:quant-ph/0610099 [quant-ph]}
  \BibitemShut {NoStop}%
\bibitem [{\citenamefont {{Evenbly}}\ and\ \citenamefont
  {{Vidal}}(2011)}]{Evenbly2011T1106.1082}%
  \BibitemOpen
  \bibfield  {author} {\bibinfo {author} {\bibfnamefont {G.}~\bibnamefont
  {{Evenbly}}}\ and\ \bibinfo {author} {\bibfnamefont {G.}~\bibnamefont
  {{Vidal}}},\ }\href {\doibase 10.1007/s10955-011-0237-4} {\bibfield
  {journal} {\bibinfo  {journal} {Journal of Statistical Physics}\ }\textbf
  {\bibinfo {volume} {145}},\ \bibinfo {pages} {891} (\bibinfo {year}
  {2011})},\ \Eprint {http://arxiv.org/abs/1106.1082} {arXiv:1106.1082
  [quant-ph]} \BibitemShut {NoStop}%
\bibitem [{\citenamefont {{Pastawski}}\ \emph {et~al.}(2015)\citenamefont
  {{Pastawski}}, \citenamefont {{Yoshida}}, \citenamefont {{Harlow}},\ and\
  \citenamefont {{Preskill}}}]{Pastawski2015H1503.06237}%
  \BibitemOpen
  \bibfield  {author} {\bibinfo {author} {\bibfnamefont {F.}~\bibnamefont
  {{Pastawski}}}, \bibinfo {author} {\bibfnamefont {B.}~\bibnamefont
  {{Yoshida}}}, \bibinfo {author} {\bibfnamefont {D.}~\bibnamefont {{Harlow}}},
  \ and\ \bibinfo {author} {\bibfnamefont {J.}~\bibnamefont {{Preskill}}},\
  }\href {\doibase 10.1007/JHEP06(2015)149} {\bibfield  {journal} {\bibinfo
  {journal} {Journal of High Energy Physics}\ }\textbf {\bibinfo {volume}
  {2015}},\ \bibinfo {eid} {149} (\bibinfo {year} {2015})},\ \Eprint
  {http://arxiv.org/abs/1503.06237} {arXiv:1503.06237 [hep-th]} \BibitemShut
  {NoStop}%
\bibitem [{\citenamefont {{Hayden}}\ \emph {et~al.}(2016)\citenamefont
  {{Hayden}}, \citenamefont {{Nezami}}, \citenamefont {{Qi}}, \citenamefont
  {{Thomas}}, \citenamefont {{Walter}},\ and\ \citenamefont
  {{Yang}}}]{Hayden2016H1601.01694}%
  \BibitemOpen
  \bibfield  {author} {\bibinfo {author} {\bibfnamefont {P.}~\bibnamefont
  {{Hayden}}}, \bibinfo {author} {\bibfnamefont {S.}~\bibnamefont {{Nezami}}},
  \bibinfo {author} {\bibfnamefont {X.-L.}\ \bibnamefont {{Qi}}}, \bibinfo
  {author} {\bibfnamefont {N.}~\bibnamefont {{Thomas}}}, \bibinfo {author}
  {\bibfnamefont {M.}~\bibnamefont {{Walter}}}, \ and\ \bibinfo {author}
  {\bibfnamefont {Z.}~\bibnamefont {{Yang}}},\ }\href {\doibase
  10.1007/JHEP11(2016)009} {\bibfield  {journal} {\bibinfo  {journal} {Journal
  of High Energy Physics}\ }\textbf {\bibinfo {volume} {2016}},\ \bibinfo {eid}
  {9} (\bibinfo {year} {2016})},\ \Eprint {http://arxiv.org/abs/1601.01694}
  {arXiv:1601.01694 [hep-th]} \BibitemShut {NoStop}%
\bibitem [{\citenamefont {Akhtar}\ \emph {et~al.}(2023)\citenamefont {Akhtar},
  \citenamefont {Hu},\ and\ \citenamefont
  {You}}]{akhtar2023measurementinduced}%
  \BibitemOpen
  \bibfield  {author} {\bibinfo {author} {\bibfnamefont {A.~A.}\ \bibnamefont
  {Akhtar}}, \bibinfo {author} {\bibfnamefont {H.-Y.}\ \bibnamefont {Hu}}, \
  and\ \bibinfo {author} {\bibfnamefont {Y.-Z.}\ \bibnamefont {You}},\
  }\href@noop {} {\enquote {\bibinfo {title} {Measurement-induced criticality
  is tomographically optimal},}\ } (\bibinfo {year} {2023}),\ \Eprint
  {http://arxiv.org/abs/2308.01653} {arXiv:2308.01653 [quant-ph]} \BibitemShut
  {NoStop}%
\bibitem [{\citenamefont {Basteiro}\ \emph {et~al.}(2022)\citenamefont
  {Basteiro}, \citenamefont {Di~Giulio}, \citenamefont {Erdmenger},
  \citenamefont {Karl}, \citenamefont {Meyer},\ and\ \citenamefont
  {Xian}}]{Basteiro_2022}%
  \BibitemOpen
  \bibfield  {author} {\bibinfo {author} {\bibfnamefont {P.}~\bibnamefont
  {Basteiro}}, \bibinfo {author} {\bibfnamefont {G.}~\bibnamefont {Di~Giulio}},
  \bibinfo {author} {\bibfnamefont {J.}~\bibnamefont {Erdmenger}}, \bibinfo
  {author} {\bibfnamefont {J.}~\bibnamefont {Karl}}, \bibinfo {author}
  {\bibfnamefont {R.}~\bibnamefont {Meyer}}, \ and\ \bibinfo {author}
  {\bibfnamefont {Z.-Y.}\ \bibnamefont {Xian}},\ }\href {\doibase
  10.21468/scipostphys.13.5.103} {\bibfield  {journal} {\bibinfo  {journal}
  {SciPost Physics}\ }\textbf {\bibinfo {volume} {13}} (\bibinfo {year}
  {2022}),\ 10.21468/scipostphys.13.5.103}\BibitemShut {NoStop}%
\bibitem [{\citenamefont {You}\ \emph {et~al.}(2018)\citenamefont {You},
  \citenamefont {Yang},\ and\ \citenamefont {Qi}}]{You_2018}%
  \BibitemOpen
  \bibfield  {author} {\bibinfo {author} {\bibfnamefont {Y.-Z.}\ \bibnamefont
  {You}}, \bibinfo {author} {\bibfnamefont {Z.}~\bibnamefont {Yang}}, \ and\
  \bibinfo {author} {\bibfnamefont {X.-L.}\ \bibnamefont {Qi}},\ }\href
  {\doibase 10.1103/physrevb.97.045153} {\bibfield  {journal} {\bibinfo
  {journal} {Physical Review B}\ }\textbf {\bibinfo {volume} {97}} (\bibinfo
  {year} {2018}),\ 10.1103/physrevb.97.045153}\BibitemShut {NoStop}%
\bibitem [{\citenamefont {Ryu}\ and\ \citenamefont
  {Takayanagi}(2006)}]{Ryu_2006}%
  \BibitemOpen
  \bibfield  {author} {\bibinfo {author} {\bibfnamefont {S.}~\bibnamefont
  {Ryu}}\ and\ \bibinfo {author} {\bibfnamefont {T.}~\bibnamefont
  {Takayanagi}},\ }\href {\doibase 10.1103/physrevlett.96.181602} {\bibfield
  {journal} {\bibinfo  {journal} {Physical Review Letters}\ }\textbf {\bibinfo
  {volume} {96}} (\bibinfo {year} {2006}),\
  10.1103/physrevlett.96.181602}\BibitemShut {NoStop}%
\bibitem [{\citenamefont {Anand}\ \emph {et~al.}(2023)\citenamefont {Anand},
  \citenamefont {Hauschild}, \citenamefont {Zhang}, \citenamefont {Potter},\
  and\ \citenamefont {Zaletel}}]{Anand2023genMERA}%
  \BibitemOpen
  \bibfield  {author} {\bibinfo {author} {\bibfnamefont {S.}~\bibnamefont
  {Anand}}, \bibinfo {author} {\bibfnamefont {J.}~\bibnamefont {Hauschild}},
  \bibinfo {author} {\bibfnamefont {Y.}~\bibnamefont {Zhang}}, \bibinfo
  {author} {\bibfnamefont {A.~C.}\ \bibnamefont {Potter}}, \ and\ \bibinfo
  {author} {\bibfnamefont {M.~P.}\ \bibnamefont {Zaletel}},\ }\href {\doibase
  10.1103/PRXQuantum.4.030334} {\bibfield  {journal} {\bibinfo  {journal} {PRX
  Quantum}\ }\textbf {\bibinfo {volume} {4}},\ \bibinfo {pages} {030334}
  (\bibinfo {year} {2023})}\BibitemShut {NoStop}%
\bibitem [{\citenamefont {{Cowsik}}\ \emph {et~al.}(2023)\citenamefont
  {{Cowsik}}, \citenamefont {{Ippoliti}},\ and\ \citenamefont
  {{Qi}}}]{cowsik_entanglementmetric_2023}%
  \BibitemOpen
  \bibfield  {author} {\bibinfo {author} {\bibfnamefont {A.}~\bibnamefont
  {{Cowsik}}}, \bibinfo {author} {\bibfnamefont {M.}~\bibnamefont
  {{Ippoliti}}}, \ and\ \bibinfo {author} {\bibfnamefont {X.-L.}\ \bibnamefont
  {{Qi}}},\ }\href {\doibase 10.48550/arXiv.2307.15689} {\bibfield  {journal}
  {\bibinfo  {journal} {arXiv e-prints}\ ,\ \bibinfo {eid} {arXiv:2307.15689}}
  (\bibinfo {year} {2023})},\ \Eprint {http://arxiv.org/abs/2307.15689}
  {arXiv:2307.15689 [quant-ph]} \BibitemShut {NoStop}%
\bibitem [{\citenamefont {{Van Kirk}}\ \emph {et~al.}(2024)\citenamefont {{Van
  Kirk}}, \citenamefont {{Kunjummen}}, \citenamefont {{Hu}}, \citenamefont
  {{Teng}}, \citenamefont {{Cain}},\ and\ \citenamefont
  {{Kokail}}}]{harvard_umd_multiscale}%
  \BibitemOpen
  \bibfield  {author} {\bibinfo {author} {\bibfnamefont {K.}~\bibnamefont {{Van
  Kirk}}}, \bibinfo {author} {\bibfnamefont {J.}~\bibnamefont {{Kunjummen}}},
  \bibinfo {author} {\bibfnamefont {H.-Y.}\ \bibnamefont {{Hu}}}, \bibinfo
  {author} {\bibfnamefont {Y.}~\bibnamefont {{Teng}}}, \bibinfo {author}
  {\bibfnamefont {M.}~\bibnamefont {{Cain}}}, \ and\ \bibinfo {author}
  {\bibfnamefont {C.}~\bibnamefont {{Kokail}}},\ }\href@noop {} {\bibfield
  {journal} {\bibinfo  {journal} {to appear}\ } (\bibinfo {year}
  {2024})}\BibitemShut {NoStop}%
\bibitem [{\citenamefont {{You}}\ and\ \citenamefont
  {{Gu}}(2018)}]{You2018E1803.10425}%
  \BibitemOpen
  \bibfield  {author} {\bibinfo {author} {\bibfnamefont {Y.-Z.}\ \bibnamefont
  {{You}}}\ and\ \bibinfo {author} {\bibfnamefont {Y.}~\bibnamefont {{Gu}}},\
  }\href {\doibase 10.1103/PhysRevB.98.014309} {\bibfield  {journal} {\bibinfo
  {journal} {\prb}\ }\textbf {\bibinfo {volume} {98}},\ \bibinfo {eid} {014309}
  (\bibinfo {year} {2018})},\ \Eprint {http://arxiv.org/abs/1803.10425}
  {arXiv:1803.10425 [quant-ph]} \BibitemShut {NoStop}%
\bibitem [{\citenamefont {{Kuo}}\ \emph {et~al.}(2020)\citenamefont {{Kuo}},
  \citenamefont {{Akhtar}}, \citenamefont {{Arovas}},\ and\ \citenamefont
  {{You}}}]{Kuo2020M1910.11351}%
  \BibitemOpen
  \bibfield  {author} {\bibinfo {author} {\bibfnamefont {W.-T.}\ \bibnamefont
  {{Kuo}}}, \bibinfo {author} {\bibfnamefont {A.~A.}\ \bibnamefont {{Akhtar}}},
  \bibinfo {author} {\bibfnamefont {D.~P.}\ \bibnamefont {{Arovas}}}, \ and\
  \bibinfo {author} {\bibfnamefont {Y.-Z.}\ \bibnamefont {{You}}},\ }\href
  {\doibase 10.1103/PhysRevB.101.224202} {\bibfield  {journal} {\bibinfo
  {journal} {\prb}\ }\textbf {\bibinfo {volume} {101}},\ \bibinfo {eid}
  {224202} (\bibinfo {year} {2020})},\ \Eprint
  {http://arxiv.org/abs/1910.11351} {arXiv:1910.11351 [cond-mat.dis-nn]}
  \BibitemShut {NoStop}%
\bibitem [{\citenamefont {{Akhtar}}\ and\ \citenamefont
  {{You}}(2020)}]{Akhtar2020M2006.08797}%
  \BibitemOpen
  \bibfield  {author} {\bibinfo {author} {\bibfnamefont {A.~A.}\ \bibnamefont
  {{Akhtar}}}\ and\ \bibinfo {author} {\bibfnamefont {Y.-Z.}\ \bibnamefont
  {{You}}},\ }\href {\doibase 10.1103/PhysRevB.102.134203} {\bibfield
  {journal} {\bibinfo  {journal} {\prb}\ }\textbf {\bibinfo {volume} {102}},\
  \bibinfo {eid} {134203} (\bibinfo {year} {2020})},\ \Eprint
  {http://arxiv.org/abs/2006.08797} {arXiv:2006.08797 [cond-mat.dis-nn]}
  \BibitemShut {NoStop}%
\bibitem [{\citenamefont {Weingarten}(1978)}]{Weingarten_Asymptotic_1978}%
  \BibitemOpen
  \bibfield  {author} {\bibinfo {author} {\bibfnamefont {D.}~\bibnamefont
  {Weingarten}},\ }\href {\doibase 10.1063/1.523807} {\bibfield  {journal}
  {\bibinfo  {journal} {Journal of Mathematical Physics}\ }\textbf {\bibinfo
  {volume} {19}},\ \bibinfo {pages} {999} (\bibinfo {year} {1978})},\ \Eprint
  {http://arxiv.org/abs/https://pubs.aip.org/aip/jmp/article-pdf/19/5/999/19149375/999\_1\_online.pdf}
  {https://pubs.aip.org/aip/jmp/article-pdf/19/5/999/19149375/999\_1\_online.pdf}
  \BibitemShut {NoStop}%
\bibitem [{\citenamefont {Collins}\ and\ \citenamefont
  {{\'{S}}niady}(2006)}]{Collins_Integration_2006}%
  \BibitemOpen
  \bibfield  {author} {\bibinfo {author} {\bibfnamefont {B.}~\bibnamefont
  {Collins}}\ and\ \bibinfo {author} {\bibfnamefont {P.}~\bibnamefont
  {{\'{S}}niady}},\ }\href {\doibase 10.1007/s00220-006-1554-3} {\bibfield
  {journal} {\bibinfo  {journal} {Communications in Mathematical Physics}\
  }\textbf {\bibinfo {volume} {264}},\ \bibinfo {pages} {773} (\bibinfo {year}
  {2006})}\BibitemShut {NoStop}%
\bibitem [{\citenamefont {Qi}\ \emph {et~al.}(2017)\citenamefont {Qi},
  \citenamefont {Yang},\ and\ \citenamefont {You}}]{qi_holographic_2017}%
  \BibitemOpen
  \bibfield  {author} {\bibinfo {author} {\bibfnamefont {X.-L.}\ \bibnamefont
  {Qi}}, \bibinfo {author} {\bibfnamefont {Z.}~\bibnamefont {Yang}}, \ and\
  \bibinfo {author} {\bibfnamefont {Y.-Z.}\ \bibnamefont {You}},\ }\href
  {\doibase 10.1007/JHEP08(2017)060} {\bibfield  {journal} {\bibinfo  {journal}
  {Journal of High Energy Physics}\ }\textbf {\bibinfo {volume} {2017}},\
  \bibinfo {pages} {60} (\bibinfo {year} {2017})}\BibitemShut {NoStop}%
\bibitem [{\citenamefont {Bartolucci}\ \emph {et~al.}(2023)\citenamefont
  {Bartolucci}, \citenamefont {Birchall}, \citenamefont {Bombin}, \citenamefont
  {Cable}, \citenamefont {Dawson}, \citenamefont {Gimeno-Segovia},
  \citenamefont {Johnston}, \citenamefont {Kieling}, \citenamefont {Nickerson},
  \citenamefont {Pant}, \citenamefont {Pastawski}, \citenamefont {Rudolph},\
  and\ \citenamefont {Sparrow}}]{bartolucci_fusion-based_2023}%
  \BibitemOpen
  \bibfield  {author} {\bibinfo {author} {\bibfnamefont {S.}~\bibnamefont
  {Bartolucci}}, \bibinfo {author} {\bibfnamefont {P.}~\bibnamefont
  {Birchall}}, \bibinfo {author} {\bibfnamefont {H.}~\bibnamefont {Bombin}},
  \bibinfo {author} {\bibfnamefont {H.}~\bibnamefont {Cable}}, \bibinfo
  {author} {\bibfnamefont {C.}~\bibnamefont {Dawson}}, \bibinfo {author}
  {\bibfnamefont {M.}~\bibnamefont {Gimeno-Segovia}}, \bibinfo {author}
  {\bibfnamefont {E.}~\bibnamefont {Johnston}}, \bibinfo {author}
  {\bibfnamefont {K.}~\bibnamefont {Kieling}}, \bibinfo {author} {\bibfnamefont
  {N.}~\bibnamefont {Nickerson}}, \bibinfo {author} {\bibfnamefont
  {M.}~\bibnamefont {Pant}}, \bibinfo {author} {\bibfnamefont {F.}~\bibnamefont
  {Pastawski}}, \bibinfo {author} {\bibfnamefont {T.}~\bibnamefont {Rudolph}},
  \ and\ \bibinfo {author} {\bibfnamefont {C.}~\bibnamefont {Sparrow}},\ }\href
  {\doibase 10.1038/s41467-023-36493-1} {\bibfield  {journal} {\bibinfo
  {journal} {Nature Communications}\ }\textbf {\bibinfo {volume} {14}},\
  \bibinfo {pages} {912} (\bibinfo {year} {2023})}\BibitemShut {NoStop}%
\end{thebibliography}%

\onecolumngrid
\newpage
\appendix

\section{Optimality of $d^k$ scaling \label{app:optimality}}

Here we show that the asymptotic scaling of sample complexity as $\sim d^k$, when it applies to {\it all} Pauli operators supported withing a segment of length $k$, is optimal. 
Let us consider a segment $A$.
From Ref.~\cite{Ippoliti2024L2307.15011} we have that 
\begin{equation}
    d^{|A|} W_A = \sum_{P: \text{supp}(P)\subseteq A} w_{\mathcal{E}_\sigma}(P) = 1 + \sum_{P\neq I: \text{supp}(P)\subseteq A} w_{\mathcal{E}_\sigma}(P).
\end{equation}
Using the fact that the EF is at most 1, we immediately get the bound 
\begin{equation}
    (d^{2|A|}-1)\min_{P:\text{supp}(P)\subseteq A} w_{\mathcal{E}_\sigma}(P) 
    \leq \sum_{P\neq I: \text{supp}(P)\subseteq A} w_{\mathcal{E}_\sigma}(P) 
    \leq d^{|A|}-1
\end{equation}
and thus 
\begin{equation}
    \min_{P:\text{supp}(P)\subseteq A} w_{\mathcal{E}_\sigma}(P) \leq \frac{1}{d^{|A|}+1}
    \implies 
    \max_{P:\text{supp}(P)\subseteq A} \| P\|_{\mathcal{E}_\sigma}^2 > d^{|A|}.
\end{equation}
So there exists at least one operator supported in $A$ with squared shadow norm larger than $d^{|A|}$. Usually this worst-case operator is just the one of maximum weight, $k=|A|$. 
This gives the desired bound for contiguous operators, $\|P\|_{\mathcal{E}_\sigma}^2 > d^k$. 
In special measurement ensembles~\cite{Ippoliti2024C2305.10723} the worst-case operators may be of lower weight, so the shadow norm of the $k=|A|$ operator may be smaller than $d^k$. In particular it is possible to achieve the scaling $(d-1/d)^k$ via randomized GHZ-basis measurements on certain operators. But this scaling, as shown above, cannot extend to all operators supported in a given contiguous subsystem.

\section{Recursive Structure of Shadow Norm in Tree Circuit }
\label{sec:recursion_derivation}
\subsection{Derivation of Recursive Relation}
\begin{figure*}[htbp]
    \centering
    \includegraphics[width=1.0\textwidth]{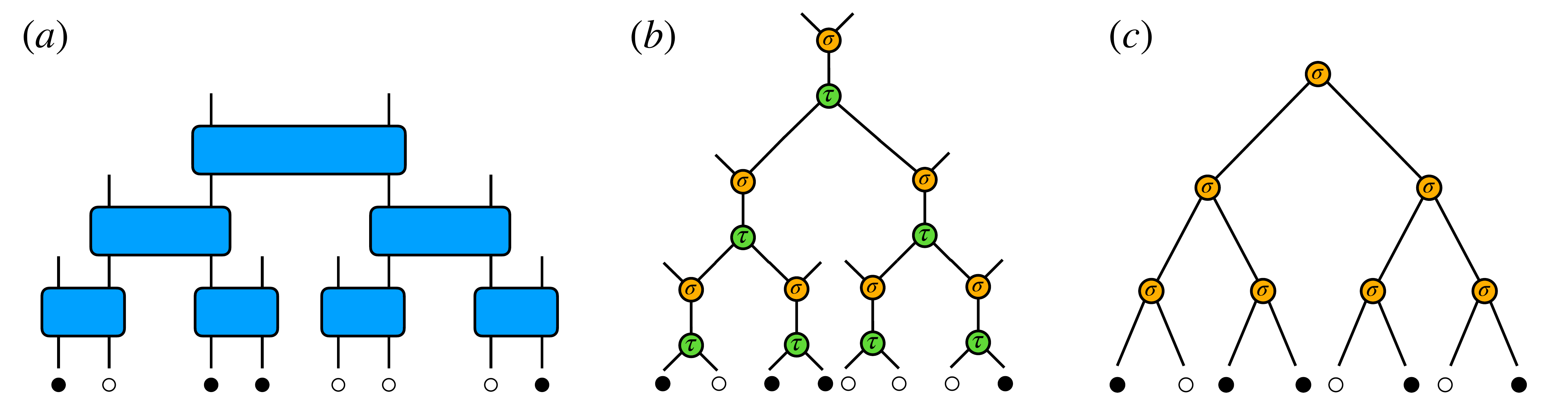}
    \caption{Illustration plot of the tree circuit. (a) The tree circuit defined on qudit system. (b) By using the Weingarten function, we map the tree circuit to a tree defined on a hexagonal lattice. Each $\sigma$ and $\tau$ is an element of $S_2$ permutation group. (c) By integrating all inner degree $\tau$, we get a tree with  $\sigma$ alone. Here we give an arbitrary example of the boundary condition. We use $\bullet/\circ$ to represent qudits inside/outside the support.}
    \label{fig:illustration_tree}
\end{figure*}
In this section, we provide a detailed derivation of the recursive relation of Pauli learning rates. For a Pauli string $P$ with its support labeled as $A$ (i.e., the region in which $P$ acts), we have the following useful relation \cite{Bu2024C2202.03272}
\be
w_{\mathcal{E}_{\sigma}}(P)= \frac{(-1)^{\left|A\right|}}{(d^2-1)^{\left|A\right|}}\sum_{B\subseteq A}(-d)^{\left| B\right|}W_{\mathcal{E}_{\sigma}}(B).\label{eq:PLR&EF}
\ee
Here $|A|$ denotes the size of the support $A$. $B$ takes all possible subsets of $A$. $W_{\mathcal{E}_{\sigma}}(B):=\mathbb{E}_{\sigma \in\mathcal{E}_{\sigma}}\Tr_{B} (\Tr_{\bar{B}}\sigma)^2$ is the entanglement feature \cite{You2018E1803.10425,Kuo2020M1910.11351,Akhtar2020M2006.08797} of subsystem $B$, which is the main task to calculate Pauli learning rate. 

For a specific $B$, the entanglement feature can be calculated through utilizing the tree structure of the circuit. In Fig.~\ref{fig:illustration_tree}(a), we give an illustration of a $m$ layer tree circuit defined on $2^m$ qudits. At each layer, half of the qubits are discarded by projective measurements. We require that all unitary gates (blue blocks) in the tree are chosen Haar randomly, where the average $\mathbb{E}_{\sigma\in \mathcal{E}_{\sigma}} (\sigma\otimes \sigma^{\dagger})^2$ can be calculated using the Weingarten function \cite{Weingarten_Asymptotic_1978,Collins_Integration_2006}. We can rewrite $W_{\mathcal{E}_{\sigma}}(B)$ as $\Tr\left[C_{B,\bar{B}}\mathbb{E}_{\mathcal{E}_{\sigma}}(\sigma\otimes \sigma^{\dagger})^2\right]$ with $C_{B,\bar{B}}$ as a boundary operator to restore the trace over $B$ and $\bar{B}$. After this step, the previous tree circuit is mapped to a tree graph defined on a hexagonal lattice as shown by Fig.~\ref{fig:illustration_tree}(b). Here each vertex $\sigma(\tau)$ represents an element of permutation group $S_2$. Every vertical bond in Fig.~\ref{fig:illustration_tree}(b) contributes a factor $Wg_{d^2}(\sigma^{-1}\tau)$ and every inclined bond between two vertices contributes a factor $\braket{\sigma}{\tau}=d^{2-C(\sigma^{-1}\tau)}$. Here $C(\sigma)$ represents the number of transposition of element $\sigma$. 
The boundary term $C_{B,\bar{B}}$ also assigns an element of $S_2$ on each leave --- qudits within $B$ are assigned the swap element due to the term $(\tr_{\bar{B}}\sigma)^2$, while those not in $B$ are assigned with the identity element since they are simply traced out. Using this notation, the entanglement feature $W_{\mathcal{E}_{\sigma}}(B)$ is the sum of weights of all possible configurations of $\sigma$ and $\tau$ with each weight obtained by multiplying all factors inside the tree. For each specific configuration of $\sigma$ and $\tau$, it is convenient to further sum over the degrees defined on vertices labeled by $\tau$. Now we have a tree lattice with only $\sigma$ vertices left as shown in Fig.~\ref{fig:illustration_tree}(c). We assign a factor $\eta$ to each vertex in the tree. This factor $\eta$ depends on $\sigma$ and the two vertices $\sigma'$ and $\sigma''$ connecting to it from the bottom. Specifically, $\eta$ takes the form
\begin{equation}
    \eta(\sigma;\sigma',\sigma'') = \left\{\begin{aligned}
        a&\quad \text{if $\sigma'\ne\sigma''$,}\\
        1&\quad\text{if $\sigma=\sigma'=\sigma''$,}\\
        0&\quad\text{if $\sigma\ne\sigma'$ and $\sigma'=\sigma''$,}
    \end{aligned}\right.\label{eq:eta_def}
\end{equation}
with $a=d/(d^2+1)$. Then the weight of this configuration is the product of $\eta$ on each vertices.

Following the above argument, for any given tree --- either a subtree or the whole tree --- we can define a vector
\be
\begin{pmatrix}
\bar{W}_\mathds{1}\\\\
\bar{W}_{c}
\end{pmatrix},
\ee
where $\bar{W}_{\sigma}$ represents the sum of weights of all configurations which have $\sigma$ at the top of this tree. $\mathds{1}$ and $c$ represent identity operator and swap operator respectively. The entanglement feature is obtained by the sum of this two components. Note that this vector is only determined by the part of $B$ inside this subtree alone. It can be further noticed that each tree can be viewed as connecting two subtrees to a node, as shown in Fig.~\ref{fig:illustration_tree}(c). Using factor $\eta$ on this node, we can write down the recursive relation of the entanglement feature. Labeling the two sub-trees as $s_1$ and $s_2$, and the new tree they form as $s$, we have
\be
\begin{pmatrix}
\bar{W}_{\mathds{1}}^{s}\\\\\bar{W}_{c}^{s}
\end{pmatrix}
=
\begin{pmatrix}
1&a&a&0\\
0&a&a&1
\end{pmatrix}
\begin{pmatrix}
\bar{W}^{s_1}_{\mathds{1}}\\\\ \bar{W}^{s_1}_{c}
\end{pmatrix}
\otimes 
\begin{pmatrix}
\bar{W}^{s_2}_{\mathds{1}}\\\\ \bar{W}^{s_2}_{c}
\end{pmatrix}.\label{eq:EF_recursion}
\ee
Starting with the particle/hole configuration on the bottom, we can calculate the entanglement feature of any subtree recursively.

Using Eq.~\ref{eq:PLR&EF}, we can define another vector associated with ``subtree" Pauli learning rate of string $P$. Denoting the support of $P$ on this subtree as $A^s$, and the part of $B\subseteq A$ inside the tree as $B^s$, we have
\be
\begin{pmatrix}
\bar{w}^s_{\mathds{1}}(A)\\\\
\bar{w}_{c}^s(A)
\end{pmatrix}
=
\begin{pmatrix}
\frac{(-1)^{\left|A^s\right|}}{(d^2-1)^{\left|A^s\right|}}\sum_{B\subseteq A}(-d)^{\left|B^s\right|}\bar{W}^s_{\mathds{1}}(B)\\\\
\frac{(-1)^{\left|A^s\right|}}{(d^2-1)^{\left|A^s\right|}}\sum_{B\subseteq A}(-d)^{\left|B^s\right|}\bar{W}^s_{c}(B).
\end{pmatrix}\label{eq:w_bar_def}
\ee
In the above definition, $\bar{w}^s(A)$ and $\bar{W}^s(B)$ depend solely on $A^s$ and $B^s$, respectively. It can be seen from Eq.~\ref{eq:EF_recursion} that the $\bar{w}$ vector we introduce satisfies the following recursive relation,
\be
\begin{pmatrix}
    \bar{w}_{\mathds{1}}^s(A)\\\\
    \bar{w}_{c}^s(A)
\end{pmatrix}
=\begin{pmatrix}
1&a&a&0\\
0&a&a&1
\end{pmatrix}
\begin{pmatrix}
    \bar{w}_{\mathds{1}}^{s_1}(A)\\\\
    \bar{w}_{c}^{s_1}(A)
\end{pmatrix}\otimes 
\begin{pmatrix}
    \bar{w}^{s_2}_{\mathds{1}}(A)\\\\
    \bar{w}^{s_2}_{c}(A)
\end{pmatrix}
\ee
When $s$ represents the entire system, the Pauli learning rate can be obtained by summing the two components of the vector. When $s$ represents a subtree with only 2 qudits, the corresponding $\bar{w}$ vector can be directly obtained from Eq.~\ref{eq:eta_def} and Eq.~\ref{eq:w_bar_def}. Then the $\bar{w}$ vector of any subtree can be obtained. It turns out that for a  2-qudit subtree with all qudits outside the support (referred to as hole-like and denoted by $\circ$), the corresponding $\bar{w}$ vector is
\be
\begin{pmatrix}
    1\\\\0
\end{pmatrix}.
\ee
While for a 2-qudit subtree with at least one qudit inside the support (referred to as particle-like and denoted by $\bullet$), the initial vector is
\be
\begin{pmatrix}
    (-1)/(d^4-1)\\\\
    d^2/(d^4-1)
\end{pmatrix}.
\ee
By using Eq.~\ref{eq:PLR_recursion} alongside these initial conditions, the Pauli learning rate can be calculated recursively.

\subsection{Comparison between Contiguous Tree and Shallow Circuit}

Here we derive the critical boundary size distinguishing between the performance of the contiguous tree and the shallow circuit. Previous work indicates that for shallow circuit with optimal depth, the squared shadow norm can be approximated by $kd^k$ with $k$ being the support size \cite{Ippoliti_2023}. To simplify the calculation, we ignore a $O(1)$ correction which comes from the coupling between the support and the rest of the system. When the shadow norm in the tree is smaller compared to the shallow circuit, 
\be
kd^{k}\ge \frac{(d^2-1)^{k}}{d^{k}}\exp(-Q(d)k).
\ee
When the above equation takes the equality, the solution can be expressed by the Lambert-$W$ function,
\be
k^{\ast} = \frac{W_{0/-1}\left(Q(d)+\ln\left(d^2/(d^2-1)\right)\right)}{\left(Q(d)+\ln\left(d^2/(d^2-1)\right)\right)}. 
\ee
The Lambert-$W$ function has two solutions when $-1/e<Q(d)+\ln\left(d^2/(d^2-1)\right)<0$. This condition is guaranteed by the fact 
\begin{align}
-\frac{1}{e}<&\frac{\ln(1+2ag^0)}{2}+\frac{\ln(1+2ag^1)}{2}+\ln\frac{d^2}{d^2-1}\nonumber\\<&Q(d)+\ln\frac{d^2}{d^2-1}\nonumber\\
<&\frac{\ln(1+2ag^0)}{2}+\frac{\ln(1+2ag^1)}{4}+\frac{\ln(1+2ag^2)}{8}+\ln\frac{d^2}{d^2-1}\nonumber\\
<&0.
\end{align}
When the size of the support is between the solutions given by these two branches, the tree circuit gives a smaller shadow norm. In fact, the $W_0$ branch gives a value $O(1)$, so the upper bound given by $W_{-1}$ is more important. When $d=2$, the $W_{-1}$ branch gives $k^{\ast}\approx 6.4$, which matches with Fig.~\ref{fig:tree_shallow_comparison}. In the limit $d$ goes to infinity, the $W_{-1}$ branch gives 
\begin{align}
 k^{\ast}\sim&-\frac{\ln(-\ln(-Q(d)-\ln(d^2/(d^2-1))))}{Q(d)+\ln(d^2/(d^2-1))}\nonumber\\
 &\quad+\frac{\ln(-Q(d)-\ln(d^2/(d^2-1)))}{Q(d)+\ln(d^2/(d^2-1)}.
\end{align}
Combining with the leading order approximation of $Q(d)+\ln\left(d^2/(d^2-1)\right)$ (expanding the series in Eq.~\ref{eq:Qd_def}), it shows that the critical system size behaves like $6d^3\ln d+2d^3\ln\ln d$.

\section{Applying holographic tensor networks by fusion \label{app:fusion}}

Here we briefly discuss the physical implementation of holographic tensor networks for classical shadows. The issue that arises, relative to conventional randomized measurement schemes, is that the holographic networks cannot be trivially expressed as a sequence of unitary operations followed by (or interspersed with~\cite{Ippoliti2024L2307.15011,akhtar2023measurementinduced}) projective measurements. It is thus not obvious {\it a priori} how to physically apply a holographic network (as a linear operator) to a system of interest. 

\begin{figure}
    \centering
    \includegraphics[width=0.8\textwidth]{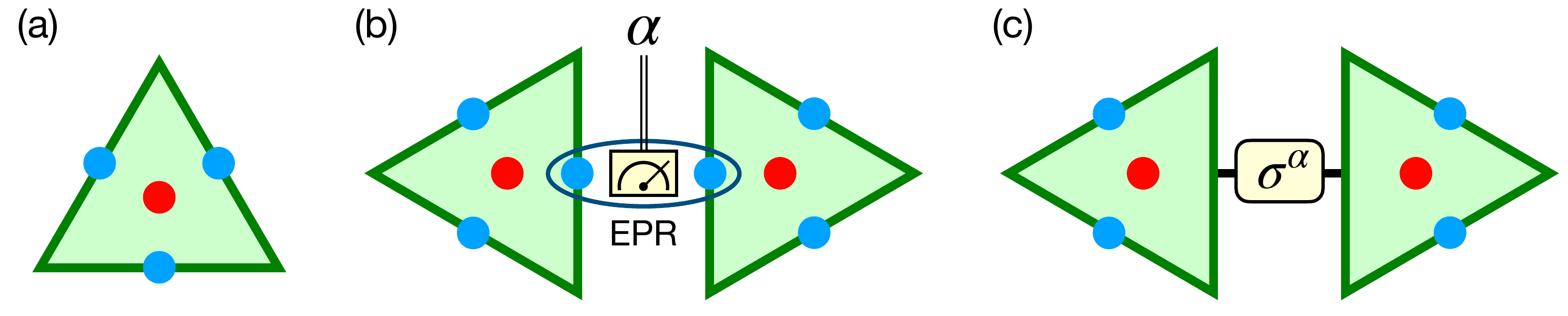}
    \caption{Creating holographic RTNs by fusion. In this schematic we focus on the $\{p,q\} = \{3,7\}$ tiling.
    (a) We prepare random $(p+1)$-qudit states (triangle) on auxiliary qudits. A central qudit (red dot) will serve as a ``bulk'' degree of freedom, and the $p$ qudits on the sides (blue) will be used to glue the tensors together. 
    (b) We take two such states and perform a two-qudit measurement in the EPR basis (e.g., we measure $XX$ and $ZZ^\dagger$) on two of the side qudits as shown. We obtain an outcome $\alpha$ that labels a state in the EPR basis, $(1/\sqrt{d}) \sum_{j=0}^{d-1} (\sigma^\alpha \ket{j})\otimes \ket{j}$. 
    (c) This process fuses the two tensors by creating a virtual bond as shown. A Pauli matrix $\sigma^\alpha$ acts on the virtual bond; it can be absorbed into either tensor if the distribution over $(p+1)$-qudit states is Pauli-invariant. Iterating this fusion process allows us to create the entire $\{p,q\}$-tiling RTN and to connect it to the boundary system to perform holographic shadow tomography. }
    \label{fig:fusion}
\end{figure}

A general route to achieve this is by {\it fusion}~\cite{qi_holographic_2017,bartolucci_fusion-based_2023}. 
We form the desired tensors for a $\{p,q\}$ hyperbolic tiling as random states on $(p+1)$ auxiliary qudits, Fig.~\ref{fig:fusion}(a); we then fuse them via EPR measurements on appropriately chosen pairs of qudits, Fig.~\ref{fig:fusion}(b). This realizes the desired tensor network up to unwanted Pauli matrices on the measured bonds, Fig.~\ref{fig:fusion}(c). However, as long as the tensor ensemble is locally-scrambled (in fact Pauli-invariant~\cite{Bu2024C2202.03272}, which is a weaker condition), the unwanted Pauli matrix $\sigma^\alpha$ can be absorbed into either of the nearby tensors, $T_{i_1\dots i_p j} (\sigma^\alpha)_{jk} \equiv T'_{i_1\dots i_p k}$; this change in the tensor $T\mapsto T'$ defines another valid snapshot $\sigma'$ that has the {\it same} prior probability as the original one $\sigma$ (where the EPR measurement outcomes are all identity): $p(\sigma') = p(\sigma)$. As a consequence, this has no effect on the shadow tomography protocol. 

This protocol requires a number of auxiliary qudits proportional to the volume of the hyperbolic space. Since in hyperbolic spaces the volume is proportional to the length of the boundary, the number of auxiliary qudits needed is $O(N)$.

\section{Derivation of Statistical Model for Pauli Learning Rate \label{sec:PLRderivation}}
In this section, we first derive the statistical model for Pauli learning rate from tensor diagrams through an example to provide some intuition. Then, we provide a more rigorous mathematical derivation using the previously known statistical model of the entanglement feature. 

In the following derivation, we will distinguish the bulk legs (the legs connecting different random tensors in the bulk) and the boundary legs (the dangling legs from the random tensors on the boundary). We will denote the bond dimension of the bulk and boundary legs as $D_e$ and $D_\partial$ separately, such that $J=\frac{1}{2}\ln D_e$ and 

Consider a RTN state $\sigma=$ \parfig{0.05}{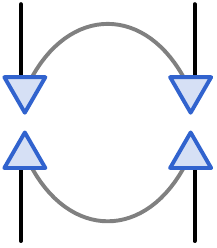}, and suppose we are interested in calculating the Pauli learning rate $w_{\mathcal{E}_\sigma}(P\otimes I)$ of a single Pauli P supported on the first boundary region. Following \eqnref{eq:PLR}, the denominator is
\eqs{
    \mathbb{E}_{\sigma \sim p(\sigma)}(\text{Tr}~\sigma)^2 &= \parfig{0.1}{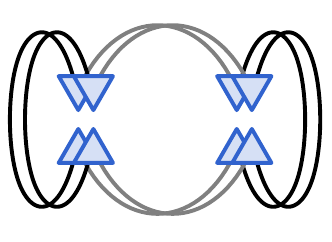} \\&= \parfig{0.1}{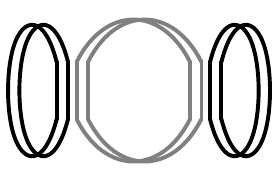} + \parfig{0.1}{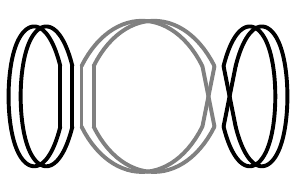} +\parfig{0.1}{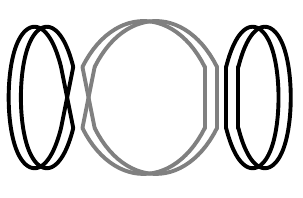}+ \parfig{0.1}{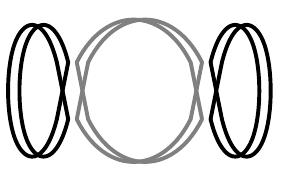}\\
    &= D_e^2 D_\partial^4 + 2 D_e D_\partial^3 + D_e^2 D_\partial^2\\
    &= e^{4J+8h}+2e^{2J+6h}+e^{4J+4h}
}
and the numerator is 
\eqs{\label{eq:num}
    \mathbb{E}_{\sigma \sim p(\sigma)}(\text{Tr}~P \otimes I\sigma)^2 &=\parfig{0.1}{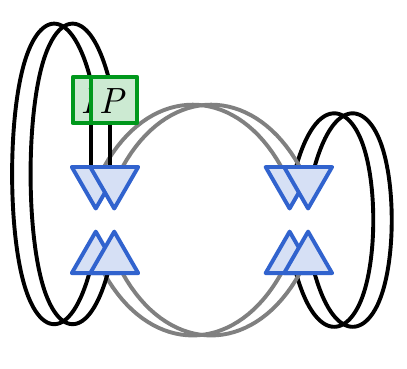} \\&= \parfig{0.1}{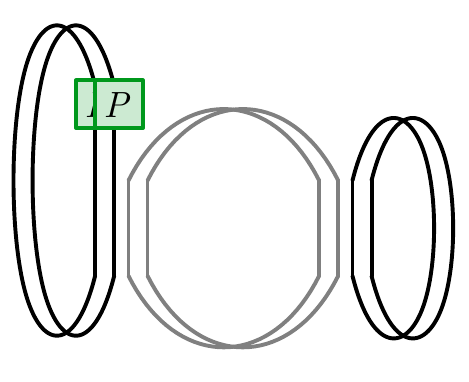} + \parfig{0.1}{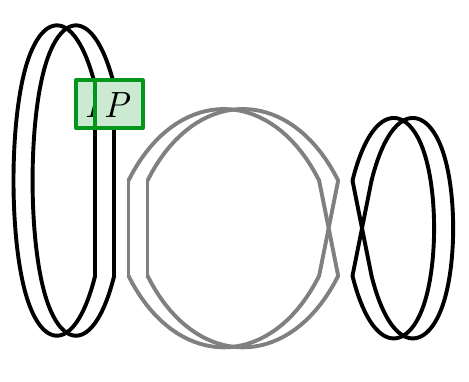} +\parfig{0.1}{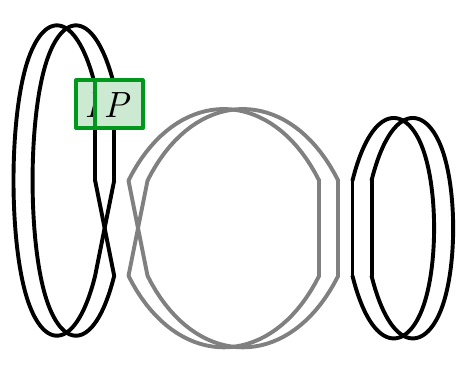}+ \parfig{0.1}{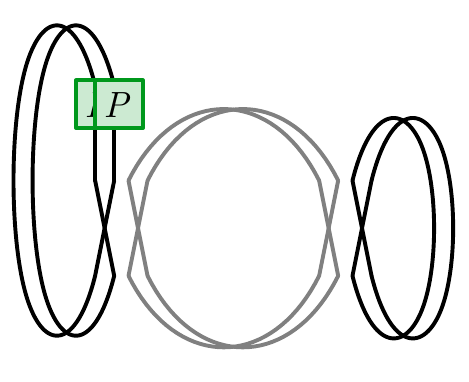}\\
    &=0+0+D_e D_\partial^3 +D_e^2D_\partial^2\\
    &=e^{2J+6h}+e^{4J+4h}
    }
Note that the first two diagrams in \eqref{eq:num} give zero contribution because the first vertex is taken to be spin +1 and the trace of a nontrivial Pauli operator is zero. Therefore, for a vertex with boundary legs that fall within the support of the Pauli operators to contribute non-trivially, it must be assigned a spin of -1. This requirement is mathematically represented as $\prod\limits_{dv \in \supp P} \delta_{\sigma_v, -1}$ in the numerator of \eqref{eq:PLR stat model}. We can directly check that the result obtained from the tensor diagram calculation is consistent with \eqref{eq:PLR stat model}:
\begin{align}
    w_{\mathcal{E}_\sigma}(P\otimes I)\simeq {e^{2J+6h}+e^{4J+4h} \over e^{4J+8h}+2e^{2J+6h}+e^{4J+4h}}={e^{-J}+e^{J-2h}\over e^{J+2h}+2e^{-J}+e^{J-2h}}={\sum\limits_{\sigma_1,\sigma_2} \delta_{\sigma_1,-1}e^{J\sigma_1 \sigma_2+h(\sigma_1+\sigma_2)}\over \sum\limits_{\sigma_1,\sigma_2} e^{J\sigma_1 \sigma_2+h(\sigma_1+\sigma_2)}}
\end{align}

Now, we present a more rigorous mathematical derivation from the well-established statistical model of entanglement feature. According to \cite{You_2018}, the statistical model for the entanglement feature is: 
\eqs{
    W_{\mathcal{E}_\sigma}(A) \simeq {\mathbb{E}_{\sigma \sim p(\sigma)}\text{Tr}~\sigma^{\otimes 2}{\hat{\tau}(A)} \over \mathbb{E}_{\sigma \sim p(\sigma)}\text{Tr}~\sigma^{\otimes 2}}={\sum\limits_{[\sigma_v]} e^{\sum_{e \in \mathcal{E}} J \prod_{v \in \partial e} \sigma_v + h\sum_{v \in V_{\partial}}\sigma_v \tau_v(A)} \over \sum\limits_{[\sigma_v]} e^{\sum_{e \in \mathcal{E}} J \prod_{v \in \partial e} \sigma_v + h\sum_{v \in V_{\partial}}\sigma_v}}={\mathcal{Z}[\tau(A)]\over \mathcal{Z}[\tau(\emptyset)]}
\label{eq:EFmodel}}
where $\tau_v(A)=1$ if $v \notin A$ and $\tau_v(A)=-1$ if $v \in A$.  Following the formalism in \cite{akhtar2023measurementinduced}
, we can construct the entanglement feature state
\begin{align}
    \ket{W_{\CE_\sigma}}=\sum_A W_{\CE_\sigma}(A)\ket{A}
\end{align} 
Then, the Pauli learning rate can be obtained by
\be w_{\CE_\sigma}(P) = \braket{P}{W_{\CE_\sigma}} \ee
where 
\be 
    \ket{P}:=\prod_{i\in\supp P}{D_{\partial}X_i-I_i \over D_{\partial}^2-1} \ket{\bm{0}} 
\ee
Thus, 
\begin{equation}
    \begin{split}
        w_{\mathcal{E}_{\sigma}}(P) &=\bra{\bm{0}}\prod\limits_{i\in\supp P} {D_{\partial}X_i-I_i \over D_{\partial}^2-1}\sum_A W_{\CE_\sigma}(A) \ket{A}\\
    &=\sum_A W_{\CE_\sigma}(A) \bra{\bm{0}}\prod\limits_{i\in\supp P} {D_{\partial}X_i-I_i \over D_{\partial}^2-1} \prod\limits_{j\in A} X_j \ket{\bm{0}}\\
    &=\frac{(-1)^{\left|\supp P\right|}}{(D_{\partial}^2-1)^{\left|\supp P\right|}}\sum\limits_{A\subseteq \supp P}(-D_{\partial})^{\left| A\right|} W_{\CE_\sigma}(A)\\
    &=\frac{(-1)^{\left|\supp P\right|}}{(D_{\partial}^2-1)^{\left|\supp P\right|}} {\sum\limits_{A\subseteq \supp P}(-D_{\partial})^{\left| A\right|}\mathcal{Z}[\tau(A)] \over \mathcal{Z}[\tau(\emptyset)]}\\
    \end{split}
\label{eq:PLRderiv}
\end{equation}
Note that
\begin{equation}
\label{eq:num}
    \begin{split}
        \sum\limits_{A\subseteq \supp P}(-D_{\partial})^{\left| A\right|}\mathcal{Z}[\tau(A)]
        &=\sum\limits_{A\subseteq \supp P}(-D_{\partial})^{\left| A\right|}\sum\limits_{[\sigma_v]} e^{\sum\limits_{e \in \mathcal{E}} J \prod\limits_{v \in \partial e} \sigma_v + h\sum\limits_{v \in V_{\partial}}\sigma_v \tau_v(A)}\\
        &=\sum\limits_{[\sigma_v]} \sum\limits_{A\subseteq \supp P}(-1)^{\left| A\right|} e^{\sum\limits_{e \in \mathcal{E}} J \prod\limits_{v \in \partial e} \sigma_v} e^{h\(\sum\limits_{v \in V_{\partial}}\sigma_v \tau_v(A)+2|A|\)}\\
        &=\sum\limits_{[\sigma_v]} e^{\sum\limits_{e \in \mathcal{E}} J \prod\limits_{v \in \partial e} \sigma_v+h \sum\limits_{v \in V_{\partial}}\sigma_v} \sum\limits_{A\subseteq \supp P}(-1)^{\left| A\right|} D_\partial^{2 \sum\limits_{v \in A}\delta_{\sigma_v,-1}}
    \end{split}
\end{equation}
where in the last equality, we used
\begin{align*}
    e^{h\(\sum\limits_{v \in V_{\partial}}\sigma_v \tau_v(A)+2|A|\)}=e^{h \sum\limits_{v \in V_{\partial}} \sigma_v}e^{2h \sum\limits_{v \in A} (1-\sigma_v))}=e^{h \sum\limits_{v \in V_{\partial}} \sigma_v}e^{4h \sum\limits_{v \in A} \delta_{\sigma_v,-1}}=e^{h \sum\limits_{v \in V_{\partial}} \sigma_v}D_\partial^{2 \sum\limits_{v \in A} \delta_{\sigma_v,-1}}
\end{align*}
Substituting \eqnref{eq:num} into \eqnref{eq:PLRderiv} yields
\begin{equation}
    \begin{split}
         w_{\mathcal{E}_{\sigma}}(P) 
         &={\sum\limits_{[\sigma_v]} e^{\sum\limits_{e \in \mathcal{E}} J \prod\limits_{v \in \partial e} \sigma_v+h \sum\limits_{v \in V_{\partial}}\sigma_v} \frac{(-1)^{\left|\supp P\right|}}{(D_{\partial}^2-1)^{\left|\supp P\right|}}\sum\limits_{A\subseteq \supp P}(-1)^{\left| A\right|}D_\partial^{2 \sum\limits_{v \in A}\delta_{\sigma_v,-1}} \over \mathcal{Z}[\tau (\emptyset)]}\\
         &={\sum\limits_{[\sigma_v]} \prod\limits_{d v \in \supp P} \delta_{\sigma_v,-1}e^{\sum\limits_{e \in \mathcal{E}} J \prod\limits_{v \in \partial e} \sigma_v + h\sum\limits_{v \in V_{\partial}}\sigma_v} \over \mathcal{Z}[\tau (\emptyset)]}\\
    &={\sum\limits_{[\sigma_v]} \prod\limits_{d v \in \supp P} \delta_{\sigma_v,-1}e^{\sum\limits_{e \in \mathcal{E}} J \prod\limits_{v \in \partial e} \sigma_v + h\sum\limits_{v \in V_{\partial}}\sigma_v} \over \sum\limits_{[\sigma_v]} e^{\sum\limits_{e \in \mathcal{E}} J \prod\limits_{v \in \partial e} \sigma_v + h\sum\limits_{v \in V_{\partial}}\sigma_v}}
    \end{split}
\end{equation}
where in the second equality, we used
\begin{align*}
    \frac{(-1)^{\left|\supp P\right|}}{(D_{\partial}^2-1)^{\left|\supp P\right|}} \sum\limits_{A\subseteq \supp P}(-1)^{\left| A\right|}D_{\partial}^{2\sum \limits_{v \in A}\delta_{\sigma_v, -1}}= \prod \limits_{v \in \supp P} \delta_{\sigma_v, -1}
\end{align*}

\section{Derivation of \eqnref{eq:EEtoBC}\label{sec:EEtoBCderivation}}
Taking $J=h={1 \over 2} \ln d$ in \eqnref{eq:EFmodel},
\eqs{
W_{\CE_\sigma}(A) \simeq {\sum\limits_{[\sigma_v]} d^{{1 \over 2}\(\sum_{e \in \mathcal{E}} \prod_{v \in \partial e} \sigma_v + \sum_{v \in V_{\partial}}\sigma_v \tau_v(A)\)} \over \sum\limits_{[\sigma_v]} d^{{1 \over 2}\(\sum_{e \in \mathcal{E}} \prod_{v \in \partial e} \sigma_v + \sum_{v \in V_{\partial}}\sigma_v\)}}
}
Then, taking the $d\rightarrow \infty$ asymptotic limit, only the leading order term in the denominator and numerator contributes. Similar to the case for PLR, the leading order term in the denominator comes from the configuration with all vertices having spin 1, which gives a contribution of $d^{2 n_b}$. The leading order term in the numerator comes from either the configuration with all vertices having spin 1 or the configuration with a minimal domain wall such that vertices with connection to $A$ have spin -1, depending on whether the length of the boundary region is smaller than the minimal domain wall length (ie. the length of geodesic bonding the boundary region). For most boundary regions A that are not too small (ie. $|A|>3$ in \{3,7\} and \{5,4\} holographic RTN), the length of the boundary region is larger than the length of geodesic bonding the boundary region, so
\eq{e^{-S(A)}=W_{\CE_\sigma}(A)\simeq d^{-\bulkC|A|}}

\section{Derivation of $c_{eff}$ in continuous space \label{sec:eff}}
From the metric in \eqnref{eq:PDmetric},
\eq{ds={2R \sqrt{d\rho^2+\rho^2d\theta^2} \over 1-\rho^2}}
The length of an arc corresponding to an angle $\phi$ of a circle with Euclidean radius $\rho$ is
\eq{L(\rho)=\int_0^{\phi} {2R \rho \over 1-\rho^2}d\theta ={2 \phi R \rho \over 1-\rho^2}\label{eq:CE}}
Inverting this equation and keeping only the positive solution yields
\eq{\rho(L)={-\phi R + \sqrt{L^2+\phi^2 R^2}\over L}\label{eq:rho}}
From Eq.~(4) in \cite{Basteiro_2022}, the geodesic connecting the two points $w_1=(\rho_1,\phi_1)$ and $w_2=(\rho_2,\phi_2)$ on the Poincar\'e disc is
\eq{d(w_1, w_2)=R\arcosh\(1+{2(\rho_1^2+\rho_2^2-2\rho_1\rho_2\cos{(\phi_1-\phi_2}) \over (1-\rho_1^2)(1-\rho_2^2)}\)}
Thus, the geodesic $d_H$ connecting the ends of the arc at $(\rho, 0)$ and $(\rho, \phi)$ is:
\eq{d_L=R\arcosh{\(1+{4 \rho^2 (1-\cos{\phi})\over (1-\rho^2)^2}\)}\label{eq:dL}}
Substituting $\rho(L)$ from \eqref{eq:rho} into \eqref{eq:dL} yields
\eq{d_L=R\arcosh\({1+{L^2(1-\cos{\phi})\over\phi^2 R^2}}\)=2 R\ln{\({L \over \phi R}\sin{{\phi \over 2}}+\sqrt{{L^2 \over \phi^2 R^2}\sin^2{{\phi \over 2}}+1}\)}}
For $\phi \simeq \pi$, ${{L \over \phi R}\sin{{\phi \over 2}}\simeq {L \over \phi R}={2 \rho \over 1- \rho^2}}$. For $\rho$ is close to 1, ie. the arc is taken from a large circle that almost covers the whole disc, ${2 \rho \over 1- \rho^2}\gg 1$. In this case, 
\eq{d_L \simeq 2R \ln{{2L \over \phi R}}}
Using \eqnref{eq:rC}, we obtain
\eq{c_{\text{eff}} \simeq d_L/\ln{L} \simeq 2R}
\end{document}